\documentclass{fairmeta}
\usepackage{amsmath,amssymb,mathtools,amsthm}
\usepackage{algorithm}
\usepackage[noend]{algpseudocode}
\usepackage{enumitem,siunitx,todonotes,xspace,pifont,array}
\setcitestyle{numbers,square}

\theoremstyle{plain}
\newtheorem{theorem}{Theorem}[section]
\newtheorem{proposition}[theorem]{Proposition}
\newtheorem{lemma}[theorem]{Lemma}
\newtheorem{corollary}[theorem]{Corollary}

\theoremstyle{definition}

\theoremstyle{plain}
\providecommand{\Description}[2][]{\ignorespaces}

\newcommand{\M}{\mathcal{M}}
\newcommand{\R}{\mathbb{R}}

\newcommand{\E}{\mathbb{E}}
\newcommand{\Prob}{\mathbb{P}}

\newcommand{\grad}{\operatorname{grad}}
\newcommand{\Hess}{\operatorname{Hess}}

\newcommand{\supp}{\operatorname{supp}}
\newcommand{\dist}{\operatorname{dist}}
\newcommand{\osc}{\operatorname{osc}}
\newcommand{\Vol}{\operatorname{Vol}}
\newcommand{\tr}{\operatorname{tr}}
\newcommand{\diag}{\operatorname{diag}}

\newcommand{\defeq}{\coloneqq}

\newcommand{\ouralg}{\ensuremath{\mathsf{PRISM}}\xspace}

\title{Riemannian Optimization on Unknown Manifolds via Stochastic Proximal Refinement}

\author[1,*]{Ruixiang Wu}
\author[2,*]{Jiahao Ai}
\author[1]{Tinko Sebastian Bartels}
\author[3]{Xuanzhuo Liu}
\author[1]{Yifei Wu}
\author[1]{Yu Mao}
\author[1]{Tongxin Li}
\author[1]{Costas Courcoubetis}
\affiliation[1]{The Chinese University of Hong Kong, Shenzhen}
\affiliation[2]{University of Pennsylvania}
\affiliation[3]{Stanford University}
\contribution[*]{These authors contributed equally to this research.}

\abstract{
We study Riemannian optimization on an unknown manifold accessed only through feasible samples. The central challenge is to turn feasible samples into optimization steps, \textit{i.e.,} the samples are drawn from the unknown feasible manifold, but do not directly provide the tangent spaces, projections, or retractions required by classic Riemannian methods. Moreover, regions with low objective values may be sparsely represented in the data, so directly sampling from a distribution fitted to these observations can require many draws to find better feasible solutions. We introduce \textbf{P}roximal \textbf{R}efinement on
\textbf{I}mplicit \textbf{S}mooth \textbf{M}anifolds~(\ouralg), which turns
feasible-sample access into an iterative optimization procedure by repeatedly
sampling an objective-aware proximal distribution conditioned on the current
iterate. Theoretically, we show that \ouralg recovers the effect of Riemannian-gradient refinement, leading to provable objective descent, feasibility~(manifold adherence), and convergence toward first-order stationarity. We demonstrate the effectiveness of our approach through synthetic experiments illustrating its geometric properties and applications to data-driven building control and diverse real world optimization tasks, highlighting its potential as a general framework for data-driven optimization across scientific and engineering domains.

\textit{Project Website:} \url{www.wuruixiang.me/prism}
}

\hypersetup{pdftitle={Riemannian Optimization on Unknown Manifolds via Stochastic Proximal Refinement},
  pdfauthor={Ruixiang Wu, Jiahao Ai, Tinko Sebastian Bartels, Xuanzhuo Liu, Yifei Wu, Yu Mao, Tongxin Li, Costas Courcoubetis}}

\begin{document}
\maketitle

\section{Introduction} \label{sec:intro}

We study optimization on an unknown manifold $\M$:
\begin{equation}
    \min_{x\in\M} f(x),
    \qquad \M\subset\R^D,
    \label{eq:problem}
\end{equation}
where the objective function $f : \R^D \to \R$ can be evaluated, but the optimizer accesses $\M$ only through a finite collection of feasible observations. No defining equations, parameterization, or geometric operation of the manifold are provided.

This setting is motivated by problems in which feasible data are easier to obtain than an explicit constraint model. Measurements, simulations, and previous system running logs can provide feasible examples even when the constraints governing them are unknown to the optimizer. The collected data points carry information beyond their objective values--they also reveal structure shared by feasible points. For example, image reconstruction uses examples of natural images to recover signals from incomplete measurements~\cite{pmlr-v70-bora17a}. In engineering inverse design, observed pairs of design parameters and physical responses provide information about their compatibility, enabling the search for improved designs~\cite{wu2024compositional}. Similarly, robot trajectory planning uses previously observed behavior to guide the search for trajectories that achieve a specified goal while respecting system dynamics~\cite{pmlr-v162-janner22a}. These applications motivate a common optimization setting: a performance criterion is available, while information about feasible solutions is only supplied through data points.
 
The same feasible set may support different tasks with different objectives. However, the available observations can only reflect how the data were collected and may not concentrate in regions favored by the current objective. Even regions containing optimal solutions may therefore be sparsely represented. Selecting an observed point with the best objective value provides a feasible candidate, but restricts the search to previously observed points. The challenge is to use the information in these observations to construct improved feasible points.

When the manifold is explicitly known, Riemannian optimization provides a principled way to the searching process~\cite{absil2008optimization,Boumal_2023}. For example, Riemannian gradient descent~(RGD) projects the objective gradient onto the tangent space and maps the resulting step back to the manifold through a retraction. These operations allow the method to decrease the objective while preserving feasibility and ensure its convergence to stationary points. However, in many real world scenarios, the optimizer can only receive feasible samples rather than an explicit description of the manifold. The tangent projections and retractions required by RGD are therefore unavailable, while an ordinary ambient gradient step may simply leave the manifold.

\begin{figure}[t]
    \centering
    \includegraphics[width=0.9\linewidth]{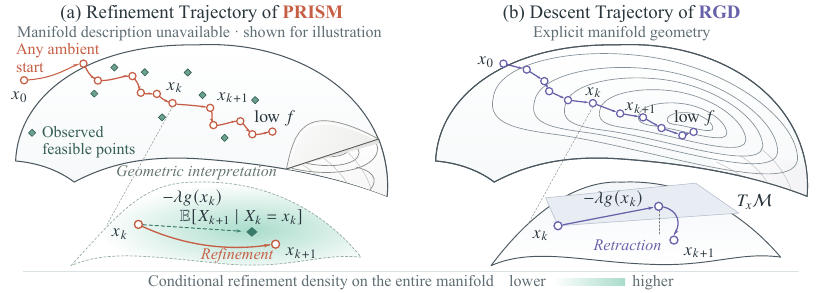}
    \caption{\textbf{Illustration of~\ouralg's iterative refinement on an unknown manifold.} Starting from an ambient point, \ouralg generates successive iterates through conditional sampling (left). Small diamonds represent feasible observations, while the peeled cover reveals the underlying manifold, whose description is unavailable to the optimizer. The lower inset illustrates a refinement distribution supported on the entire manifold, with color intensity indicating its density. \ouralg's mean displacement reveals a Riemannian gradient step. The right panel illustrates RGD's tangent step and retraction using explicitly available geometry for comparison.}
    \label{fig:demo}
\end{figure}

The challenge is to obtain comparable iterative optimization progress from this limited representation of the manifold. Learning from feasible observations can provide a way to propose new candidates, but those candidates need not remain on the manifold or improve the objective. Furthermore, each approximate update determines the starting point of the next, so its error can accumulate and affect subsequent optimization progress. A satisfactory framework must therefore connect the accuracy of its statistical approximation to both the feasibility of individual updates and the stationarity attained through repeated optimization steps. This motivates the following question:

\begin{quote}
\itshape
Can feasible samples support iterative optimization on an unknown manifold with guarantees comparable to those of classic Riemannian gradient descent?
\end{quote}

Generative approaches have demonstrated the potential of optimization from observed data~\cite{kong2024diffusion,krishnamoorthy2023diffusion}. Our focus is to connect conditional sampling to the feasibility and optimization progress of successive refinement steps.

\textbf{Contributions.} In this paper, we answer the question above in the affirmative by providing a novel optimization framework that uses consecutive refinement call to optimize the objective function on an unknown manifold, while preserving the key properties of RGD in expectation.

Our first contribution is \textbf{P}roximal \textbf{R}efinement on \textbf{I}mplicit \textbf{S}mooth \textbf{M}anifolds (\ouralg). At each iteration, the algorithm samples a new point from an objective-dependent proximal distribution anchored at the current iterate. The sampled point then becomes the anchor of the next refinement. This construction produces a sequence that gradually decreases the objective value without requiring explicit manifold equations, tangent projections, or retractions. Figure~\ref{fig:demo} illustrates the update mechanism of~\ouralg compared with classic RGD.

The recursive call of this kernel raises a further issue: each sample changes the point at which the next refinement kernel is initiated. Our second contribution establishes optimization guarantees for this recursion. Using RGD as a benchmark, we show that under our model assumptions~\ouralg admits a Riemannian-gradient drift, an expected-descent bound, and a stationarity guarantee. The results give a geometric interpretation of~\ouralg and are informally presented in the Table~\ref{tab:intro-rgd}. We further extend our analysis to manifolds with boundary, establishing guarantees of~\ouralg for constrained stationarity in Section~\ref{sec:with_boundary_theorems}.

\begin{table}[ht]
\centering
\small
\setlength{\tabcolsep}{4pt}
\renewcommand{\arraystretch}{1.65}
\caption{Informal Riemannian optimization properties of~\ouralg.}
\label{tab:intro-rgd}
\begin{tabular}{@{}>{\raggedright\arraybackslash}p{0.31\linewidth}p{0.24\linewidth}p{0.41\linewidth}@{}}
\toprule
Property & RGD & \ouralg \\
\midrule
Ambient displacement~(Thm~\ref{thm:mean-step})& $x^+ = x -\lambda g(x) +O(\lambda^2)$ & $\E \, X^+ - x=-\lambda g(x)+O(\lambda^2+\lambda\tau + \varepsilon)$ \\
One-step descent~(Thm~\ref{thm:descent}) & $f(x^+) \le f(x) -c\lambda\|g(x)\|^2$
 & $
 \E \, f (p^+)\le f (p)-c\lambda\|g(p)\|^2 +O(\lambda\tau+ \varepsilon^2/\lambda)$ \\
Stationarity~(Thm~\ref{thm:finite-time})
 & $\|g(x_\infty)\| = 0$
 & $ \E\, \|g(p_\infty)\| = O(\varepsilon / \lambda)$ \\
\bottomrule
\end{tabular}
\par\smallskip
\parbox{0.98\linewidth}{\tiny
Here $g=\operatorname{grad}_{\M}f$ and $\lambda$ denote the Riemannian gradient and the stepsize respectively. $\tau$ and $\varepsilon$ capture kernel's temperature and precision, which will be specified in Section~\ref{sec:method}. Let $p = P(x)$, $p_k=P(X_k)$, where $P$ denotes a nearest point projection onto $\M$. These feasible representatives are used only in the analysis. Our algorithms cannot access $P$ during execution. In the one-step comparisons, the current point $x$ is fixed,
and $\E X^+$ averages only over the next sampled point. In iterative notation, this corresponds to the conditional mean
$\E[X_{k+1}\mid X_k=x]$ for the same sampling law and parameters.}
\end{table}

Finally, in Section~\ref{sec:experiments}, we demonstrate the effectiveness of~\ouralg through three case studies: a Grassmann optimization problem, building energy control, and six data-driven optimization tasks from Design-Bench~\cite{trabucco2022designbench}. The Grassmann optimization experiment illustrates RGD-like refinement, manifold adherence, and stationarity guarantee consistent with our theoretical results. The building control task further illustrate the practical utility and broad applicability of our framework across real world scenario. In the final study, comparisons with nine optimization baselines yield the best observed average rank for~\ouralg under a common final design budget. Together, these results highlight the advantages of our refinement framework when feasible structure must be inferred from data and its practical applicability across scientific and engineering domains.

\textbf{Related work.} Our work connects optimization from observed data with geometric and sampling-based methods for an iterative optimization process.

\noindent \textit{Data-driven and sampling-based optimization.} Offline optimization uses observed decision points and their objective values to search for the next candidate point. Conservative objective models reduce the overestimation that can arise when optimizing learned surrogates beyond the observed data~\cite{trabucco2021conservative}. Generative approaches incorporate the structure of observed designs into candidate generation: CbAS adapts a sampling distribution toward desired properties~\cite{brookes2019conditioning}, while DDOM learns a diffusion model conditioned on objective values~\cite{krishnamoorthy2023diffusion}. These approaches connect objective improvement with statistical information about candidates. Closely related to our setting, DiffOPT combines a learned data distribution with an
objective-dependent Boltzmann factor to optimize under unknown feasibility constraints~\cite{kong2024diffusion}. Building on the idea of using learned distributions to guide optimization,~\ouralg develops a recursive proximal refinement rule whose progress can be characterized geometrically.

\noindent \textit{Proximal methods in optimization:} The proximal construction is also related to sampling methods for approximating optimization operations. HJ-Prox approximates proximal points through weighted Monte Carlo
averages~\cite{osher2023hamilton}, and subsequent work establishes convergence guarantees for operator-splitting methods using these approximations~\cite{di2026operator}. These results connect sampled proximal approximations to iterative optimization. \ouralg studies a proximal refinement kernel supported on an unknown manifold and uses a draw from that kernel as the next iterate.

\noindent \textit{Learning to optimize.} Learning to optimize uses experience with optimization problems to construct corresponding iterative solvers~\cite{chen2022learning}. Representative approaches train
recurrent networks to generate updates from gradient information~\cite{andrychowicz2016learning} or formulate optimizer design as policy learning~\cite{li2017learning}. These methods demonstrate how learned structure can guide successive optimization steps. \ouralg incorporates such structure through the distribution of feasible observations within a prescribed proximal refinement rule. This formulation provides a mathematical target for the proximal refinement kernel and connects its accuracy to the progress of the resulting optimization sequence.

\noindent \textit{Optimization on learned manifolds.} Riemannian optimization supplies the geometric foundation for our analysis. Classical methods use tangent projections and retractions to obtain descent and stationarity
guarantees~\cite{absil2008optimization,Boumal_2023}, while statistical manifold estimation quantifies the recovery of manifolds and tangent spaces from observations~\cite{10.1214/18-AOS1685}. Score-based landing and denoising Riemannian gradient methods connect these perspectives by approximating manifold operations through learned denoisers and their derivatives~\cite{kharitenko2026landing}. Their feasibility and stationarity guarantees require initialization within a tubular neighborhood of the manifold. \ouralg obtains related guarantees through conditional sampling without explicitly reconstructing these geometric operators. During the execution of our framework, the initial point can be chosen anywhere in the ambient space. This provides greater flexibility in initialization, and does not require a starting point that already lie near the manifold.
\section{Problem Formulation and Motivating Applications} \label{sec:problem}

In this section, we formulate optimization over an unknown feasible manifold and state the geometric and sampling assumptions used in our analysis. We then present motivating applications and recall classical Riemannian gradient descent as a benchmark for the optimization properties of~\ouralg.

\subsection{Problem Formulation} \label{sec:problem:formulation}

\noindent\textit{Notation conventions.} Throughout this paper, $\|\cdot\|$ and $\langle\cdot,\cdot\rangle$ denote the Euclidean norm and inner product, respectively. For a nonempty set $A\subseteq\R^D$, we write $\dist(x,A)\defeq\inf_{y\in A}\|x-y\|$ for the ambient Euclidean distance from $x$ to $A$. We write $\nu_n\Rightarrow\nu$ for weak convergence of probability measures.

We now specify the information available for the optimizer to solve the problem in~\eqref{eq:problem}. The objective $f: \R^D \to \R$ can be evaluated, but the feasible set $\M$ is unknown. The optimizer receives a finite collection of observations drawn independently from a fixed but unknown probability distribution $\mu$.
\begin{equation*}
    \mathcal D_n
    =
    \{Y^{(1)},\ldots,Y^{(n)}\},
    \qquad
    Y^{(i)}\overset{\mathrm{iid}}{\sim}\mu.
\end{equation*}
The distribution $\mu$ describes how feasible observations are generated. The optimizer only receives $\mathcal{D}_n$ but does not know $\mu$ and has no membership oracle, tangent projector, retraction, or parameterization of the manifold. We first state basic conditions on the optimization problem and the population distribution.
\newtheorem{assumption}[theorem]{Assumption}
\begin{assumption}\label{ass:regularity:basic}
The set $\M\subset\R^D$ is nonempty and compact. The objective function $f$ is continuous on $\M$, and $\mu$ is a Borel probability measure with $\supp \mu=\M$. move $\mu$ to problem setting.
\end{assumption}

Compactness and continuity ensure that problem~\eqref{eq:problem} admits a minimizer. The support condition connects the observation model to the feasible set, which ensures every neighborhood of every feasible point receives positive probability mass. To characterize optimization progress through Riemannian geometry, we impose additional smoothness conditions on $\M$, $f$, and the sampling density.

\begin{assumption}\label{ass:regularity:geometry}
The set $\M \subseteq \R^D$ is an embedded $C^3$ submanifold of dimension $m\geq1$, without boundary\footnote{\tiny This means that we consider a manifold $\M$ with no outer edges or borders. A compact manifold without boundary is called a \textit{closed} manifold~\cite{spivak1970comprehensive}.}.
\end{assumption}

This assumption provides the tangent spaces and local geometric expansions used in our analysis. Compact smooth manifolds cover many modern optimization scenarios, including spheres, Stiefel and orthogonal manifolds, Grassmann and projective manifolds, and more general flag manifolds~\cite{absil2008optimization, Boumal_2023, lee2018introduction, ye2022optimization}. These geometries are relevant in synchronization, matrix factorization, and structured representation learning~\cite{liu2023resync, boumal2015low, cherian2020representation}. They also arise in contemporary machine learning areas: recent parameter-efficient fine-tuning methods for large language models, for example, optimize orthogonal LoRA factors directly on the Stiefel manifold and exploit orthogonal subspaces to structure adapter updates~\cite{park2025riemannian, park2025riemannianlora, park2026acceleration, chang-etal-2026-sos}. These examples are $C^\infty$ and therefore satisfy the required $C^3$ regularity. In Section~\ref{sec:with_boundary_theorems} we further extend our analysis to manifolds with boundary.

Smooth geometry alone does not control how the objective or the sampling mass varies near a feasible point. The local analysis therefore requires regularity of both $f$ and $\mu$.

\begin{assumption}\label{ass:regularity:smooth}
In addition to the geometric regularity in Assumption~\ref{ass:regularity:geometry}, we assume that $f$ is $C^2$ on an open neighborhood of $\M$, and $\mu$ has a strictly positive $C^1$ density with respect to the induced Riemannian volume measure $\Vol_{\M}$:
\begin{equation*}
        \mu(dy)=\rho(y)\Vol_{\M}(dy),
    \qquad
    \rho\in C^1(\M),
    \qquad
    \rho(y)>0
    \quad\text{for all }y\in\M.
\end{equation*}
\end{assumption}

Twice differentiability of $f$ permits local expansions of the objective along an optimization step. Strict positivity of $\rho$ rules out zero sampling density at feasible points, while its $C^1$ regularity controls local variation of probability mass. These conditions allow the analysis to relate the distribution of nearby observations to the local geometry and objective. Intrinsic sampling densities bounded above and away from zero are standard in statistical manifold estimation~\cite{10.1214/18-AOS1685}. Smooth density assumptions also appear in analyses of optimization over learned manifolds~\cite{kharitenko2026landing}. Since $\mu$ is a probability measure, $ \int_{\M}\rho\,d\Vol_{\M}=1$. Moreover, compactness of $\M$, together with continuity and strict positivity of $\rho$, implies the existence of constants $\rho_{\min},\rho_{\max}>0$ such that
\begin{equation*}
        0<\rho_{\min}
    \leq \rho(y)
    \leq \rho_{\max}
    <\infty,
    \qquad y\in\M.
\end{equation*}
These bounds above provide uniform control of local sampling mass. The assumptions on $\M$ and $\mu$ are imposed only for the theoretical analysis, the optimizer is not supplied with an explicit description of $\M$, knowledge of $\mu$, or the associated regularity constants.
Its only access to $\M$ is through the observations in $\mathcal D_n$.

\subsection{Motivating Applications} \label{sec:problem:applications}

In this section, we present several applications that motivate our problem design. For each application, we specify the objective and the underlying manifold $\M$.

\noindent\textbf{Control and planning with unknown dynamics.} A system behavior manifold describes the action--state trajectories that a dynamical system can realize~\cite{polderman1998introduction}. Within a fixed environment, each point on this manifold represents an action sequence together with its resulting state trajectory. This perspective provides a natural connection between planning and optimization over a manifold. Fix an environment configuration $c$ and an initial state $x_0$, let $U=(u_0,\ldots,u_{H-1})$ and $X=(x_1,\ldots,x_H)$ denote an action sequence and its resulting state observation. Their compatibility can be expressed through an unknown response map $G_c$: $\M_c \defeq \{(U,X):U\in\mathcal U,\ X=G_c(U)\}$. In model-predictive control~(MPC), each planning round needs to optimize a finite-horizon action sequence by evaluating its predicted consequences. This process requires a dynamical model and repeated evaluations of candidate trajectories. When the system dynamics are unknown or computationally demanding, obtaining or frequently calling the dynamical model can be difficult. Nevertheless, recorded system interactions can provide observations of feasible action--state trajectories. Data manifold hypothesis suggests that many real-world high-dimensional observations concentrate approximately on a manifold with much lower intrinsic dimension~\cite{fefferman2016testing, loaiza2024deep}. This motivates solving each MPC subproblem directly over the (unknown) system behavior manifold $\M_c$. Therefore, the planning problem can be formulated into the optimization problem below.
\begin{equation*}
    \min_{(U,X)\in\M_c} f(U,X),
    \qquad
    f(U,X)
    \defeq
    \sum_{t=0}^{H-1}\ell_t(x_t,u_t)
    +
    \ell_{\mathrm T}(x_H),
\end{equation*}

\noindent \textbf{Image reconstruction from incomplete measurements.} Consider reconstructing an image from measurements $b=Ax^\star+\xi$, where $A$ is a known sensing operator and $\xi$ represents measurement noise. This setting includes compressed sensing and reconstruction from missing pixels. Since there might be several reconstructed images that agree with the measurements, successful reconstruction requires information about the structure of images in the target domain. Learned generative priors provide one way to incorporate
such information~\cite{pmlr-v70-bora17a}. Let $\mathcal M_{\mathrm{img}}\subset\mathbb R^D$ denote the unknown image manifold. The reconstruction problem seeks to find $x\in \M_{\rm img}$ that minimizes $f_{A, b}(x) = \|Ax - b\|^2$. The objective function is accessible, while a collection of clean images provides observations of the feasible image structure.

\noindent\textbf{Counterfactual explanations.} A counterfactual explanation describes how an input could be modified to obtain a different prediction from a trained classifier. The aim is to find a nearby plausible alternative that helps explain the model's decision. However, finding a small change alone is insufficient, modifying individual features without accounting for their relationships can produce an unrealistic input. This motivates restricting the search to inputs consistent with the structure of observed data~\cite{pmlr-v238-tsiourvas24a}. Let $\M_{\mathrm{data}}\subset\mathbb R^D$ denote the unknown manifold of plausible inputs. Given an original input $x_0$, let
$p_{\mathrm{tar}}(x)$ be the probability that the classifier assigns to the desired target class. We can formulate the search as minimizing $f(x) = -\log p_{\mathrm{tar}}(x) + \alpha\|x-x_0\|^2$ over $x\in\M_{\mathrm{data}}$.

The motivating examples above illustrate the potential of~\ouralg across diverse tasks. These applications share a common structure in which the objective function is explicit and can be evaluated, while only finite observations provide information about the feasible set. Related numerical examples appear in our \hyperref[sec:experiment1]{optimization over a Grassmann manifold (Section~\ref*{sec:experiment1})}, \hyperref[sec:experiment_building_control]{building energy control (Section~\ref*{sec:experiment_building_control})} and \hyperref[sec:experiment2]{offline design study using a learned objective (Section~\ref*{sec:experiment2})}. We next recall the classic Riemannian gradient descent method to specify the optimization properties that an iterative refinement procedure aims to achieve.

\subsection{Classic Riemannian Gradient Descent as an Optimization Benchmark} \label{sec:problem:geometry}

To formally characterize optimization progress of~\ouralg, we use classic RGD as a benchmark. When the manifold is explicitly known, RGD produces an optimization trajectory that follows a descent direction, decreases the objective, and approaches first-order stationarity. In this section, we recall how these properties arise, and use them as reference for the optimization progress of~\ouralg.

For each $x\in\M$, let $T_x\M$ denote the tangent space of $\M$ at $x$, and let $\Pi_x:\R^D\rightarrow T_x\M$ denote the orthogonal projection onto the tangent space. Since $f$ is $C^2$ on a neighborhood of $\M$, its Riemannian gradient on $\M$ is
\begin{equation} \label{eq:riemannian-gradient}
    g(x) \defeq \grad_{\M} f(x) = \Pi_x \nabla f(x),
    \qquad x\in\M.
\end{equation}
Whenever $g(x) \neq 0$, moving in the direction $-g(x)$ with suitable stepsize $\lambda$ decreases the objective value to first order. However, a tangent step may not remain on a curved manifold. A retraction $R_x$ maps a tangent displacement back to $\M$ while preserving its first order effect. Therefore, classic RGD takes the form:
\begin{equation} \label{eq:rgd-reference}
    x_{k+1}
    =
    R_{x_k}\!\left(-\lambda_k g(x_k)\right),
    \qquad
    \lambda_k>0.
\end{equation}
The Riemannian gradient determines the descent direction, and the retraction ensures feasibility. These operations also quantify an optimization progress. Under Assumption~\ref{ass:regularity:smooth}, with a smooth retraction and sufficiently small stepsize, RGD produces the next iterate satisfies:
\begin{equation*}
    x_{k + 1} = x_k - \lambda g(x_k) + O(\lambda^2)  \text{ and } f(x_{k + 1}) \leq f(x_k) - c \lambda \|g(x_k) \|^2,
\end{equation*}
for a constant $c > 0$~\cite{absil2008optimization, Boumal_2023}. The first relation captures the direction of the update, while the second ensures objective decrease whenever the Riemannian gradient is positive. For a sufficiently small fixed stepsize, summing the second relation over iteration $k$ and using the lower bound of $f$ gives $\|g(x_k)\| \to 0$. Therefore, first-order stationarity on a manifold without boundary is captured by
\begin{equation}
    g(x)=0.
    \label{eq:interior-stationarity}
\end{equation}

For the extension to manifold with boundary, the stationarity condition needs to take directions that remain feasible into consideration. Let $\mathcal T_{\M}(x)$ denote the contingent tangent cone of feasible
first-order directions at $x$. The corresponding constrained first-order condition is
\begin{equation}
    \langle \nabla f(x),v\rangle \ge 0,
    \qquad
    \forall\,v\in\mathcal T_{\M}(x).
    \label{eq:boundary-stationarity}
\end{equation}
At an interior point, $\mathcal T_{\M}(x)=T_x\M$, so \eqref{eq:boundary-stationarity} reduces to \eqref{eq:interior-stationarity}. We use this constrained version of stationarity in Section~\ref{sec:with_boundary_theorems}.

In our setting, the tangent projections and the retractions required by~\eqref{eq:rgd-reference} are unavailable. In the next section, we present our optimization framework~\ouralg, which recovers the properties possessed by classic RGD analogously without accessing explicit information about the manifold.

\section{Proximal Refinement on Implicit Smooth Manifolds} \label{sec:method}

\begin{figure}[t]
    \centering
    \captionsetup{skip=4pt}
    \includegraphics[width=\linewidth]{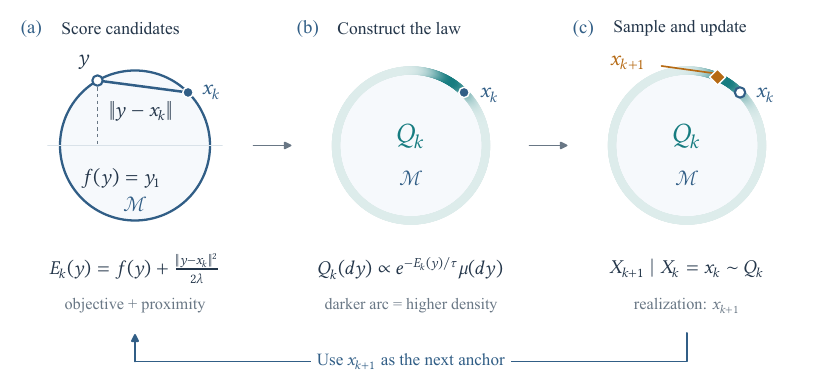}
    \caption{\textbf{Ideal refinement on the unit circle.}
    (a) Score with $\smash{E_k=E^f_{\lambda,X_k}}$; (b) Reweight $\mu$ to obtain $\smash{Q_k=Q^f_{\lambda,\tau}(x_k,\cdot)}$; (c) Sample $X_{k+1}\mid X_k=x_k\sim Q_k$ and use its
realization $x_{k+1}$ as the next anchor, at which the energy
and conditional law are recomputed.
Panels (b) and (c) display the same density of $X_{k+1}$, with
darker arcs indicating higher density.
Here $\mu$ is uniform, $f(y)=y_1$.}
    \label{fig:refinement-mechanism}
\end{figure}

In this section, we present \ouralg, a framework for iteratively minimizing $f$ over an unknown manifold $\M$, whose stochastic updates recover the first-order effect of the Riemannian gradient update~\eqref{eq:rgd-reference}. Section~\ref{sec:method_energy} motivates a proximal criterion whose local behavior is connected to the Riemannian gradient update~\eqref{eq:rgd-reference}. Section~\ref{sec:method_kernel} then turns this deterministic criterion
into a stochastic refinement law defined through the distribution of feasible
observations. Finally, Section~\ref{sec:method_alg} applies this law recursively, using each sampled point to define the next refinement step. The resulting framework~(Algorithm~\ref{alg:prism}) is designed to recover Riemannian first-order
refinement without explicit access to $\M$ or its geometric operators.
Figure~\ref{fig:refinement-mechanism} illustrates these three stages for an ideal sampler.

\subsection{A Proximal Criterion for Iterative Refinement} \label{sec:method_energy}

The tangent displacement underlying the Riemannian gradient update~\eqref{eq:rgd-reference} admits an equivalent characterization:
\begin{equation} \label{eq:tangent-step-model}
    v_\lambda(x)
    \defeq
    \operatorname*{arg\,min}_{v\in T_x\M}
    \left\{
        \underbrace{f(x) + \langle g(x),v\rangle}_{\text{first-order expansion}}+\frac{\|v\|^2}{2\lambda}
    \right\}
    =-\lambda g(x),
    \qquad \lambda>0,
\end{equation}
which makes the role of the quadratic term explicit: it controls the magnitude of the update, with $\lambda$ playing the role of a stepsize. However, evaluating~\eqref{eq:tangent-step-model} requires the Riemannian gradient and tangent space, neither of which is available in our setting. We therefore seek an analogous criterion that can be expressed directly in terms of candidate points and objective values. Motivated by~\eqref{eq:tangent-step-model}, we retain the quadratic
displacement penalty but replace the first-order expansion term with the actual objective evaluated at feasible candidates. For any point $x\in\R^D$, define the \textit{proximal energy}
\begin{equation}
    E_{\lambda,x}^{f}(y)
    \mathrel{:=}
    f(y)+\frac{\|y-x\|^2}{2\lambda},
    \qquad y\in\M,
    \label{eq:prox-energy}
\end{equation}
and the corresponding \textit{proximal minimizing set}
\begin{equation}
    \mathcal P_\lambda^f(x)
    \mathrel{:=}
    \operatorname*{arg\,min}_{y\in\M}E_{\lambda,x}^{f}(y).
    \label{eq:prox-map}
\end{equation}
At iteration $k$, $x=X_k$ represents the current iterate and $y\in\M$ is a candidate for the next iterate. When the current iterate is feasible, minimizing the \textit{proximal energy} also has a descent interpretation like~\eqref{eq:tangent-step-model}. Since $y=x$ is admissible in~\eqref{eq:prox-map}, every $y^+\in\mathcal P_\lambda^f(x)$ satisfies
\begin{equation*}
    f(y^+)
    \le
    f(x)-\frac{\|y^+-x\|^2}{2\lambda}
    \le f(x),
    \qquad x\in\M.
\end{equation*}
Hence an exact proximal update cannot increase the objective, and a nontrivial displacement must be accompanied by a corresponding objective decrease. Compactness and continuity ensure that $\mathcal P_\lambda^f(x)$ is nonempty for every $x\in\R^D$, although the minimizer need not be unique.

\subsection{From Proximal Minimization to a Stochastic Refinement Law} \label{sec:method_kernel}

The deterministic criterion in~\eqref{eq:prox-map} describes the update we would like to make, but cannot be evaluated directly because $\M$ is unknown. Restricting the candidate set to \(\mathcal D_n\) allows selection among observed points, but cannot generate new candidates or achieve an objective value below the best observed value. Instead, we use the data-generating distribution $\mu$ as a statistical description of the unknown feasible manifold. Although $\mu$ itself is explicitly unknown, the samples in $\mathcal D_n$ are drawn from it and therefore provide information about which regions of the ambient space are feasible. We seek a distribution $\nu$ for the next iterate that favors low proximal energy while
remaining close to this feasible-data distribution. This leads to the regularized problem
\begin{equation}
    \min_{\nu\in\mathcal P(\M)}
    \left\{
        \int_{\M}E_{\lambda,x}^{f}(y)\,\nu(dy)
        +\tau\operatorname{KL}(\nu\|\mu)
    \right\},
    \qquad \tau>0,
    \label{eq:variational}
\end{equation}
following the classical belief updating framework~\cite{bissiri2016general}. The first term favors candidates that improve the proximal criterion, while the second discourages the update from moving probability too aggressively away from regions represented by $\mu$. The parameter $\tau$ controls this balance, smaller $\tau$ places greater emphasis on proximal improvement, whereas larger $\tau$ keeps the update more strongly tied to the feasible-data distribution. The optimization problem~\eqref{eq:variational} has the unique solution
\begin{equation} \label{eq:kernel}
    Q_{\lambda,\tau}^{f}(x,dy)
    \defeq
    \frac{\exp[-E_{\lambda,x}^{f}(y)/\tau]}
         {Z_{\lambda,\tau}^{f}(x)}\,\mu(dy),
    \text{ where }
    Z_{\lambda,\tau}^{f}(x)
    \defeq
    \int_{\M}
    \exp[-E_{\lambda,x}^{f}(z)/\tau]\,\mu(dz).
\end{equation}
Hence $Q_{\lambda,\tau}^{f}$ reweights the feasible-data distribution $\mu$ according to the current \textit{proximal energy}. Candidates with smaller $E_{\lambda,x}^{f}$ receive greater probability, while the reference law $\mu$ retains the feasible structure learned from the observations. Proposition~\ref{prop:landing} formally establishes this variational
characterization and shows that~\eqref{eq:kernel} is well defined for every $x\in\R^D$ and has support exactly $\M$. Although $\mu$ is unknown, a range of well-studied methods is available to approximate the refinement law from $\mathcal{D}_n$. Established tools for kernel density estimation~\cite{parzen1962estimation}, score-based generative modeling~\cite{song2021sde}, and Monte Carlo sampling~\cite{hastings1970montecarlo} provide practical implementations to estimate~\eqref{eq:kernel}. We provide examples of these approaches in Appendix~\ref{app:statistical} and~\ref{app:hatQ_implementations}. Next, we provide some essential properties of the refinement kernel $Q^f_{\lambda, \tau}$.

The temperature $\tau$ determines how closely the stochastic refinement
approaches the deterministic proximal update. For every fixed $x\in\R^D$ and $\lambda>0$, if $Y_\tau\sim Q_{\lambda,\tau}^{f}(x,\cdot)$, Proposition~\ref{prop:proximal-concentration} shows that
\begin{equation*}
    \operatorname{dist}
    \bigl(Y_\tau,\mathcal P_\lambda^f(x)\bigr)
    \xrightarrow[\tau\downarrow0]{\mathbb P}0.
    \label{eq:prism-concentration-main}
\end{equation*}
Thus, as the temperature decreases, an ideal sample is more likely to lie arbitrarily close to an exact proximal minimizer. This result identifies the set which $Q_{\lambda,\tau}^{f}$ converges to, but it does not specify how probability is distributed when this set contains multiple minimizers. To characterize this allocation, consider the case where $\mathcal P_\lambda^f(x)=\{y_1,\ldots,y_J\}$ consists of finitely many isolated, nondegenerate proximal minimizers. At each $y_j$, the Riemannian Hessian of the \textit{proximal energy} is positive semidefinite, nondegeneracy rules out zero-curvature tangent directions, and hence  $H_j \defeq \operatorname{Hess}_{\M} E_{\lambda,x}^{f}(y_j)\succ 0$. Then Proposition~\ref{prop:branches} shows that as $\tau \downarrow 0$
\begin{equation*}
    Q_{\lambda,\tau}^{f}(x,\cdot)
    \Rightarrow
    \sum_{j=1}^{J}\omega_j\delta_{y_j}, \text{where }
    \omega_j = 
    \frac{\rho(y_j)(\det H_j)^{-1/2}}
     {\sum_{\ell=1}^{J}
     \rho(y_\ell)(\det H_\ell)^{-1/2}}.
\end{equation*}
Hence the stochastic refinement law can preserve multiple feasible solutions while using both their statistical representation under $\mu$ and their local curvature to determine how frequently they are selected. These properties characterize a single refinement step. To turn such objective-aware feasible selections into an optimization procedure, \ouralg repeats this construction: each sampled point becomes the current iterate, and
the proximal refinement law is recomputed relative to that new point. We formalize this iterative procedure in the next section.

\subsection{Recursive Proximal Refinement} \label{sec:method_alg}

The refinement law in~\eqref{eq:kernel} specifies a one-step update for a
given current iterate. \ouralg\ turns this local rule into an optimization
procedure by repeatedly using the sampled point as the current iterate for
the next refinement step. At iteration $k$, its ideal recursion is
\begin{equation*}
    X_{k+1}\mid X_k
    \sim
    Q_{\lambda_k,\tau_k}^{f}(X_k,\cdot).
    \label{eq:ideal-update}
\end{equation*}
Thus, while the feasible-data distribution $\mu$ remains fixed, the \textit{proximal energy} is recomputed at the newly sampled iterate after every update. The resulting sequence is therefore adaptive to its optimization trajectory rather than repeatedly sampling from a fixed objective-weighted distribution~(see~\cite{kong2024diffusion, krishnamoorthy2023diffusion}). 

An implementation replaces $Q$ with a conditional sampler $\widehat Q$ constructed from $\mathcal D_n$. Once the data and fitted sampler are fixed, each call generates
\begin{equation}
    X_{k+1}\mid X_k
    \sim
    \widehat Q_{\lambda_k,\tau_k}^{f}(X_k,\cdot).
    \label{eq:implemented-recursion}
\end{equation}
Algorithm~\ref{alg:prism} summarizes this implemented iteration.

\begin{algorithm}[t]
\caption{\ouralg: Stochastic Proximal Refinement}
\label{alg:prism}
\begin{algorithmic}[1]
\Require Conditional sampler $\widehat Q$ constructed from $\mathcal D_n$, ambient initialization $X_0\in\R^D$, refinement budget $N\ge1$,
positive parameter pairs $(\lambda_k,\tau_k)_{k=0}^{N}$
\State Draw $X_1\sim \widehat Q_{\lambda_0,\tau_0}^{f}(X_0,\cdot)$
\For{$k=1,\ldots,N$}
    \State Draw
    $X_{k+1}\sim \widehat Q_{\lambda_k,\tau_k}^{f}(X_k,\cdot)$
\EndFor
\State \Return $X_{N+1}$, optionally together with
$(X_1,\ldots,X_{N+1})$
\end{algorithmic}
\end{algorithm}

The next section studies how the resulting stochastic updates recover Riemannian first-order refinement and establishes their descent and stationarity properties.

\paragraph{Use of AI tools.}
AI tools were used to polish the writing, assist with proof checking, and improve figure clarity and presentation.

\section{Optimization Guarantees} \label{sec:main_results}

This section theoretically establishes how the sampling updates of~\ouralg lead to Riemannian optimization progress. We first show that the execution of \ouralg recovers Riemannian-gradient steps in expectation, and establish descent and stationarity guarantees. In Section~\ref{sec:riemannian-descent}, and~\ref{sec:stationarity-guarantees}, we consider the regime, where the manifold $\M$ is without boundary, and then extend the stationarity analysis to manifolds with boundary in Section~\ref{sec:with_boundary_theorems}.

As specified in Section~\ref{sec:method_alg}, Algorithm~\ref{alg:prism} calls an approximate conditional sampler $\widehat Q$. Its discrepancy from the ideal refinement kernel $Q$ can  affect both feasibility and optimization progress. We measure this discrepancy through the following accuracy requirement.

\begin{assumption}\label{ass:oracle}
There exists an error $\varepsilon \ge 0$ such that, for every call $k\ge0$ and every input $x$ of the
implemented iteration,
\begin{equation}\label{eq:oracle}
 W_2\!\left(
   \widehat Q^f_{\lambda_k,\tau_k}(x,\cdot),
   Q^f_{\lambda_k,\tau_k}(x,\cdot)
 \right)\le \varepsilon,
\end{equation}
where $W_2$ denotes the $2$-Wasserstein distance induced by the
ambient Euclidean distance.
\end{assumption}
This assumption is attainable in practice, Appendix~\ref{app:statistical} provides a concrete example that an estimated kernel constructed from the dataset $\mathcal{D}_n$ satisfies~\eqref{eq:oracle}. Since the ideal kernel $Q$ produces feasible points~(see Proposition~\ref{prop:landing}), controlling its approximation error also controls the feasibility of the samples from $\widehat Q$. The following proposition makes this connection precise.

\begin{proposition} \label{prop:feasibility}
Under Assumption~\ref{ass:oracle}, for any $x\in\R^D$ and $\lambda,\tau>0$, let $\widehat Y\sim\widehat Q^f_{\lambda,\tau}(x,\cdot)$. Then
\begin{equation*}
 \E \dist(\widehat Y,\M)^2\le \varepsilon^2.
\end{equation*}
\end{proposition}

The proposition links conditional-sampling accuracy to feasibility at every iteration. We next turn to optimization properties of~\ouralg, beginning with the relation between the mean sample displacement and the Riemannian gradient.

\subsection{Riemannian Descent from Conditional Samples} \label{sec:riemannian-descent}

To understand how conditional sampling produces optimization progress, in the following Theorem~\ref{thm:mean-step} we compare the updates in~\eqref{eq:implemented-recursion} with the Riemannian-gradient step in~\eqref{eq:rgd-reference}.

\begin{theorem}\label{thm:mean-step}
Under Assumptions~\ref{ass:regularity:basic},~\ref{ass:regularity:geometry},~\ref{ass:regularity:smooth}, and~\ref{ass:oracle}, let $\widehat Y\sim\widehat Q^f_{\lambda,\tau}(x,\cdot)$, there exist $\lambda_{\rm loc}\in(0,1]$ and a constant $C<\infty$ such that, for every $x\in\M$, $0<\lambda\le\lambda_{\rm loc}$, $0<\tau\le\bar\tau$,  the mean displacement can be written as
\begin{equation*}
 \E \, \widehat Y = x - \lambda g(x)+b,
 \qquad \|b\|\le C\lambda(\lambda+\tau)+\varepsilon,
\end{equation*}
where $g(x)$ is the Riemannian gradient defined in~\eqref{eq:riemannian-gradient}. Moreover,
\begin{equation}\label{eq:response-second-moment}
 \E\|\widehat Y-x\|^2 \le C\left(\lambda^2\|g(x)\|^2+\lambda\tau+\varepsilon^2\right).
\end{equation}
\end{theorem}

The same sufficiently small bound $\lambda_{\rm loc}$ is used in Theorems~\ref{thm:mean-step}--\ref{thm:finite-time}; its construction is given in Appendix~\ref{app:loc-local-analysis}.

Theorem~\ref{thm:mean-step} shows that~\ouralg approximates a Riemannian-gradient step in expectation without explicitly computing the gradient or tangent projector. The quadratic term in~\eqref{eq:prox-energy} controls the step size, while~\eqref{eq:response-second-moment} bounds the mean-square displacement, including the contributions of temperature $\tau$ and sampling error $\varepsilon$. At a stationary point,~\eqref{eq:response-second-moment} reduces to $\E\|\widehat Y-x\|^2\le C(\lambda\tau+\varepsilon^2)$, which motivates cooling the temperature $\tau$ to reduce motion near stationary points.

We next extend these displacement estimates into an objective-descent guarantee for the iteration in Algorithm~\ref{alg:prism}. To express progress in terms of the feasible objective, fix a Borel measurable nearest-point selection $P:\R^D\to\M$ and define
\begin{equation} \label{eq:near_point_projection}
    P(z)\in\operatorname*{arg\,min}_{y\in\M}\|z-y\|,
    \qquad p_k=P(X_k).
\end{equation}
Such a selection exists by Lemma~\ref{loc:lem:geometry}, and the nearest point is unique in a sufficiently small neighborhood of $\M$. The following theorem establishes expected descent produced by Algorithm~\ref{alg:prism}.

\begin{theorem} \label{thm:descent}
Under the conditions of Theorem~\ref{thm:mean-step}, there exist $r>0$ and constants $C_1, C_2 > 0$ such that the nearest-point projection $P$ is unique on $ \{x\in\R^D:\dist(x,\M)<r\}$. For any $x$ in this neighborhood,
$0<\lambda\le\lambda_{\rm loc}$, and
$0<\tau\le\bar\tau$, suppose
$\widehat Y\sim\widehat Q^f_{\lambda,\tau}(x,\cdot)$. Writing $p=P(x)$, we have
\begin{equation}\label{eq:local-descent}
 \E f(P(\widehat Y))
 \le f(p)-C_1\lambda\|g(p)\|^2
          +C_2\left(\lambda\tau+\frac{\varepsilon^2}{\lambda}\right).
\end{equation}
\end{theorem}

In~\eqref{eq:local-descent}, the negative term $-C_1 \lambda \|g(p)\|^2$ measures the optimization progress, while temperature $\tau$ and sampling error $\varepsilon$ determine its resolution. In particular, expected descent is certified when $\|g(p)\|^2$ exceeds a constant multiple of $\tau+(\varepsilon/\lambda)^2$. Together, Theorems~\ref{thm:mean-step} and~\ref{thm:descent} establish the local analogy with Riemannian gradient descent: the stochastic refinement has a negative-gradient drift in expectation and produces corresponding expected objective decrease without explicitly computing the manifold gradient or projection. We next study what repeated application of these local updates implies for first-order stationarity.

\subsection{Convergence to Stationarity} \label{sec:stationarity-guarantees}

In this section, we extend the one-step descent to the full optimization trajectory analysis. Summing it over successive refinement calls, the intermediate objective terms cancel, leaving only the difference between the initial and final objective values. Since $\M$ is compact and $f$ is continuous, the minimum $f_\star$ is attained, so this total objective decrease is bounded by $\Delta_1=\E f(p_1)-f_\star$. Consequently, the accumulated squared Riemannian gradients are also controlled by $\Delta_1$. This yields a stationarity guarantee analogous to classic Riemannian gradient descent.

The first refinement call is treated separately since the initial point $X_0$ may be any point in the ambient space. After the first call, we analyze the feasible representatives $p_k=P(X_k)$ introduced above. The following theorem quantifies the resulting average first-order stationarity.

\begin{theorem} \label{thm:finite-time}
Under the conditions of Theorem~\ref{thm:mean-step}, for $k\ge1$, fix $0<\lambda_k=\lambda\le\lambda_{\rm loc}$ and any temperature sequence $\{\tau_k\}_{k\ge1}$ satisfying $0<\tau_k\le\bar\tau$. For every $N\ge1$, a constant $C$ independent of $N,\lambda,\varepsilon$ and
$\{\tau_k\}$ satisfies
\begin{equation*}
 \frac1N\sum_{k=1}^N\E\|g(p_k)\|^2
 \le C\left[
      \frac{\Delta_1}{N\lambda}
      +\frac1N\sum_{k=1}^N\tau_k
      +\left(\frac \varepsilon \lambda\right)^2\right].
\end{equation*}
\end{theorem}
Theorem~\ref{thm:finite-time} extends the gradient-like interpretation of a single update to the full optimization trajectory. The contribution of the initial objective gap $\Delta_1$ decreases as $1/N$, while the average temperature $1/N \sum_{k = 1}^{N} \tau_k$ and the relative sampling error $\varepsilon / \lambda$ determine the remaining stationarity accuracy. The bound therefore provides an insightful choice of temperature schedule. For example, choosing
\begin{equation} \label{eq:temperature-choice}
    \tau_k=\frac{\tau_0}{(k+1)^{1+\zeta}}, \qquad \zeta>0,
\end{equation}
gives $N^{-1}\sum_{k=1}^N\tau_k=O(N^{-1})$. Together with fixed $\lambda$, this yields
\begin{equation*}
    \frac1N \sum_{k=1}^N\E\|g(p_k)\|^2
    =
    O(N^{-1}) + O(\varepsilon^2).
\end{equation*}
Thus, increasing the number of refinement steps drives the contribution of initial optimality gap and temperature to zero, while a fixed sampling accuracy leaves a residual. In particular, for any prescribed temperature sequence satisfying $0<\tau_k\le\bar\tau$ and $\tau_k\downarrow0$, such as the schedule in~\eqref{eq:temperature-choice}, compactness of $\M$ and continuity of $g$ allow a measurable selection of an accumulation point $p_\infty$ of the sequence $\{p_k\}_{k \geq 1}$ satisfying
\begin{equation*}
    \| g(p_\infty) \|^2 = \liminf_{k \to \infty} \| g(p_k) \|^2 \qquad \text{almost surely}.
\end{equation*}
Fatou's lemma and Theorem~\ref{thm:finite-time} therefore give
\begin{equation*}
    \E\|g(p_\infty)\|^2
    =
    \E\left[\liminf_{k\to\infty}\|g(p_k)\|^2\right]\le
    \liminf_{N\to\infty}
    \frac1N\sum_{k=1}^{N}\E\|g(p_k)\|^2
    \le C\left(\frac{\varepsilon}{\lambda}\right)^2.
\end{equation*}
Cauchy--Schwarz further yields
$\E\|g(p_\infty)\|\le\sqrt C\,\varepsilon/\lambda$.
The accumulation point construction is detailed in Appendix~\ref{app:proof-finite-time}.


\subsection{Constrained Stationarity at Boundaries} \label{sec:with_boundary_theorems}

The preceding analysis measures first-order progress through the Riemannian gradient. At a boundary of $\M$, a nonzero gradient may not indicate an available feasible improvement since its negative direction may point outside $\M$. We therefore keep the refinement rule~\eqref{eq:implemented-recursion} unchanged and measure the optimization progress through the improvement available to a feasible descent step. In this subsection, we retain Assumptions~\ref{ass:regularity:basic}--\ref{ass:regularity:smooth}, except that $\M$ in Assumption~\ref{ass:regularity:geometry} may have a smooth boundary. Compactness of $\M$ and the $C^2$ regularity of $f$ on a neighborhood of $\M$ imply that $f|_{\M}$ is Lipschitz in Euclidean distance; let $L_f>0$ be a corresponding Lipschitz constant. 

\begin{figure}[t]
    \centering
    \captionsetup{skip=5pt}
    \includegraphics[width=\linewidth]{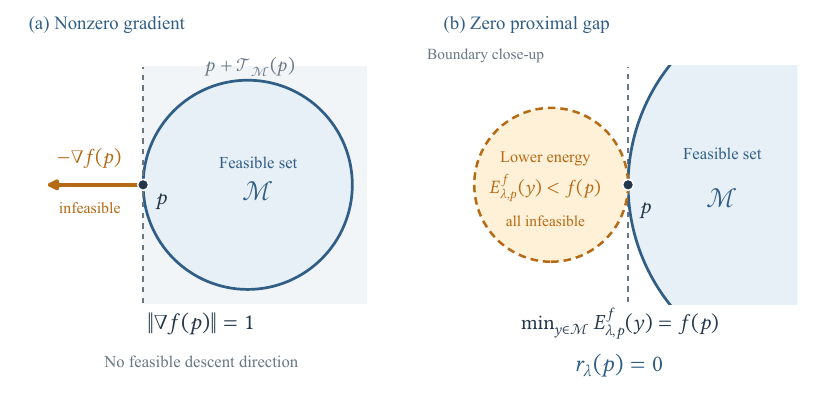}
    \caption{\textbf{Boundary geometry and zero proximal gap.}
Let $\M$ be the closed unit disk, $f(y)=y_1$, $p=(-1,0)$.
(a) The outward normal $n(p)=-\nabla f(p)$ points outside
the feasible tangent cone.
The shadowy halfplane represents $p+\mathcal T_{\M}(p)$, with $\mathcal T_{\M}(p)=\{w\in\R^2\mid w_1\ge0\}$; (b) A magnified view shows the strict sublevel
$E^f_{\lambda,p}(y)<f(p)$ of
$E^f_{\lambda,p}(y)=f(y)+\|y-p\|^2/(2\lambda)$,
using the same energy formula on the ambient plane.
It is an open amber disk centered at
$c_p=p-\lambda(1,0)$ with radius $\lambda$,
disjoint from the blue feasible set $\M$.
Its boundary touches $\M$ only at $p$, so
$r_\lambda(p)=0$ although $\|\nabla f(p)\|=1$.
The vertical dashed lines show the common supporting tangent.}
    \label{fig:boundary-stationarity}
\end{figure}

For a feasible point $p$, define the proximal gap
\begin{equation*}
    r_\lambda(p)
    \defeq 
    f(p)-\min_{y\in\M}
    \left\{
        f(y)+\frac{\|y-p\|^2}{2\lambda}
    \right\},
    \qquad p\in\M.
\end{equation*}
This gap is nonnegative because $y=p$ is admissible. If $r_\lambda(p)=0$, no feasible candidate improves this penalized objective,
which implies
\begin{equation*}
    \langle\nabla f(p),w\rangle\ge0,
    \qquad w\in\mathcal T_{\M}(p).
\end{equation*}
Thus a zero proximal gap certifies the constrained first-order condition. The proximal gap also connects constrained stationarity to objective
decrease. For an exact proximal update
$y^+\in\mathcal P_\lambda^f(p)$, its definition gives
\begin{equation*}
    f(p)-f(y^+)
    =
    r_\lambda(p)
    +\frac{\|y^+-p\|^2}{2\lambda}.
\end{equation*}
Figure~\ref{fig:boundary-stationarity} illustrates the tangent cone condition and an exact zero proximal gap at a boundary point.
Thus the objective decrease controls both the proximal gap and the squared displacement of the update. Summing this relation over successive iterations yields a telescoping difference between the initial and final objective values. Since the objective is bounded below, the total decrease is finite. The proximal gaps $r_\lambda$ and squared displacements are nonnegative, so their accumulated sum is bounded and both quantities tend to zero. This connects repeated proximal descent to constrained stationarity. The following theorem extends this argument to the stochastic updates of~\ouralg.

\begin{theorem} \label{thm:boundary}
Under the assumptions of this subsection and Assumption~\ref{ass:oracle}, consider Algorithm~\ref{alg:prism} with $\lambda_k=\lambda>0$ for every $k\ge1$ and any positive temperature sequence $\{\tau_k\}_{k\ge1}$. Assumption~\ref{ass:oracle} applies to every sampling call, including initialization. There exists $C_\lambda<\infty$, depending only on $\lambda$, $m$, $L_f$, the diameter of $\M$, and the local mass constants $v,r_0$ in~\eqref{eq:bd-local-mass}, such that, for every $k\ge1$,
\begin{equation}\label{eq:boundary-descent}
    \E\bigl[f(p_{k+1})-f(p_k)\bigr]
    \le
    -\E\left[
        r_\lambda(p_k)
        +\frac{\|p_{k+1}-p_k\|^2}{2\lambda}
    \right]
    +C_\lambda\left[
        \tau_k(1+|\log\tau_k|)
        +\varepsilon
    \right].
\end{equation}
Consequently, summing~\eqref{eq:boundary-descent} over
any $N\ge1$ iterations yields
\begin{equation}
    \frac1N\sum_{k=1}^{N}
    \E\left[
        r_\lambda(p_k)
        +\frac{\|p_{k+1}-p_k\|^2}{2\lambda}
    \right]
    \le
    \frac{\Delta_1}{N} + C_\lambda\left[
        \frac1N\sum_{k=1}^{N}
        \tau_k(1+|\log\tau_k|)
        +\varepsilon
    \right].
\label{eq:boundary-rate}
\end{equation}
\end{theorem}

Theorem~\ref{thm:boundary} extends the optimization guarantee to manifolds with boundary by replacing the gradient norm with feasible proximal improvement. The initial objective gap contributes at order $N^{-1}$, while temperature $\tau$ and sampling error $\varepsilon$ determine the remaining accuracy. Choosing the same temperature schedule in~\eqref{eq:temperature-choice} makes $\sum_{k=1}^{\infty}\tau_k(1+|\log\tau_k|)$ finite. Consequently,
\begin{equation*}
    \frac1N\sum_{k=1}^{N}\E r_\lambda(p_k)
    \le
    \frac1N\sum_{k=1}^{N}\E \left[ r_\lambda(p_k) + \frac{\|p_{k+1}-p_k\|^2}{2\lambda}\right ]
    \le
    \frac{\Delta_1+C_{\lambda,\tau_0,\zeta}}{N}
    +C_\lambda \varepsilon,
\end{equation*}
where the constants are independent of $N$ and $\varepsilon$. Under the exact sampling~(i.e., $\varepsilon = 0$) regime, multiplying~\eqref{eq:boundary-rate} by $N$
and letting $N\to\infty$, Tonelli's theorem gives
\begin{equation*}
    \E\left[
        \sum_{k=1}^{\infty}
            r_\lambda(p_k)
    \right] \leq
    \E\left[
        \sum_{k=1}^{\infty}
        \left(
            r_\lambda(p_k)
            +\frac{\|p_{k+1}-p_k\|^2}{2\lambda}
        \right)
    \right]
    \le \Delta_1+C_{\lambda,\tau_0,\zeta}<\infty.
\end{equation*}
The nonnegative sum is therefore finite almost surely. In particular, $r_\lambda(p_k)\to0$ almost surely.

On the event of probability one where $r_\lambda(p_k)\to0$, continuity of $r_\lambda$ implies that every accumulation point $p_\infty$ of $\{p_k\}$ satisfies $r_\lambda(p_\infty)=0$. Equivalently,
\begin{equation*}
    f(p_\infty)
    \le
    f(y)+\frac{\|y-p_\infty\|^2}{2\lambda},
    \qquad y\in\M,
\end{equation*}
which yields the constrained first-order condition
\begin{equation*}
    \langle\nabla f(p_\infty),w\rangle\ge0,
    \qquad w\in\mathcal T_{\M}(p_\infty).
\end{equation*}
Thus, under exact sampling and the temperature schedule~\eqref{eq:temperature-choice}, almost surely every accumulation point of $\{p_k\}$ satisfies the constrained first-order stationarity condition. We next evaluate~\ouralg through numerical experiments.

\section{Numerical Studies}\label{sec:experiments}

In this section, we evaluate~\ouralg through three experiments. In Section~\ref{sec:experiment1}, we use~\ouralg to solve a optimization problem over a high dimensional Grassmann manifold, aim to examin the Riemannian gradient like refinement as we demonstrated in our theoretical results. Section~\ref{sec:experiment_building_control} considers data-driven building temperature control, assessing whether~\ouralg can work on a real life control scenario. Finally, Section~\ref{sec:experiment2} compares~\ouralg with established optimization methods on six Design-Bench
tasks to assess its applicability across diverse continuous and
discrete design problems from scientific and engineering areas.

\subsection{Optimization on Grassmann Manifold}\label{sec:experiment1}

In this section, we evaluate~\ouralg on an optimization task over a Grassmann manifold. The Grassmann manifold $\mathrm{Gr}(r,d)$ consists of the $r$-dimensional linear subspaces of $\R^d$, representing a subspace independently of the basis used to describe it. It provides a natural search space for principal component analysis (PCA) and dimensionality reduction~\cite{absil2008optimization}. Optimization over Grassmann manifolds also arises in recovering missing entries of low-rank matrices~\cite{boumal2015low} and learning compact video representations for human action recognition~\cite{cherian2020representation}. These applications motivate testing whether~\ouralg can refine a subspace representation using only feasible examples.

\textbf{Problem setting.} We construct a PCA task on the $16$-dimensional Grassmann manifold $\mathrm{Gr}(2,10)$, with objective
\begin{equation}\label{eq:exp-grassmann-objective}
 \min_{P\in\M=\mathrm{Gr}(2,10)} f(P),\qquad
 f(P)=\tr\!\big((I-P)C(I-P)^\top\big)
     =\tr(C)-\tr(PC),
\end{equation}
where $P$ is a rank-two orthogonal projector and $C$ is a prescribed covariance matrix for a synthetic Gaussian population. As in classic PCA, $\tr (C)$ represents the total variance, while $\tr(PC)$ is the variance retained by the chosen subspace. Minimizing $f(P)$ therefore finds the representation that preserves the most variance, equivalently minimizing the reconstruction error.

\begin{figure}[ht]
 \centering
 \includegraphics[width=0.95\linewidth]{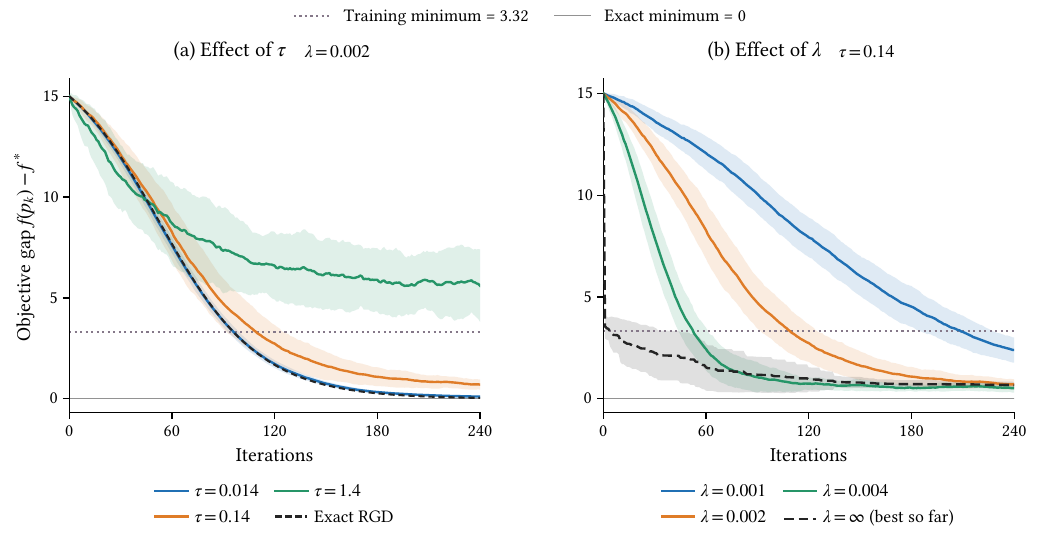}
 \caption{Optimization on Grassmann manifold $\mathrm{Gr}(2,10)$. (a) Fixed $\lambda=0.002$, varying $\tau$, compared with exact RGD (with access to the manifold) with stepsize $0.002$. (b) Fixed $\tau=0.14$, varying $\lambda$. Curves with finite $\lambda$ value report current objective gaps, while $\lambda=\infty$ reports the gap of the best sampled response so far. The dotted line marks the minimum over the training dataset $\mathcal{D}_n$, and zero corresponds to the exact optimum. For each parameter choice, we run $50$ trajectories of $240$ iterations, and plot their mean, shaded regions denote standard deviation across trajectories.}
 \label{fig:exp-grassmann}
\end{figure}

\textbf{Experimental results.} Figure~\ref{fig:exp-grassmann}(a) compares our refinement framework~\ouralg with exact Riemannian gradient descent~(RGD), both using the stepsize $\lambda=0.002$. At $\tau = 0.014$, the mean objective gap of our refinement method closely follows the objective gap trajectory produced by RGD. Smaller $\tau$ yields lower terminal objective gaps and smaller across-run standard deviations. The terminal mean objective gaps are approximately $0.105$, $0.700$, and $5.608$ for $\tau=0.014$, $0.14$, and $1.4$, respectively. For the two smaller temperatures, all $50$ trajectories terminate below the minimum training objective gap, demonstrating refinement beyond the observed examples.

Classic RGD determines its descent direction from the first-order tangent model in~\eqref{eq:tangent-step-model}, which approximates objective changes through the local gradient. In contrast,~\ouralg uses the full objective in the \textit{proximal energy}~\eqref{eq:prox-energy}, evaluating candidate updates through their actual objective values. The alignment of~\ouralg and exact RGD observed at small $\tau$ qualitatively illustrates the gradient-like behavior described by Theorem~\ref{thm:mean-step}.

Figure~\ref{fig:exp-grassmann}(b) fixes $\tau = 0.14$ and varies $\lambda$. For finite $\lambda$, we plot the current objective gap. The figure shows faster initial descent as $\lambda$ increases, consistent with its local stepsize interpretation in Theorem~\ref{thm:mean-step}. Setting $\lambda=\infty$ removes the quadratic term from the \textit{proximal energy}~\eqref{eq:prox-energy}, making the sampling kernel $ \widehat Q^f$ independent of the current iterate. This turns the refinement process into sampling-based optimization, related to generative approaches such as DDOM~\cite{krishnamoorthy2023diffusion}. For this unanchored baseline, we plot the lowest objective gap among sampled responses so far. We compare the methods under the same budget of returned samples. Over iterations $70$--$125$,~\ouralg with $\lambda=0.004$ and the unanchored best-so-far baseline attain comparable mean objective gaps, with substantially smaller standard deviation. We focus on this regime because the curve for $\lambda = \infty$ reports a best-so-far gap, as independent samples accumulate, its standard deviation will eventually vanish. The comparison illustrates the benefit of quadratic anchoring under a limited sampling budget, where comparable mean accuracy is achieved with less output variability.

\subsection{Data-Driven Building Energy Control}
\label{sec:experiment_building_control}

Section~\ref{sec:experiment1} examines gradient-like refinement on a Grassmann manifold with geometry available for diagnostics. We next evaluate~\ouralg in a building-control simulation based on BOPTEST's residential building simulator with hydronic floor heating and an air-to-water heat pump~\cite{blum2021building}. We examine whether a refinement kernel constructed from finite operating records can support temperature tracking with performance close to a controller that queries the underlying physical model. All controllers use a two-hour prediction horizon, apply one half-hour command, and operate over a 48-hour deployment.

\textbf{Problem formulation.}
We consider a 48-hour building-control task with $K=96$ control intervals, each lasting 30 minutes. Let $z_t$ denote the full building and equipment state, $u_t\in[0,1]$ the compressor-modulation command, and $d_t$ the weather and occupancy inputs during interval $t$. The physical system satisfies
\begin{equation*}
 (z_{t+1},y_t)=\mathcal F(z_t,u_t,d_t),
 \qquad t=0,\ldots,K-1,
\end{equation*}
where $y_t$ contains the measured responses during that interval. Specifically, observations are recorded every 15 minutes, so we group the two observations associated with each command as
\begin{equation*}
     y_t=(y_{t,1},y_{t,2}),\qquad
 y_{t,q}=
 (T_{t,q},T_{t,q}^{\mathrm{sup}},T_{t,q}^{\mathrm{ret}},P_{t,q}),
 \quad q\in\{1,2\}.
\end{equation*}
Here $T_{t,q}$ is operative room temperature,
$T_{t,q}^{\mathrm{sup}}$ and $T_{t,q}^{\mathrm{ret}}$ are supply-
and return-water temperatures, and $P_{t,q}$ is interval-average
electrical power in kW. Thus, $z_t$ denotes the full physical state,
whereas $y_t$ denotes the recorded temperature and power outputs.

The controller seeks to minimize the cumulative objective
\begin{equation} \label{eq:exp-building-objective}
 f(u,y)=
 \Delta t\sum_{t=0}^{K-1}\sum_{q=1}^{2}
 \bigl[p_{t,q}P_{t,q}
       +10(T_{t,q}-r_{t,q})^2\bigr] +0.01\sum_{t=0}^{K-1}(u_t-u_{t-1})^2 +20(T_{K-1,2}-r_{K-1,2})^2,
\end{equation}
where $\Delta t=0.25$ hours, $p_{t,q}$ is the electricity price,
$r_{t,q}$ is the target temperature, and $u_{-1}$ is the command
preceding deployment. The objective penalizes electricity expenditure,
temperature-tracking error, command variation, and terminal tracking
error. Room temperature is required to remain within
$[18,26]\,{}^\circ\mathrm C$.

We address this task using model predictive control (MPC).
At decision time $t$, MPC optimizes a sequence of commands over
a prediction horizon $H_t=\min\{4,K-t\}$, executes only its first
command, and repeats the optimization after observing the new state.
In this experiment, all controllers observe the full simulator state
$z_t$ and receive exact forecasts of the external inputs.
Using hats to denote planned commands and predicted responses,
the optimization at time $t$ is
\begin{align}
 \min_{\widehat u,\widehat y,\widehat z}\quad&
 \Delta t\sum_{\tau=t}^{t+H_t-1}\sum_{q=1}^{2}
 \bigl[p_{\tau,q}\widehat P_{\tau,q}
       +10(\widehat T_{\tau,q}-r_{\tau,q})^2\bigr]
+0.01\sum_{\tau=t}^{t+H_t-1}
       (\widehat u_\tau-\widehat u_{\tau-1})^2
 \nonumber\\
 &+20\bigl(\widehat T_{t+H_t-1,2}
          -r_{t+H_t-1,2}\bigr)^2
 \label{eq:exp-building-mpc}\\
 \text{subject to}\quad&
 (\widehat z_{\tau+1},\widehat y_\tau)
 =\mathcal F(\widehat z_\tau,\widehat u_\tau,d_\tau), \quad \widehat z_t=z_t, \quad
   \widehat u_{t-1}=u_{t-1},
 \nonumber
\end{align}
for $\tau=t,\ldots,t+H_t-1$ and $q\in\{1,2\}$.
The controller executes $u_t=\widehat u_t$ from the returned plan
and replans at the next decision time.

When the physical dynamics $\mathcal F$ are unavailable to the
optimizer, \ouralg can approximately solve this planning problem
using a conditional reference law $\hat{Q}_{\lambda, \tau}$ constructed from recorded
control--response trajectories. The reference approximates the
relationship between commands and responses under the current state
and forecast inputs. The refinement kernel then combines this
reference with the known objective, operational constraints, and
proximal term to produce a plan. Thus, \ouralg-MPC uses learned
trajectory relationships in place of direct evaluations of the
physical dynamics during planning.

A sampling-based alternative is the cross-entropy method within
MPC, termed CEM-MPC~\cite{pinneri2021sample, walker2026sample}. Here, the dynamics are accessible
through a black-box simulator rather than supplied as explicit
equations. At each decision time, CEM samples candidate command
sequences, obtains their responses through simulator rollouts,
and updates its sampling distribution using the best-performing
candidates. Both methods therefore address
\eqref{eq:exp-building-mpc}, but \ouralg reuses information learned
from operating records, whereas CEM queries the simulator during
online planning.

We compare \ouralg-MPC and CEM-MPC under the same simulator-query budget, used for collecting offline training
trajectories for \ouralg-MPC or evaluating candidate trajectories during online planning for CEM-MPC. Once its reference model $\mathcal{Q}_{\lambda, \tau}$ is fitted, \ouralg-MPC reuses it throughout
deployment without further simulator queries during planning, whereas CEM-MPC distributes its allowance across deployment decisions. We additionally include Reference MPC, which optimizes
using the same exact physical dynamics as CEM-MPC but employs
a larger numerical search without the matched budget constraint.
We also compare against the lowest-objective feasible first-48-hour
trajectory among the collected records.
Appendix~\ref{app:exp_building} provides the implementation and
budget-accounting details.

\begin{figure}[t]
 \centering
 \includegraphics[width=0.95\linewidth]{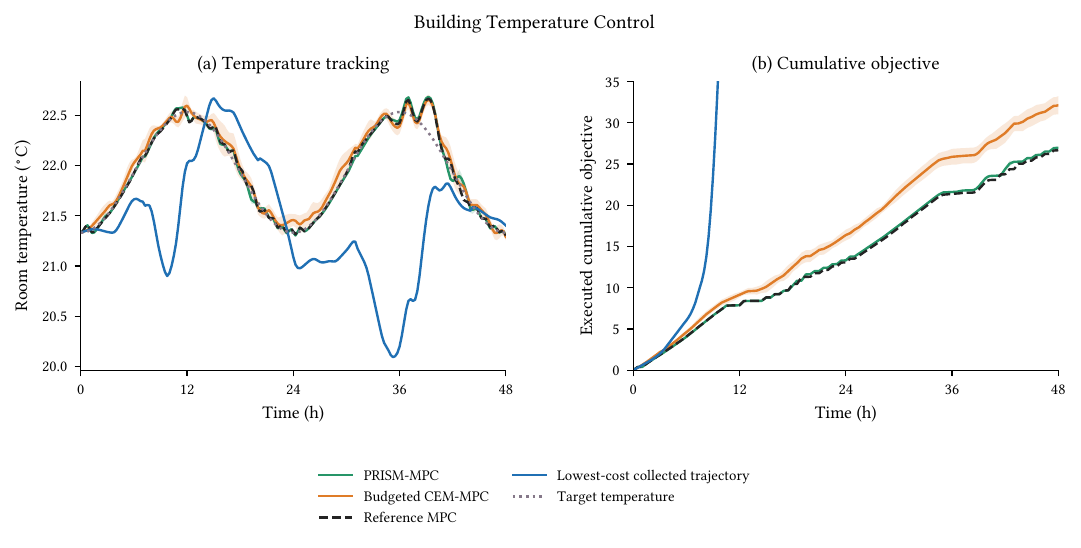}
 \caption{Building temperature control.
 (a) Executed room temperatures and the target.
 (b) Cumulative objective~\eqref{eq:exp-building-objective}.
 Green and orange curves show means across five \ouralg-MPC and CEM-MPC runs, shading denotes $\pm$ standard deviation.
 Dashed black shows Reference MPC, and blue shows the lowest-cost collected trajectory.}
 \label{fig:exp-building-control}
\end{figure} 

\textbf{Experimental results.}
Figure~\ref{fig:exp-building-control}(a) shows that \ouralg-MPC closely follows the target and Reference MPC, while the collected trajectory has substantially larger tracking deviations. In panel~(b), \ouralg-MPC attains total objective $26.969\pm0.054$, compared with $32.128\pm1.102$ for budgeted CEM-MPC and $26.672$ for Reference MPC. Its mean objective is $16.06\%$ below budgeted CEM-MPC and $1.12\%$ above the numerical reference. All \ouralg-MPC runs improve on the lowest-cost collected trajectory. These results support effective reuse of finite operating data in this real world scenario.

\subsection{Offline Design Optimization} \label{sec:experiment2}

In this experiment, we further assess the applicability of~\ouralg across diverse data-driven optimization tasks arising from scientific and engineering domains using six Design-Bench benckmarks~\cite{trabucco2022designbench}, where objective guidance is learned from offline labels.
We use $f=-\widehat r$, the negative reward of the respective tasks, in the proximal energy~\eqref{eq:prox-energy} and perform four outer anchor updates, following Algorithm~\ref{alg:prism}.
The locality--temperature balance in Theorems~\ref{thm:mean-step} and~\ref{thm:descent} motivates this construction. The samplers below implement it for the different task representations.

\textbf{Setup and evaluation.}
For each task, we select \ouralg's sampling implementation and parameters, and each baseline's approximation models, using oracle feedback over ten discovery seeds, then freeze these choices before evaluation on five disjoint confirmation seeds.
Each baseline can use its author-default models or the compatible reward and reference models available to \ouralg.
All methods use the same offline datasets and oracles and submit 256 designs per task and confirmation seed.
Table~\ref{tab:designbench-matched} reports the mean and standard deviation of each batch's maximum oracle score.
Implementation details and baseline references are given in Appendix~\ref{app:experiments2}.

\begin{table}[t]
  \centering
  \caption{Results on five disjoint confirmation seeds, with 256 submitted designs per task and seed. Each cell gives the mean batch-maximum oracle score and its population standard deviation. Bold and underlined means mark the highest and second-highest distinct means (higher score is better). Average rank uses unrounded means across the six tasks and ten displayed methods, with midranks for exact ties (lower rank is better). $^*$ChEMBL scores are divided by $10^3$ for readability and invalid designs receive the offline dataset's minimum label (293,000) before taking the batch maximum. Validity uses the official decoder and a nonempty RDKit 2022.09.5 molecular parse.}
\label{tab:designbench-matched}
  \begingroup
  \setlength{\tabcolsep}{2.5pt}
  \renewcommand{\arraystretch}{1.2}
  \scriptsize
  \begin{tabular*}{\linewidth}{@{\extracolsep{\fill}}l*{7}{r}@{}}
    \toprule
    \textbf{Method} & \textbf{TFBind8} & \textbf{TFBind10} & \textbf{Supercond.} & \textbf{Ant} & \textbf{DKitty} & \shortstack[r]{\textbf{ChEMBL}$^*$\\$(\times10^3)$} & \shortstack[r]{\textbf{Average}\\\textbf{rank} $\downarrow$} \\
    \midrule
    \ouralg & \shortstack[r]{$\mathbf{0.9936}$\\$\pm 0.0048$} & \shortstack[r]{$0.8766$\\$\pm 0.1733$} & \shortstack[r]{$\underline{96.28}$\\$\pm 6.43$} & \shortstack[r]{$517.33$\\$\pm 7.99$} & \shortstack[r]{$270.75$\\$\pm 3.83$} & \shortstack[r]{$\underline{396.022}$\\$\pm 0.308$} & $\mathbf{3.333}$ \\
    \midrule
    DiffOPT & \shortstack[r]{$0.9529$\\$\pm 0.0349$} & \shortstack[r]{$0.8385$\\$\pm 0.1401$} & \shortstack[r]{$\mathbf{97.99}$\\$\pm 3.90$} & \shortstack[r]{$530.43$\\$\pm 20.17$} & \shortstack[r]{$261.24$\\$\pm 7.23$} & \shortstack[r]{$394.901$\\$\pm 1.011$} & $\underline{4.333}$ \\
    COMs & \shortstack[r]{$0.7960$\\$\pm 0.1019$} & \shortstack[r]{$\mathbf{1.0152}$\\$\pm 0.0000$} & \shortstack[r]{$88.91$\\$\pm 8.96$} & \shortstack[r]{$538.93$\\$\pm 15.94$} & \shortstack[r]{$\underline{280.77}$\\$\pm 6.67$} & \shortstack[r]{$\mathbf{396.176}$\\$\pm 0.000$} & $4.417$ \\
    DDOM & \shortstack[r]{$0.9222$\\$\pm 0.0348$} & \shortstack[r]{$0.7902$\\$\pm 0.0761$} & \shortstack[r]{$94.65$\\$\pm 8.44$} & \shortstack[r]{$\underline{553.72}$\\$\pm 13.62$} & \shortstack[r]{$\mathbf{282.72}$\\$\pm 10.59$} & \shortstack[r]{$395.407$\\$\pm 0.000$} & $4.500$ \\
    GP-qEI & \shortstack[r]{$0.8309$\\$\pm 0.1095$} & \shortstack[r]{$\mathbf{1.0152}$\\$\pm 0.0000$} & \shortstack[r]{$95.05$\\$\pm 1.69$} & \shortstack[r]{$463.83$\\$\pm 0.00$} & \shortstack[r]{$209.47$\\$\pm 0.00$} & \shortstack[r]{$\mathbf{396.176}$\\$\pm 0.000$} & $5.083$ \\
    Gradient ascent & \shortstack[r]{$\underline{0.9810}$\\$\pm 0.0111$} & \shortstack[r]{$\mathbf{1.0152}$\\$\pm 0.0000$} & \shortstack[r]{$92.96$\\$\pm 3.13$} & \shortstack[r]{$98.41$\\$\pm 84.47$} & \shortstack[r]{$198.30$\\$\pm 16.17$} & \shortstack[r]{$375.079$\\$\pm 41.041$} & $5.667$ \\
    CMA-ES & \shortstack[r]{$0.9702$\\$\pm 0.0169$} & \shortstack[r]{$\underline{1.0127}$\\$\pm 0.3034$} & \shortstack[r]{$89.42$\\$\pm 2.14$} & \shortstack[r]{$\mathbf{4936.69}$\\$\pm 3256.37$} & \shortstack[r]{$5.32$\\$\pm 1.72$} & \shortstack[r]{$327.440$\\$\pm 42.181$} & $5.667$ \\
    MINs & \shortstack[r]{$0.9439$\\$\pm 0.0210$} & \shortstack[r]{$0.7950$\\$\pm 0.1751$} & \shortstack[r]{$93.57$\\$\pm 3.48$} & \shortstack[r]{$502.23$\\$\pm 23.36$} & \shortstack[r]{$272.20$\\$\pm 8.29$} & \shortstack[r]{$329.688$\\$\pm 45.144$} & $6.000$ \\
    CbAS & \shortstack[r]{$0.9413$\\$\pm 0.0266$} & \shortstack[r]{$0.7215$\\$\pm 0.0739$} & \shortstack[r]{$80.98$\\$\pm 6.85$} & \shortstack[r]{$472.86$\\$\pm 16.47$} & \shortstack[r]{$232.31$\\$\pm 12.60$} & \shortstack[r]{$389.495$\\$\pm 1.236$} & $7.500$ \\
    REINFORCE & \shortstack[r]{$0.9647$\\$\pm 0.0174$} & \shortstack[r]{$0.7173$\\$\pm 0.1720$} & \shortstack[r]{$88.57$\\$\pm 0.87$} & \shortstack[r]{$-149.80$\\$\pm 77.24$} & \shortstack[r]{$-208.88$\\$\pm 239.19$} & \shortstack[r]{$366.149$\\$\pm 36.881$} & $8.500$ \\
    \bottomrule
  \end{tabular*}
  \endgroup
\end{table}

\textbf{Sampling implementations.}
The empirical and smoothed constructions in Appendix~\ref{app:statistical} motivate preserving observed support when appropriate and smoothing it when new designs are needed.
For TFBind8, learned reverse diffusion with Euler discretization shares denoising information across the comparatively densely observed short-sequence space and can propose unobserved sequences~\cite{vincent2011connection,song2021sde}.
The same multiscale mechanism suits Superconductor's continuous descriptors, bounding the denoiser's predictions to the training coordinate box limits coordinate extrapolation.
TFBind10 has sparser sequence coverage.
Direct categorical Metropolis transitions~\cite{hastings1970montecarlo} use valid-symbol substitutions and Hamming proximity, avoiding continuous optimization through discrete decoding.
Its final portfolio combines sampled designs ranked by the surrogate with high-scoring offline designs.

Ant uses a Gaussian kernel density estimate (KDE) of the reference law.
Its explicit density permits Metropolis-adjusted Langevin sampling (MALA)~\cite{rosenblatt1956remarks,roberts1996exponential}.
Decreasing the bandwidth is intended to allow broader movement before concentrating near observed morphologies, following the multiscale intuition of annealed score sampling~\cite{song2019generative}.
DKitty instead uses small-step, learned-score unadjusted Langevin sampling (ULA). Its continuous coordinates permit local objective gradients, while score and proximity forces regularize the refinement~\cite{roberts1996exponential}.
ChEMBL's small pool makes exact empirical sampling inexpensive.
It uses the weighting form in~\eqref{eq:stat-categorical}, with the learned objective, preserving observed molecular sequences and restricting the method to retrieval.

\textbf{Experimental results.}
\ouralg obtains the best observed average rank, 3.333, compared with 4.333 for DiffOPT, 4.417 for COMs, and 4.500 for DDOM.
It has the highest mean on TFBind8 and the second-highest on Superconductor and ChEMBL, with weaker rankings on TFBind10, Ant, and DKitty.
CMA-ES's large Ant score is associated with strong extrapolation: every submitted design lies outside at least one offline coordinate range.
On ChEMBL, its discovery scores equal those of the anchor-only control, so this result does not necessarily demonstrate an improvement over retrieval. These finite-budget output scores complement the trajectory diagnostics in Section~\ref{sec:experiment1} and temperature tracking task in Section~\ref{sec:experiment_building_control}. They assess a task-specific sampling policy without certifying Assumption~\ref{ass:oracle} or the stationarity bounds. Appendix~\ref{app:experiments2} distinguishes the relevant approximation errors.
\section{Concluding Remarks}

This paper developed~\ouralg for optimization on unknown manifolds using feasible observations, established its theoretical guarantees, and evaluated its performance through numerical studies.

First, we introduced~\ouralg, a framework that uses feasible observations to optimize an explicit objective on an unknown manifold. At each step, the method samples from a proximal distribution defined by the objective and the distance from the current iterate, then uses the sample as the next iterate. It requires no explicit manifold equations, tangent projections, or retractions.

Second, under the stated assumptions, our analysis relates the mean update to Riemannian gradient descent and bounds the effects of temperature and sampling error on manifold adherence, expected objective descent at feasible representatives, and stationarity. An analysis based on the proximal gap extends the stationarity guarantees to manifolds with boundary.

Third, we evaluated~\ouralg in three numerical studies. The Grassmann manifold experiment illustrates the predicted geometric behavior and the effect of penalizing distance from the current iterate. The building control study demonstrates improvement over the best observed feasible solution. Across six offline design tasks, \ouralg achieves the best observed average rank among the evaluated methods.

An important direction for future work is to make the refinement framework adaptive to uneven data coverage on the manifold. Because the reference distribution need not be uniform, the refinement trajectory may enter regions with relatively few observations, where the approximate conditional sampler $\widehat Q$ may be less accurate. Online rules should jointly adjust the temperature $\tau_k$ and step size $\lambda_k$ to limit the effects of sampling error on manifold adherence and expected objective improvement. Developing practical accuracy indicators and establishing guarantees for these adaptive updates would strengthen the framework in settings with limited and unevenly distributed observations.

\bibliographystyle{ACM-Reference-Format}
\bibliography{paper}
\appendix

\section{Structural Properties and Proximal-energy Estimates} \label{app:kernel_properties}

This appendix establishes the structural properties of our kernel $Q_{\lambda,\tau}^{f}$ introduced in Section~\ref{sec:method}. We first verify that it defines a valid stochastic transition and characterize it through the entropy-regularized variational problem defined in~\eqref{eq:variational}. We then study its low-temperature behavior: the kernel concentrates on the \textit{proximal minimizing set} $ \mathcal P_\lambda^f(x)$ as $\tau\downarrow0$, and, when several isolated nondegenerate minimizers coexist, its limiting mass can be characterized explicitly.


\subsection{The Proximal Refinement Kernel} \label{app:kernel_variational}

Before analyzing the behavior of the refinement law, we verify that $Q_{\lambda,\tau}^{f}$ is mathematically well defined for every ambient point $x \in \R^D$ and explain why its specific form arises from~\eqref{eq:variational}.

\begin{proposition} \label{prop:landing}
Under Assumption~\ref{ass:regularity:basic}, for every $\lambda,\tau>0$, $Z_{\lambda,\tau}^{f}(x)$ is finite and strictly positive for every $x\in\R^D$. Moreover, $Q_{\lambda,\tau}^{f}(x,\cdot)$ defined in~\eqref{eq:kernel}
is a Borel probability kernel from $\R^D$ to $\M$, satisfies
\begin{equation*}
    \supp Q_{\lambda,\tau}^{f}(x,\cdot)=\M,
\end{equation*}
and depends continuously on $x$ in total variation. For each fixed $x\in\R^D$, $Q_{\lambda,\tau}^{f}(x,\cdot)$ is the unique minimizer of~\eqref{eq:variational}.
\end{proposition}

\begin{proof}
    Fix $x\in\R^D$ and $\lambda,\tau>0$, and abbreviate
    \begin{equation*}
        E=E_{\lambda,x}^{f},\qquad Z=Z_{\lambda,\tau}^{f}(x).
    \end{equation*}
    Due to compactness of $\M$ and continuity of $f$, the continuous energy $E$ has finite extrema $E_{\min}$ and $E_{\max}$ on
$\M$. Since $\mu$ is a probability measure,
\begin{equation*}
    0<e^{-E_{\max}/\tau} \le Z\le e^{-E_{\min}/\tau}<\infty.
\end{equation*}
Thus $q(y)=e^{-E(y)/\tau}/Z$ is a probability density with respect to $\mu$, bounded above and below by positive constants for fixed $x$, $\lambda$, and $\tau$. Consequently $Q=q\mu$ and $\mu$ are mutually absolutely continuous. Therefore 
\begin{equation*}
    \supp Q= \supp \mu = \M.
\end{equation*}
We next establish continuity with respect to $x$. Let $\{x_j\}_{j\ge1}\subset\R^D$ be any sequence converging to $x$, and write $E_j=E_{\lambda,x_j}^{f}$. Since $\lambda>0$ is fixed,
\begin{equation*}
 \sup_{y\in\M}|E_j(y)-E(y)|
 \le
 \frac{\|x_j-x\|}{2\lambda}
 \left(2\sup_{y\in\M}\|y\|+\|x_j\|+\|x\|\right)
 \rightarrow0.
\end{equation*}
Thus $E_j$ converges uniformly to the bounded function $E$ on $\M$. The functions $E_j$ are consequently uniformly bounded, so $e^{-E_j/\tau}$ converges uniformly to $e^{-E/\tau}$. It follows that $Z_{\lambda,\tau}^{f}(x_j)\to Z_{\lambda,\tau}^{f}(x)>0$, and the corresponding normalized densities converge uniformly on $\M$. Therefore, the normalized densities $q_{x_j}$ converge uniformly to $q_x$, which implies
\begin{equation*}
 \|Q_{\lambda,\tau}^{f}(x_j,\cdot)-Q_{\lambda,\tau}^{f}(x,\cdot)\|_{\mathrm{TV}}
 =\frac12\int_{\M}|q_{x_j}-q_x|\,d\mu\longrightarrow0.
\end{equation*}
For every Borel set $B\subset\R^D$, the map
$x\mapsto Q_{\lambda,\tau}^{f}(x,B)$ is consequently continuous and hence
Borel measurable. Together with the probability-measure property for fixed
$x$, this proves that $Q$ is a Borel probability kernel.

We now establish the variational characterization. We write $\nu \ll \mu$ if $\nu$ is absolutely continuous with respect to $\mu$, and denote the relative entropy by $\operatorname{KL}(\nu\|\mu)=\int\log(d\nu/d\mu)d\nu$ for $\nu\ll\mu$, and by $+\infty$ otherwise. Since $E=E_{\lambda,x}^{f}$ is continuous on the compact set $\M$, it is bounded. Moreover,
\begin{equation}
    \frac{dQ}{d\mu}(y)
    =
    \frac{e^{-E(y)/\tau}}{Z}, \text{so that }\log\frac{dQ}{d\mu}(y)
    =
    -\frac{E(y)}{\tau}-\log Z.
    \label{eq:kernel-log-density}
\end{equation}
The right-hand side is bounded on $\M$. Consider a probability measure $\nu$ satisfying $\nu\ll\mu$. Since $dQ/d\mu=e^{-E/\tau}/Z$ is strictly positive on $\M$, we have $\mu\ll Q$. Together with $\nu\ll\mu$, this implies $\nu\ll Q$. By the Radon--Nikodym chain rule,
\begin{equation*}
    \frac{d\nu}{dQ}
    =
    \frac{d\nu/d\mu}{dQ/d\mu}, \text{and hence } \log\frac{d\nu}{dQ}
    =
    \log\frac{d\nu}{d\mu}
    -
    \log\frac{dQ}{d\mu}.
\end{equation*}
Integrating with respect to $\nu$ gives
\begin{equation}
    \operatorname{KL}(\nu\|Q)
    =
    \operatorname{KL}(\nu\|\mu)
    -
    \int_{\M}
    \log\frac{dQ}{d\mu}\,d\nu.
    \label{eq:kl-change-measure}
\end{equation}
Substituting~\eqref{eq:kernel-log-density} into~\eqref{eq:kl-change-measure} yields
\begin{equation*}
    \operatorname{KL}(\nu\|Q)
    =
    \operatorname{KL}(\nu\|\mu)
    +
    \frac{1}{\tau}\int_{\M}E\,d\nu
    +
    \log Z.
\end{equation*}
Rearranging,
\begin{equation*}
    \int_{\M}E\,d\nu
    +
    \tau\operatorname{KL}(\nu\|\mu)
    =
    -\tau\log Z
    +
    \tau\operatorname{KL}(\nu\|Q).
    \label{eq:kernel-entropy-identity}
\end{equation*}
Finally, since $\operatorname{KL}(\nu\|Q)\ge0,$ with equality if and only if $\nu=Q$. Therefore
\begin{equation*}
    \int_{\M}E\,d\nu
    +
    \tau\operatorname{KL}(\nu\|\mu)
    \ge
    -\tau\log Z,
\end{equation*}
and equality holds uniquely at $\nu=Q$. Hence $Q_{\lambda,\tau}^{f}(x,\cdot)$ is the unique minimizer of~\eqref{eq:variational}.
\end{proof}

Thus $Q_{\lambda,\tau}^{f}$ is a valid stochastic refinement kernel at every positive temperature $\tau$. We next show that lowering the temperature connects this stochastic law back to the deterministic proximal problem in~\eqref{eq:prox-map}.

\subsection{Concentration on Proximal Minimizers} \label{app:kernel-concentration}

At positive temperature $\tau$, $Q_{\lambda,\tau}^{f}(x,\cdot)$ has full support on $\M$ and therefore remains stochastic. The next result quantifies what happens as $\tau\downarrow0$.

\begin{proposition} \label{prop:proximal-concentration}
Under Assumption~\ref{ass:regularity:basic}, fix $x\in\R^D$ and $\lambda>0$, and write $\mathcal P=\mathcal P_\lambda^f(x)$. For every $\epsilon>0$ there are constants $C_\epsilon,c_\epsilon>0$, which depend on $x,\lambda,\epsilon$ but not on $\tau$, such that
\begin{equation*}
 Q_{\lambda,\tau}^{f}
 \left(x,\{y\in\M:\operatorname{dist}(y,\mathcal P)\ge\epsilon\}\right)
 \le C_\epsilon e^{-c_\epsilon/\tau},\qquad \tau>0.
\end{equation*}
Moreover, let $\tau_j\downarrow0$ be any sequence. Since $Q_{\lambda,\tau_j}^{f}(x,\cdot)$ is supported on $\M$, the sequence $\{Q_{\lambda,\tau_j}^{f}(x,\cdot)\}_j$ admits a weakly convergent subsequence. If $Q_{\lambda,\tau_{j_k}}^{f}(x,\cdot) \Rightarrow Q_0$, then the limiting measure is concentrated on the proximal minimizing set: $Q_0\!\left(\mathcal P_\lambda^f(x)\right)=1.$ Equivalently, $\supp (Q_0) \subseteq \mathcal P_\lambda^f(x)$.
\end{proposition}

\begin{proof}
Fix $x\in\R^D$, $\lambda>0$, and $\epsilon>0$. For brevity, write
\begin{equation*}
    E(y)\defeq E_{\lambda,x}^{f}(y),
    \qquad
    E_\star \defeq \min_{y\in\M}E(y),
    \qquad
    \mathcal P \defeq \mathcal P_\lambda^f(x)
    =\{y\in\M:E(y)=E_\star\}.
\end{equation*}
Since $E$ is continuous on $\M$, the minimum $E_\star$ is attained. Hence $\mathcal P$ is nonempty. Moreover, because $\mathcal P=E^{-1}(\{E_\star\})$, it is closed in $\M$ and therefore compact. Define the set of feasible points that stay at least $\epsilon$ away from the
proximal minimizing set:
\begin{equation*}
    F_\epsilon
    \defeq 
    \left\{
        y\in\M:
        \operatorname{dist}(y,\mathcal P)\ge\epsilon
    \right\}.
\end{equation*}
If $F_\epsilon=\varnothing$, then $Q_{\lambda,\tau}^{f}(x,F_\epsilon)=0$ for every $\tau>0$, and the concentration bound is immediate. We therefore assume $F_\epsilon\neq\varnothing$. Because $\operatorname{dist}(\cdot,\mathcal P)$ is continuous,
$F_\epsilon$ is closed in $\M$. Since $\M$ is compact, $F_\epsilon$ is compact. By construction, $F_\epsilon\cap\mathcal P=\varnothing$. Therefore the minimum of $E$ on $F_\epsilon$ is strictly larger than the global minimum $E_\star$:
\begin{equation*}
        \eta_\epsilon \defeq \min_{y\in F_\epsilon}E(y)-E_\star >0.
\end{equation*}
Choose any $y^\star\in\mathcal P$. By continuity of $E$, there exists a relatively open neighborhood $V\subset\M$ of $y^\star$ such that
\begin{equation*}
    E(y)
    \le
    E_\star+\frac{\eta_\epsilon}{4},
    \qquad y\in V.
\end{equation*}
Since $\supp \mu=\M$, every nonempty relatively open subset of $\M$ has positive $\mu$-measure, hence $\mu(V) > 0$. On $F_\epsilon$, we have $E(y)\ge E_\star+\eta_\epsilon$, so
\begin{equation*}
    \int_{F_\epsilon}
    e^{-E(y)/\tau}\,\mu(dy)
    \le
    e^{-(E_\star+\eta_\epsilon)/\tau}.
\end{equation*}
On the other hand, restricting the normalizer to $V$ yields
\begin{equation*}
    Z_{\lambda,\tau}^{f}(x) = \int_{\M}e^{-E(y)/\tau} \mu(dy) \ge \int_Ve^{-E(y)/\tau} \mu(dy) \ge \mu(V) e^{-(E_\star+\eta_\epsilon/4)/\tau}
\end{equation*}
Therefore
\begin{equation}
    Q_{\lambda,\tau}^{f}(x,F_\epsilon)
    =
    \frac{
        \int_{F_\epsilon}e^{-E(y)/\tau}\,\mu(dy)
    }{
        Z_{\lambda,\tau}^{f}(x)
    }
    \le
    \mu(V)^{-1}
    \exp\!\left(
        -\frac{3\eta_\epsilon}{4\tau}
    \right)
\label{eq:proximal-concentration}
\end{equation}
Thus the claimed exponential bound holds with $C_\epsilon=\mu(V)^{-1}, c_\epsilon= 3 \eta_\epsilon/4 >0.$

We next prove the weak-limit statement. Let $\tau_j\downarrow0$ be arbitrary. Since $\M$ is a compact metric space, every sequence of probability measures on $\M$ has a weakly convergent subsequence. Therefore, there exist a subsequence $(\tau_{j_k})$ and a probability measure $Q_0$ on $\M$ such that
\begin{equation*}
    Q_{\lambda,\tau_{j_k}}^{f}(x,\cdot) \Rightarrow Q_0.
\end{equation*}
It remains to show that $Q_0$ assigns no mass outside $\mathcal P$. For each integer $m\ge1$, define
\begin{equation*}
    G_m \defeq \left\{
        y\in\M:
        \operatorname{dist}(y,\mathcal P)>\frac1m
    \right\},
\end{equation*}
which is relatively open in $\M$. Applying~\eqref{eq:proximal-concentration} with $\epsilon=1/m$ gives $Q_{\lambda,\tau_{j_k}}^{f}(x,G_m) \rightarrow0.$ Since $G_m$ is open, the Portmanteau theorem yields
\begin{equation*}
    Q_0(G_m)
    \le
    \liminf_{k\to\infty}
    Q_{\lambda,\tau_{j_k}}^{f}(x,G_m) = 0.
\end{equation*}
Finally, because $\mathcal P$ is closed, every point
$y\in\M\setminus\mathcal P$ has strictly positive distance from
$\mathcal P$. Therefore $\M\setminus\mathcal P = \bigcup_{m=1}^{\infty}G_m$. By countable subadditivity, $Q_0(\M\setminus\mathcal P) \leq \sum_{m=1}^{\infty}Q_0(G_m) = 0$. Consequently, $Q_0(\mathcal P)=1.$ Since $\mathcal P$ is closed, this also implies
\begin{equation*}
    \supp Q_0 \subseteq \mathcal P.
\end{equation*}
\end{proof}

Proposition~\ref{prop:proximal-concentration} determines where the probability mass goes as $\tau\downarrow0$, but it does not determine how that mass is divided when $\mathcal P_\lambda^f(x)$ contains several minimizers. The next subsection resolves this question under the standard nondegeneracy condition.

\subsection{Probability Allocation Across Proximal Branches} \label{app:kernel-branches}

When several proximal minimizers coexist, Proposition~\ref{prop:proximal-concentration} only implies that when $\tau \downarrow 0$, the mass will concentrate on their union. To determine how the limiting mass is divided among these alternatives, we impose the usual nondegeneracy condition: each proximal minimizer is isolated and has a positive-definite Riemannian Hessian. This rules out flat tangent directions and permits a local quadratic approximation of the proximal energy.

\begin{proposition} \label{prop:branches}
Under all regularity assumptions in Section~\ref{sec:problem}, fix $x\in\R^D$ and $\lambda>0$. Suppose
\begin{equation*}
    \mathcal P_\lambda^f(x)
    =
    \{y_1,\ldots,y_J\},
    \qquad
    H_j
    \defeq
    \operatorname{Hess}_{\M}
    E_{\lambda,x}^{f}(y_j)
    \succ0,
    \quad j=1,\ldots,J.
\end{equation*}
Then, as $\tau\downarrow0$,
\begin{equation*}
    Q_{\lambda,\tau}^{f}(x,\cdot)
    \Rightarrow
    \sum_{j=1}^{J}\omega_j\delta_{y_j},
    \qquad
    \omega_j
    =
    \frac{\rho(y_j)(\det H_j)^{-1/2}}
    {\displaystyle
     \sum_{\ell=1}^{J}
     \rho(y_\ell)(\det H_\ell)^{-1/2}}.
\end{equation*}
Here $\det H_j$ is computed in any orthonormal basis of
$T_{y_j}\M$ and is therefore basis independent.
\end{proposition}
This result is a manifold specialization of the classical Laplace principle~\cite[Theorem~2.1]{hwang1980laplace}. The proof first derives the local Laplace contribution of each proximal minimizer and then normalizes these contributions to obtain the branch weights.

\begin{proof}
Fix $x\in\R^D$ and $\lambda>0$, and abbreviate
\begin{equation*}
    E(y)\defeq E_{\lambda,x}^{f}(y),
    \qquad
    E_\star \defeq \min_{y\in\M}E(y).
\end{equation*}
By assumption, for $j = 1, \ldots, J$, $E(y_j)=E_\star$. We first determine the contribution of a neighborhood of each minimizer to a Laplace-type integral. For every $j$, choose an orthonormal isometry
\begin{equation*}
    U_j:\R^m\longrightarrow T_{y_j}\M.
\end{equation*}
Since $\M$ is an embedded $C^3$ manifold, after reducing the neighborhood there exists a $C^3$ local parameterization
\begin{equation*}
    \psi_j:B(0,r_j)\subset\R^m\longrightarrow\M
\end{equation*}
satisfying $\psi_j(0)=y_j$, and $D\psi_j(0)=U_j$. Because the minimizers are finitely many and distinct, we can choose a sufficiently small radius $0<r<\min_{1\le j\le J}r_j$ such that the neighborhoods $V_j \defeq \psi_j(B(0,r))$ are pairwise disjoint. Since $y_j$ is a minimizer of $E(y)$, $\grad_{\M} E(y_j) = 0$. Moreover, $H_j \defeq \operatorname{Hess}_{\M}E(y_j)\succ0 $. Taylor expansion in the coordinates $\psi_j$ therefore gives
\begin{equation}
    E(\psi_j(v))
    =
    E_\star
    +\frac12 v^\top H_jv
    +R_j(v),
    \qquad
    \frac{R_j(v)}{\|v\|^2}\longrightarrow0
    \quad\text{as }v\to0.
    \label{eq:branch-local-taylor}
\end{equation}
Let $\gamma_j \defeq \lambda_{\min}(H_j)>0$. By reducing $r$ if necessary, the remainder in~\eqref{eq:branch-local-taylor} satisfies
\begin{equation*}
    |R_j(v)|
    \le
    \frac{\gamma_j}{4}\|v\|^2, \qquad \|v\|<r.
\end{equation*}
Consequently,
\begin{equation}
    E(\psi_j(v))-E_\star
    \ge
    \frac{\gamma_j}{4}\|v\|^2,
    \qquad
    \|v\|<r.
    \label{eq:branch-quadratic-lower}
\end{equation}
This lower bound will provide the Gaussian domination needed below.

We next express intrinsic volume in these coordinates. Define
\begin{equation*}
    J_j(v)
    \defeq
    \sqrt{
        \det\!\left(
            D\psi_j(v)^\top D\psi_j(v)
        \right)
    }.
\end{equation*}
Then $\Vol_{\M}(d\psi_j(v)) = J_j(v)dv$. Because $D\psi_j(0)=U_j$ is an isometry, $J_j(0)=1$. Let $a:\M\to\R$ be any continuous function and consider the contribution
\begin{equation*}
    I_{j,a}(\tau) \defeq 
    \int_{V_j}
    a(y)e^{-E(y)/\tau} \Vol_{\M}(dy) = \int_{B(0,r)}
    a(\psi_j(v))
    J_j(v)
    e^{-E(\psi_j(v))/\tau}dv.
\end{equation*}
Make the rescaling $v=\sqrt{\tau}u$:
\begin{equation} \label{eq:branch-scaled-integral}
    e^{E_\star/\tau}\tau^{-m/2}I_{j,a}(\tau) = \int_{\|u\|<r/\sqrt{\tau}}
    a(\psi_j(\sqrt{\tau}u))
    J_j(\sqrt{\tau}u)
    \exp \left[
        -\frac{
            E(\psi_j(\sqrt{\tau}u))-E_\star
        }{\tau}
    \right]du.
\end{equation}
For every fixed $u\in\R^m$, $\psi_j(\sqrt{\tau}u)\rightarrow y_j$, and $J_j(\sqrt{\tau}u)\rightarrow1$ and, from~\eqref{eq:branch-local-taylor},
\begin{equation*}
    \frac{
        E(\psi_j(\sqrt{\tau}u))-E_\star
    }{\tau}
    \rightarrow
    \frac12u^\top H_ju.
\end{equation*}
Hence the integrand in~\eqref{eq:branch-scaled-integral}, extended by zero outside its integration domain, converges pointwise to 
\begin{equation*}
    a(y_j)\exp\!\left(-\frac12u^\top H_ju\right).
\end{equation*}
We now verify domination. From~\eqref{eq:branch-quadratic-lower}, we have
\begin{equation*}
        \exp\left[
        -\frac{
            E(\psi_j(\sqrt{\tau}u))-E_\star
        }{\tau}
    \right]
    \le
    \exp\left(-\frac{\gamma_j}{4}\|u\|^2\right).
\end{equation*}
The functions $a\circ\psi_j$ and $J_j$ are bounded on the closure of the chosen coordinate neighborhood. Thus the absolute value of the integrand is bounded by $C_j \exp \left(- \gamma_j \|u\|^2 / 4 \right)$ for some finite $C_j$, and this function is integrable on $\R^m$. Dominated convergence yields
\begin{equation*} \label{eq:branch-local-laplace}
    \lim_{\tau \to 0} e^{E_\star/\tau}\tau^{-m/2}I_{j,a}(\tau) = a(y_j) \int_{\R^m} e^{-u^\top H_ju/2} du = (2\pi)^{m/2}
    \frac{a(y_j)}{\sqrt{\det H_j}}
\end{equation*}
Equivalently,
\begin{equation}
    I_{j,a}(\tau)
    =
    e^{-E_\star/\tau}(2\pi\tau)^{m/2}
    \left[
        \frac{a(y_j)}{\sqrt{\det H_j}}
        +o(1)
    \right].
    \label{eq:branch-local-asymptotic}
\end{equation}
It remains to show that points outside these neighborhoods make a negligible
contribution. Let $F \defeq \M\setminus\bigcup_{j=1}^{J}V_j$. The set $F$ is compact and contains no global minimizer of $E$. Hence, by continuity of $E$, there exists $\eta>0$ such that
\begin{equation*}
    E(y)\ge E_\star+\eta,
    \qquad y\in F.
\end{equation*}
For continuous $a$, compactness of $\M$ gives $\|a\|_\infty<\infty$, and therefore
\begin{equation*}
    \left|
        \int_F a(y)e^{-E(y)/\tau}\Vol_{\M}(dy)
    \right| \leq \|a\|_\infty\Vol_{\M}(\M)
    e^{-(E_\star+\eta)/\tau} =  o\!\left(
        e^{-E_\star/\tau}\tau^{m/2}
    \right).
\end{equation*}
The last relation follows because $\eta>0$ and exponential decay dominates polynomial growth:
\begin{equation*}
 \frac{e^{-(E_\star+\eta)/\tau}}
      {e^{-E_\star/\tau}\tau^{m/2}}
 =
 \tau^{-m/2}e^{-\eta/\tau}
 \longrightarrow0
 \qquad\text{as }\tau\downarrow0.
\end{equation*}
Thus the contribution from $F$ is negligible at the scale $e^{-E_\star/\tau}\tau^{m/2}$, and the leading terms come from the neighborhoods of the minimizers. Together with~\eqref{eq:branch-local-asymptotic}, this gives the Laplace expansion
\begin{equation}
    \int_{\M}
    a(y)e^{-E(y)/\tau}\,\Vol_{\M}(dy) =
    e^{-E_\star/\tau}(2\pi\tau)^{m/2}
    \left[
        \sum_{j=1}^{J}
        \frac{a(y_j)}{\sqrt{\det H_j}}
        +o(1)
    \right].
    \label{eq:branch-laplace-general}
\end{equation}
We now apply~\eqref{eq:branch-laplace-general} to the refinement law.
Since $\mu(dy)=\rho(y)\Vol_{\M}(dy)$, taking $a=\rho$ yields
\begin{equation}
    Z_{\lambda,\tau}^{f}(x) =
    \int_{\M}
    e^{-E(y)/\tau}\rho(y) \Vol_{\M}(dy) =
    e^{-E_\star/\tau}(2\pi\tau)^{m/2}
    \left[
        \sum_{j=1}^{J}
        \frac{\rho(y_j)}{\sqrt{\det H_j}}
        +o(1)
    \right].
    \label{eq:branch-normalizer-expansion}
\end{equation}
To identify the weak limit, let $\varphi:\M\to\R$ be bounded and continuous. By definition of
$Q_{\lambda,\tau}^{f}$,
\begin{equation*}
    \int_{\M}
    \varphi(y)Q_{\lambda,\tau}^{f}(x,dy)
    =
    \frac{
        \displaystyle
        \int_{\M}
        \varphi(y)e^{-E(y)/\tau}
        \rho(y)\,\Vol_{\M}(dy)
    }{
        Z_{\lambda,\tau}^{f}(x)
    }.
\end{equation*}
Because $\rho\varphi$ is continuous, applying~\eqref{eq:branch-laplace-general} with $a=\rho\varphi$ gives
\begin{equation}
\int_{\M}
    \varphi(y)e^{-E(y)/\tau}
    \rho(y)\,\Vol_{\M}(dy) =
    e^{-E_\star/\tau}(2\pi\tau)^{m/2}
    \left[
        \sum_{j=1}^{J}
        \frac{
            \rho(y_j)\varphi(y_j)
        }{
            \sqrt{\det H_j}
        }
        +o(1)
    \right].
    \label{eq:branch-test-numerator}
\end{equation}
Since $\rho(y_j)>0$ and $H_j\succ0$,
\begin{equation*}
    \sum_{j=1}^{J}
    \frac{\rho(y_j)}{\sqrt{\det H_j}}
    >0.
\end{equation*}
Dividing~\eqref{eq:branch-test-numerator} by \eqref{eq:branch-normalizer-expansion} therefore yields
\begin{equation*}
        \int_{\M}
    \varphi(y)Q_{\lambda,\tau}^{f}(x,dy)
    \rightarrow
    \sum_{j=1}^{J}\omega_j\varphi(y_j), \text{ where }  \omega_j
    =
    \frac{
        \rho(y_j)(\det H_j)^{-1/2}
    }{
        \displaystyle
        \sum_{\ell=1}^{J}
        \rho(y_\ell)(\det H_\ell)^{-1/2}
    }.
\end{equation*}
Finally,
\begin{equation*}
        \sum_{j=1}^{J}\omega_j\varphi(y_j)
    =
    \int_{\M}
    \varphi(y)
    \left(
        \sum_{j=1}^{J}
        \omega_j\delta_{y_j}
    \right)(dy).
\end{equation*}
Since this holds for every bounded continuous test function $\varphi$, the definition of weak convergence gives
\begin{equation*}
        Q_{\lambda,\tau}^{f}(x,\cdot)
    \Rightarrow
    \sum_{j=1}^{J}\omega_j\delta_{y_j},
    \qquad
    \tau\downarrow0.
\end{equation*}
\end{proof}

\section{Proofs of Riemannian Descent and Stationarity} \label{app:riemannian-descent}

In this appendix, we first establish uniform geometric estimates and use them to derive mean displacement and descent bounds for the ideal sampling kernel $\smash{Q_{\lambda,\tau}^{f}}$. We then quantify how these bounds change when sampling from its approximation $\smash{\widehat Q_{\lambda,\tau}^{f}}$. Finally, we combine the descent and feasibility estimates to establish stationarity of~\ouralg. Sections~\ref{app:loc-geometry}--\ref{app:proof-finite-time} use the model assumptions in Section~\ref{sec:problem}, with $\M$ compact and without boundary. Section~\ref{app:proof-boundary} uses the boundary assumptions of Section~\ref{sec:with_boundary_theorems}.

\subsection{Uniform Geometry and Analytical Representatives} \label{app:loc-geometry}

This subsection establishes geometric estimates with constants uniform over $\M$. We do not require $\M$ to be connected. For every point $p\in\M$, intrinsic distances and balls are defined within the connected component containing $p$, using lengths of paths along the manifold. All other distances are Euclidean. Since a manifold is locally connected, its connected components are open. Compactness of $\M$ therefore implies that there are finitely many such components. Compactness also provides a finite collection of coordinate neighborhoods covering $\M$. The proof below uses these neighborhoods to obtain a common local graph radius and uniform derivative bounds, valid at every point of $\M$.

\noindent \textbf{Notation and uniform constants.} Set $F=f\circ P$, where $P$ is the Borel measurable nearest-point selection fixed in Section~\ref{sec:main_results}. Its existence is proved in Lemma~\ref{loc:lem:geometry}. We use the following constants:
\begin{equation*}
    \begin{gathered}
        f_\star=\min_{y\in\M}f(y),\quad
        \Delta_f=\max_{y\in\M}f(y)-f_\star,\quad
        D_{\M}=\operatorname{diam}(\M), \quad
        G=\sup_{p\in\M}\|g(p)\|,\\
        L=\sup_{p\in\M}\|\Hess_{\M} f(p)\|,\quad
        \rho_{\min}=\min_{p\in\M}\rho(p)>0,\quad
        \rho_{\max}=\max_{p\in\M}\rho(p)<\infty.
    \end{gathered}
\end{equation*}
These quantities are finite under the compactness and regularity conditions in Section~\ref{sec:problem}. Constants used in the geometric estimations below are uniform across all the points on the manifold. The function $F$ is defined globally, and is $C^2$ near $\M$ using Assumption~\ref{ass:regularity:smooth} by Theorem~2 of~\cite{leobacher2021existence}. To analyze the sampling kernel $Q$, we first introduce local coordinates, where its defining integrals can be studied. The following lemma restates standard geometric estimates
and the corresponding bounds on the sampling density, with constants uniform over $\M$.

\begin{lemma}\label{loc:lem:coordinates}
Under Assumptions~\ref{ass:regularity:basic},~\ref{ass:regularity:geometry}, and~\ref{ass:regularity:smooth}, there exist constants $b,d_0,C>0$ and $0<a_-\le a_+<\infty$, independent of $p\in\M$, with the following properties.

For each $p\in\M$, choose an orthonormal isometry
$E_p:\R^m\to T_p\M$. A neighborhood of $p$ admits a
$C^3$ parameterization
\begin{equation}\label{loc:eq:tangent-graph}
    \psi_p(v)=p+E_pv+N_p(v),
    \qquad \|v\|<2b,
    \qquad N_p(v)\in N_p\M,
\end{equation}
where $N_p(0)=0$, $DN_p(0)=0$, and
\begin{equation}\label{loc:eq:graph-bounds}
    \|N_p(v)\| \le C\|v\|^2, \quad
    \|DN_p(v)\| \le C\|v\|,\quad
    \|D^2N_p(v)-D^2N_p(0)\| \le C\|v\|,
    \quad \|v\|<2b.
\end{equation}
Every manifold point within distance $d_0$ of $p$ belongs to the inner coordinate neighborhood:
\begin{equation*}
    \M\cap B_{\R^D}(p,d_0)
    \subseteq
    \psi_p\bigl(B_{\R^m}(0,b)\bigr).
\end{equation*}
In these coordinates, $\mu$ has density
\begin{equation}\label{loc:eq:graph-amplitude}
    a_p(v)
    =
    \rho(\psi_p(v))
    \sqrt{\det\!\left(D\psi_p(v)^\top D\psi_p(v)\right)}
\end{equation}
with respect to Lebesgue measure $dv$.
The function $a_p$ is $C^1$ and satisfies
\begin{equation*}
    a_-\le a_p(v)\le a_+,
    \qquad
    \|D\log a_p(v)\|\le C,
    \qquad \|v\|<2b.
\end{equation*}
\end{lemma}

\begin{proof}
Under the model assumptions, $\M$ is a compact embedded $C^3$ manifold without boundary. The local graph representation gives~\eqref{loc:eq:tangent-graph}, with $N_p(0)=DN_p(0)=0$. A finite coordinate cover and the inverse function theorem provide a common radius $2b>0$ and uniform bounds on $D^jN_p$ for $j=1,2,3$. Taylor's formula then gives~\eqref{loc:eq:graph-bounds}.

To obtain a uniform $d_0$, let $V_p=\psi_p(B_{\R^m}(0,b))$.
Because $\psi_p$ is defined on the larger ball $B_{\R^m}(0,2b)$,
the boundary of $V_p$ relative to $\M$ is $\psi_p(\{v:\|v\|=b\})$.
Every point on this boundary has Euclidean distance at least $b$ from $p$.
Thus any curve on $\M$ from $p$ leaving $V_p$ has length at least $b$,
so the intrinsic ball $B_{\M}(p,b)$ lies in $V_p$.
Let $K_b\subset\M\times\M$ consist of pairs in different connected components
and pairs in the same component whose intrinsic distance is at least $b$.
There are finitely many compact components, and intrinsic distance is continuous
on each component, so $K_b$ is compact and disjoint from the diagonal.
Hence there is $d_0>0$ such that $\|p-y\|\ge d_0$ for every $(p,y)\in K_b$.
It follows that $\|p-y\|<d_0$ implies $y\in B_{\M}(p,b)\subset V_p$,
which proves the asserted uniform inclusion.

The Riemannian volume formula, together with $\mu(dy)=\rho(y)\Vol_{\M}(dy)$, gives the coordinate density~\eqref{loc:eq:graph-amplitude}. Write
\begin{equation*}
    J_p(v)
    =
    \sqrt{\det\!\left(D\psi_p(v)^\top D\psi_p(v)\right)},
    \qquad
    a_p(v)=\rho(\psi_p(v))J_p(v).
\end{equation*}
Since $E_p$ is an orthonormal isometry and $DN_p(v)$ takes values in $N_p\M$,
\begin{equation*}
    D\psi_p(v)^\top D\psi_p(v)
    =
    I_m+DN_p(v)^\top DN_p(v).
\end{equation*}
The uniform bounds on $DN_p$ and $D^2N_p$ give a constant $C_J<\infty$, independent of $p$, such that
\begin{equation*}
    1\le J_p(v)\le C_J,
    \qquad
    \|DJ_p(v)\|\le C_J,
    \qquad \|v\|<2b.
\end{equation*}
The strictly positive $C^1$ function $\rho$ has a positive minimum, a finite maximum, and a bounded differential on compact $\M$. Together with the uniform bound on $D\psi_p$, these properties show that $a_p$ is $C^1$ and satisfies uniform bounds
$0<a_-\le a_p(v)\le a_+<\infty$. Finally,
\begin{equation*}
    D\log a_p(v)
    =
    \frac{D(\rho\circ\psi_p)(v)}
         {\rho(\psi_p(v))}
    +
    \frac{DJ_p(v)}{J_p(v)},
\end{equation*}
together with the uniform bounds presented above shows that $D\log a_p(v)$ is also bounded. Enlarging $C$ if necessary completes
the proof.
\end{proof}

Lemma~\ref{loc:lem:coordinates} provides the common coordinate neighborhoods and density bounds used to estimate the moments of $\smash{Q_{\lambda,\tau}^{f}}$. To compare ideal and approximate responses, we also need estimates that apply to arbitrary points in $\R^D$.

\begin{lemma}\label{loc:lem:geometry}
Under the conditions of Lemma~\ref{loc:lem:coordinates}, there exists a Borel measurable nearest-point selection $P:\R^D\to\M$ satisfying~\eqref{eq:near_point_projection}. For any such selection, there exist constants $r,C>0$ such that the nearest point is unique on
\begin{equation*}
    U_{r}(\M)
    =
    \{z\in\R^D:\dist(z,\M)<r\},
\end{equation*}
and the restriction of $P$ to $U_{r}(\M)$ is $C^2$. For all $p,y\in\M$, $n\in N_p\M$, and $z\in\R^D$,
\begin{align}
    |\langle n,y-p\rangle|
    &\le C\|n\|\|y-p\|^2,
    \label{loc:eq:normal-chord}\\
    |f(y)-f(p)-\langle g(p),y-p\rangle|
    &\le C\|y-p\|^2,
    \label{loc:eq:objective-chord}\\
    \|g(y)-g(p)\|
    &\le C\|y-p\|,
    \label{loc:eq:gradient-chord}\\
    |F(z)-f(y)-\langle g(y),z-y\rangle|
    &\le C\|z-y\|^2,
    \label{loc:eq:shadow-taylor}\\
    \|P(z)-y\|
    &\le 2\|z-y\|.
    \label{loc:eq:projection-comparison}
\end{align}
The constants are independent of the points. In particular, $f$ is Lipschitz on $\M$ with respect to Euclidean distance.
\end{lemma}

\begin{proof}
For $z\in\R^D$, let
$\Gamma(z)=\{y\in\M:\|z-y\|=\dist(z,\M)\}$.
Compactness of $\M$ makes $\Gamma(z)$ nonempty and compact.
Since $(z,y)\mapsto\|z-y\|-\dist(z,\M)$ is continuous,
the graph of $\Gamma$ is closed in $\R^D\times\M$.
The Arsenin--Kunugui selection theorem~\cite[Theorem~5.12.1]{srivastava1998borel}
therefore gives a Borel measurable map $P:\R^D\to\M$
with $P(z)\in\Gamma(z)$ for every $z$.
We fix one such selection. Near $\M$, the nearest point is unique,
so this selection agrees with the $C^2$ projection from
Theorem~2 of~\cite{leobacher2021existence}. By compactness, choose $r>0$ so that this holds throughout $U_r(\M)$. We next prove~\eqref{loc:eq:normal-chord}. Let $d_0$ be the common distance threshold from Lemma~\ref{loc:lem:coordinates}.
If $\|y-p\|<d_0$, write $y=\psi_p(v)=p+E_pv+N_p(v)$. Since $n$ and $N_p(v)$ are normal to $T_p\M$, while $E_pv$ is tangent to $T_p\M$, we have
\begin{equation*}
    \langle n,y-p\rangle
    =
    \langle n,N_p(v)\rangle,
    \qquad
    \|v\|^2
    \le \|v\|^2+\|N_p(v)\|^2
    =
    \|y-p\|^2.
\end{equation*}
Using~\eqref{loc:eq:graph-bounds}, we obtain
\begin{equation*}
    |\langle n,y-p\rangle|
    \le C\|n\|\|v\|^2
    \le C\|n\|\|y-p\|^2.
\end{equation*}
For $\|y-p\|\ge d_0$, we have
\begin{equation*}
    |\langle n,y-p\rangle|
    \le \|n\|\|y-p\|
    \le d_0^{-1}\|n\|\|y-p\|^2.
\end{equation*}
Increasing $C$ proves~\eqref{loc:eq:normal-chord} for all $p,y\in\M$. Taking the supremum over unit normal vectors also gives
\begin{equation}\label{loc:eq:normal-vector}
    \|(I_D-\Pi_p)(y-p)\|
    \le C\|y-p\|^2.
\end{equation}
We next establish the objective and gradient estimates. On $U_r(\M)$, the function $F=f\circ P$ is $C^2$.
Moreover, $DP(y)=\Pi_y$ for $y\in\M$ by Theorem C in~\cite{leobacher2021existence}, so
\begin{equation*}
    F(y)=f(y),
    \qquad
    \nabla F(y)
    =
    DP(y)^\top\nabla f(y)
    =
    \Pi_y\nabla f(y)
    =
    g(y).
\end{equation*}
Set $\delta=r/2$. The closed neighborhood $\{z:\dist(z,\M)\le\delta\}$ is compact and contained
in $U_{r}(\M)$. Consequently,
\begin{equation*}
    H
    =
    \sup_{\dist(z,\M)\le\delta}\|\nabla^2F(z)\|
    <\infty.
\end{equation*}
If $y\in\M$ and $\|z-y\|\le\delta$, every point $y+s(z-y)$, $s\in[0,1]$, satisfies
\begin{equation*}
    \dist(y+s(z-y),\M)
    \le s\|z-y\|
    \le\delta.
\end{equation*}
Thus the segment from $y$ to $z$ lies in the region where the Hessian bound applies. Taylor's theorem gives
\begin{equation*}
    |F(z)-f(y)-\langle g(y),z-y\rangle|
    \le \frac{H}{2}\|z-y\|^2.
\end{equation*}
For $\|z-y\|>\delta$, boundedness of $f$ and $g$ on $\M$ provides the required estimate. Since $P(z)\in\M$, we have
\begin{equation*}
     |F(z)-f(y)-\langle g(y),z-y\rangle| \le \Delta_f+G\|z-y\| \le
    \left(
        \frac{\Delta_f}{\delta^2}
        +\frac{G}{\delta}
    \right)\|z-y\|^2.
\end{equation*}
Combining the two cases proves
\eqref{loc:eq:shadow-taylor}.
Applying this estimate with base point $p$ and endpoint $y\in\M$, and using $F(y)=f(y)$, proves~\eqref{loc:eq:objective-chord}. For $\|y-p\|\le\delta$, the segment from $p$ to $y$
also lies in the same closed neighborhood. Therefore,
\begin{equation*}
    g(y)-g(p)
    =
    \int_0^1
    \nabla^2F(p+s(y-p))(y-p)\,ds,
\end{equation*}
which implies
\begin{equation*}
    \|g(y)-g(p)\|
    \le H\|y-p\|.
\end{equation*}
For $\|y-p\|>\delta$, we instead use
\begin{equation*}
    \|g(y)-g(p)\|
    \le 2G
    \le \frac{2G}{\delta}\|y-p\|.
\end{equation*}
This proves~\eqref{loc:eq:gradient-chord}.
Furthermore, \eqref{loc:eq:objective-chord} yields
\begin{equation*}
    |f(y)-f(p)|
    \le
    \bigl(G+C\operatorname{diam}(\M)\bigr)\|y-p\|,
\end{equation*}
establishing the Lipschitz continuity of $f$ with respect to Euclidean distance. Finally, the defining property of $P$ gives $\|P(z)-z\|\le\|y-z\|$ for every $y\in\M$.
The triangle inequality then implies
\begin{equation*}
    \|P(z)-y\|
    \le \|P(z)-z\|+\|z-y\|
    \le 2\|z-y\|,
\end{equation*}
which proves~\eqref{loc:eq:projection-comparison}. All estimates hold with a common sufficiently large
constant $C$, independently of the selection outside $U_{r}(\M)$.
\end{proof}

\subsection{Uniform localization and ideal progress} \label{app:loc-local-analysis}

This subsection analyzes a single update drawn from the ideal kernel $Q_{\lambda,\tau}^{f}(x,\cdot)$. We first establish localization and moment bounds for the sampled point, then use these bounds to derive gradient-like mean displacement and expected-descent estimates.

Fix $0<\bar\tau<\infty$. The estimates below hold uniformly over the parameter ranges and neighborhoods specified in their statements. We write $s = \lambda \tau$ for the variance scale of the sampling fluctuations.

\noindent \textbf{Uniform localization.} We first show that every proximal minimizer lies within a radius independent of the current point. We then bound sampling fluctuations around the minimizer and the probability of drawing a sample outside this neighborhood.

\begin{lemma}\label{loc:lem:localization}
Under the conditions of Lemma~\ref{loc:lem:coordinates}, there exist $r>0$, $\lambda_{\rm loc}\in(0,1]$, and positive constants such that, for $\dist(x,\M)<r$, $0<\lambda\le\lambda_{\rm loc}$, and $0<\tau\le\bar\tau$, the following statements hold. Write $p=P(x)$, $n=x-p\in N_p\M$, and $q=\|g(p)\|$. The proximal energy has a
unique global minimizer $y_\star$ and
\begin{equation}\label{loc:eq:minimizer-localization}
 \|y_\star-p\|\le C\lambda q.
\end{equation}
For $Y\sim Q^f_{\lambda,\tau}(x,\cdot)$,
\begin{equation}\label{loc:eq:local-second-moment}
 \E\|Y-y_\star\|^2\le Cs,
 \qquad
 \E\|Y-p\|^2\le C(\lambda^2q^2+s).
\end{equation}
Let $V_p=\psi_p(B_{\R^m}(0,b))$ denote the inner coordinate neighborhood of $p=P(x)$. For $Y\sim Q_{\lambda,\tau}^{f}(x,\cdot)$, the probability of sampling outside this neighborhood satisfies
\begin{equation}\label{loc:eq:graph-tail}
    \Prob\{Y\notin V_p\}
    =
    Q_{\lambda,\tau}^{f}(x,\M\setminus V_p)
    \le C s^{-m/2}e^{-c/s},
    \qquad s=\lambda\tau.
\end{equation}
For every fixed $\ell\ge0$, the same proof gives $\E\|Y-y_\star\|^\ell\le C_\ell s^{\ell/2}$.
\end{lemma}

\begin{proof}
Take $r$ smaller than the projection radius in Lemma~\ref{loc:lem:geometry}.
The anchor displacement is normal by the first-order condition for the
unique nearest point. For $y \in \M$, define
\begin{align*}
 D_x(y)&=\frac{\|y-x\|^2-\|n\|^2}{2}
 =\frac{\|y-p\|^2}{2}-\langle n,y-p\rangle,\\
 \phi(y)&=\lambda(f(y)-f(p))+D_x(y).
\end{align*}
Subtracting $f(p)+\|n\|^2/(2\lambda)$ from $E_{\lambda,x}^{f}$ leaves the law $Q^f_{\lambda, \tau}$ unchanged, and its density is therefore proportional to
$e^{-\phi(y)/s}$. By~\eqref{loc:eq:normal-chord}, decreasing $r$ gives the
global comparison
\begin{equation*}
 \tfrac14\|y-p\|^2\le D_x(y)\le\tfrac34\|y-p\|^2
 \qquad(y\in\M).
\end{equation*}
The objective Taylor estimate then implies
\begin{equation}\label{loc:eq:energy-coercivity}
 \phi(y)\ge(\tfrac14-C\lambda)\|y-p\|^2
 -\lambda q\|y-p\|.
\end{equation}
Choose $\lambda_{\rm loc}$ so the quadratic coefficient is at least $1/8$. Let $y_\star$ be any global minimizer of $\phi$ on $\M$, we have $\phi(y_\star)\le\phi(p)=0$. The lower bound above implies
\begin{equation*}
    0
    \ge \phi(y_\star)
    \ge
    \frac18\|y_\star-p\|^2
    -\lambda q\|y_\star-p\|.
\end{equation*}
If $y_\star=p$, the desired distance bound holds immediately. Otherwise, rearranging and dividing by $\|y_\star-p\|>0$ yields
\begin{equation*}
    \frac18\|y_\star-p\|
    \le\lambda q,
    \qquad\text{hence}\qquad
    \|y_\star-p\|\le8\lambda q.
\end{equation*}

To place these minimizers inside the local coordinate
neighborhood, let $d_0>0$ be the uniform distance
threshold from Lemma~\ref{loc:lem:coordinates}.
Since $q\le G$, we may further choose
$\lambda_{\rm loc}$ so that
$8\lambda_{\rm loc}G\le d_0/2$. Then every global
minimizer satisfies
\begin{equation*}
    \|y_\star-p\|
    \le8\lambda q
    \le8\lambda_{\rm loc}G
    <d_0,
\end{equation*}
and therefore belongs to
$\psi_p(B_{\R^m}(0,b))$.
Because $G$ and $d_0$ are independent of $p$,
the same choice of $\lambda_{\rm loc}$ works
uniformly over the current points under consideration.

The preceding argument shows that every global minimizer lies inside a local coordinate neighborhood of $p$. We next show that the proximal energy is strictly convex in coordinates on a smaller neighborhood containing all these minimizers. Using Lemma~\ref{loc:lem:coordinates}, define $\Phi(v)=\phi(\psi_p(v))$. Since $E_pv$ is tangent and $N_p(v)$ is normal to
$T_p\M$, we obtain
\begin{equation}\label{loc:eq:coordinate-energy}
    \Phi(v)
    =
    \lambda\bigl(f(\psi_p(v))-f(p)\bigr)
    +\frac12\bigl(\|v\|^2+\|N_p(v)\|^2\bigr)
    -\langle n,N_p(v)\rangle.
\end{equation}
Because $N_p(0)=DN_p(0)=0$, the Hessian of the last two terms at $v=0$ is
\begin{equation}\label{loc:eq:offset-metric}
    A_{p,n}
    =
    I_m-\langle n,D^2N_p(0)\rangle.
\end{equation}
Here $\langle n,D^2N_p(0)\rangle$ denotes the symmetric matrix whose $(i,j)$ entry is $\langle n,\partial_i\partial_jN_p(0)\rangle$. The uniform bound on $D^2N_p(0)$ implies
\begin{equation*}
    \|A_{p,n}-I_m\|
    \le C\|n\|
    \le Cr.
\end{equation*}
Choosing $r$ sufficiently small therefore gives
\begin{equation*}
    \frac34 I_m
    \preceq A_{p,n}
    \preceq \frac54 I_m.
\end{equation*}
Thus the distance-dependent energy has a uniformly positive definite Hessian at the coordinate origin.

We now control how the full Hessian differs from $A_{p,n}$ away from the origin. Differentiating~\eqref{loc:eq:coordinate-energy} gives
\begin{align*}
    D^2\Phi(v)-A_{p,n}
    &=
    \lambda D^2(f\circ\psi_p)(v)
    +DN_p(v)^\top DN_p(v)\\
    &\quad
    +\langle N_p(v),D^2N_p(v)\rangle
    -\langle n,D^2N_p(v)-D^2N_p(0)\rangle.
\end{align*}
The matrix-valued inner products use the same entrywise convention as in~\eqref{loc:eq:offset-metric}. The bounds $\|DN_p(v)\|\le C\|v\|$ and $\|N_p(v)\|\le C\|v\|^2$, together with the uniform bound on $D^2N_p$, control the second and third terms by $C\|v\|^2$. The last term is bounded by $C\|n\|\|v\|$ using~\eqref{loc:eq:graph-bounds}. Finally, $f\circ\psi_p$ has uniformly bounded second derivatives, so the objective term is bounded by $C\lambda$. Consequently,
\begin{equation*}
    \|D^2\Phi(v)-A_{p,n}\|
    \le C\bigl(\lambda+\|v\|^2+r\|v\|\bigr).
\end{equation*}
Choose a sufficiently small coordinate radius and then reduce $\lambda_{\rm loc}$ so that the right-hand side is at most $1/4$ throughout this coordinate ball. Combining this estimate with the bounds on $A_{p,n}$ gives fixed constants
$0<a<A<\infty$ such that
\begin{equation}\label{loc:eq:coordinate-convexity}
    aI_m\preceq D^2\Phi(v)\preceq AI_m.
\end{equation}
For example, the preceding choices allow $a=1/2$ and $A=3/2$.

Choose fixed radii $0<b_1<b_2<b_3<b$ such that $B_{\R^m}(0,b_3)$ lies inside the region where~\eqref{loc:eq:coordinate-convexity} holds. For subsequent integrals, write
\begin{equation*}
    V_i=\psi_p(B_{\R^m}(0,b_i)),
    \qquad i=1,2,3.
\end{equation*}
For any global minimizer $y_\star$, let $v_\star=\psi_p^{-1}(y_\star)$. The localization estimate gives
\begin{equation*}
    \|v_\star\|
    \le \|y_\star-p\|
    \le 8\lambda q
    \le 8\lambda_{\rm loc}G.
\end{equation*}
We may therefore reduce $\lambda_{\rm loc}$ so that every global minimizer has coordinates in $B_{\R^m}(0,b_1/2)$.

Each such point is an interior minimizer of $\Phi$ and hence satisfies $D\Phi(v_\star)=0$. The lower Hessian bound in~\eqref{loc:eq:coordinate-convexity} makes $\Phi$ strictly convex on the convex ball $B_{\R^m}(0,b_3)$, so this ball contains at most one critical point. Since every global minimizer has already been placed inside this ball, the
global minimizer on $\M$ is unique.

For any $v\in B_{\R^m}(0,b_3)$, the segment from
$v_\star$ to $v$ remains inside the same ball.
Taylor's formula and $D\Phi(v_\star)=0$ give
\begin{equation*}
    \Phi(v)-\Phi(v_\star)
    =
    \int_0^1(1-t)
    D^2\Phi\bigl(v_\star+t(v-v_\star)\bigr)
    [v-v_\star,v-v_\star]\,dt.
\end{equation*}
Applying the two Hessian bounds and using
$\int_0^1(1-t)\,dt=1/2$ yields
\begin{equation}\label{loc:eq:centered-quadratic}
    \frac a2\|v-v_\star\|^2
    \le \Phi(v)-\Phi(v_\star)
    \le \frac A2\|v-v_\star\|^2.
\end{equation}
We next establish a positive energy gap outside the inner neighborhood $V_1$. Set
\begin{equation*}
    d_1=\frac12\min\{d_0,b_1\}.
\end{equation*}
If $\|y-p\|<d_1$, Lemma~\ref{loc:lem:coordinates}
gives $y=\psi_p(v)$, and
$\|v\|\le\|y-p\|<b_1$.
Thus $y\in V_1$, which shows that
\begin{equation*}
    y\in\M\setminus V_1
    \quad\Longrightarrow\quad
    \|y-p\|\ge d_1.
\end{equation*}
Further reduce $\lambda_{\rm loc}$ so that
$\lambda_{\rm loc}G\le d_1/16$.
For $y\in\M\setminus V_1$, write
$t=\|y-p\|\ge d_1$.
Using~\eqref{loc:eq:energy-coercivity}, we obtain
\begin{equation*}
    \phi(y)
    \ge \frac18t^2-\lambda Gt
    \ge \frac18t^2-\frac1{16}t^2
    \ge \frac{d_1^2}{16}.
\end{equation*}
Since $\phi(y_\star)\le\phi(p)=0$, setting
$\eta=d_1^2/16$ gives
\begin{equation}\label{loc:eq:outside-energy-gap}
    \phi(y)-\phi(y_\star)\ge\eta,
    \qquad y\in\M\setminus V_1.
\end{equation}
The constant $\eta$ is independent of $p$ and the
parameters in the stated ranges. Since the argument
uses the global bound~\eqref{loc:eq:energy-coercivity},
it also covers points on other connected components
of $\M$.

To estimate the normalization constant, subtract the
minimum energy and define
\begin{equation*}
    \mathcal Z
    =
    \int_{\M}
    e^{-(\phi(y)-\phi(y_\star))/s}\,\mu(dy).
\end{equation*}
Subtracting this constant from the energy does not
change the normalized sampling distribution.

We already have $\|v_\star\|<b_1/2$.
Further reduce $\lambda_{\rm loc}$ so that
$\sqrt{\lambda_{\rm loc}\bar\tau}\le b_1/2$.
Since $s=\lambda\tau\le\lambda_{\rm loc}\bar\tau$,
every point with $\|v-v_\star\|<\sqrt s$ satisfies
\begin{equation*}
    \|v\|
    \le \|v_\star\|+\|v-v_\star\|
    <b_1.
\end{equation*}
Thus this entire integration region lies inside
the inner coordinate neighborhood.

Using the coordinate-density lower bound
$a_p(v)\ge a_-$ and the upper quadratic bound in
\eqref{loc:eq:centered-quadratic}, we obtain
\begin{equation}
        \mathcal Z
    \ge
    a_-\int_{\|v-v_\star\|<\sqrt s}
    e^{-A\|v-v_\star\|^2/(2s)}\,dv =
    a_-s^{m/2}
    \int_{\|u\|<1}e^{-A\|u\|^2/2}\,du
    =c_Zs^{m/2},
    \label{loc:eq:local-normalizer-lower}
\end{equation}
where the equality uses the substitution
$v=v_\star+\sqrt s\,u$.
The constant
$c_Z=a_-\int_{\|u\|<1}e^{-A\|u\|^2/2}\,du$
is strictly positive and independent of the current
point and parameters.

For the upper bound, split the integral defining
$\mathcal Z$ over $V_3$ and $\M\setminus V_3$.
On $V_3$, apply the lower quadratic bound in
\eqref{loc:eq:centered-quadratic} and the
coordinate-density upper bound $a_p(v)\le a_+$.
On $\M\setminus V_3$, the energy gap
\eqref{loc:eq:outside-energy-gap} applies because
$V_1\subset V_3$. Hence
\begin{equation*}
        \mathcal Z \le
    a_+\int_{\R^m}
    e^{-a\|v-v_\star\|^2/(2s)}\,dv
    +e^{-\eta/s} =
    a_+\left(\frac{2\pi}{a}\right)^{m/2}s^{m/2}
    +e^{-\eta/s}.
\end{equation*}
For every fixed $\beta\ge0$,
$s^{-\beta}e^{-\eta/s}$ tends to zero as
$s\downarrow0$ and is bounded on
$(0,\lambda_{\rm loc}\bar\tau]$.
Taking $\beta=m/2$ shows that
$e^{-\eta/s}\le Cs^{m/2}$ throughout this interval.
Consequently, $\mathcal Z\le C_Zs^{m/2}$
for a uniform constant $C_Z$.

We now use these estimates to bound the moments of
the displacement from $y_\star$.
The zeroth moment equals one, so consider any fixed
$\ell>0$.
Let $L_\psi$ be a uniform upper bound on
$\|D\psi_p\|$ over the coordinate neighborhoods.
For $v\in B_{\R^m}(0,b_3)$, integration along the
segment from $v_\star$ to $v$ gives
\begin{equation*}
    \|\psi_p(v)-y_\star\|
    \le L_\psi\|v-v_\star\|.
\end{equation*}
Therefore, the contribution from $V_3$ to the
unnormalized moment satisfies
\begin{align*}
    &\int_{V_3}\|y-y_\star\|^\ell
    e^{-(\phi(y)-\phi(y_\star))/s}\,\mu(dy)\\
    &\qquad\le
    a_+L_\psi^\ell
    \int_{\R^m}\|v-v_\star\|^\ell
    e^{-a\|v-v_\star\|^2/(2s)}\,dv\\
    &\qquad=
    a_+L_\psi^\ell s^{(m+\ell)/2}
    \int_{\R^m}\|u\|^\ell e^{-a\|u\|^2/2}\,du\\
    &\qquad\le C_\ell s^{(m+\ell)/2}.
\end{align*}
Here we again used $v=v_\star+\sqrt s\,u$;
the remaining Gaussian moment integral is finite.

On $\M\setminus V_3$, the displacement is at most
$D_{\M}$ and the energy gap is at least $\eta$.
The contribution from this region is consequently
at most $D_{\M}^{\ell}e^{-\eta/s}$.
Dividing the sum of the two contributions by
\eqref{loc:eq:local-normalizer-lower} gives
\begin{equation*}
    \E\|Y-y_\star\|^\ell
    \le C_\ell s^{\ell/2}
       +C_\ell s^{-m/2}e^{-\eta/s}.
\end{equation*}
To bound the second term at the same order as the
first, write
\begin{equation*}
    s^{-m/2}e^{-\eta/s}
    =
    s^{\ell/2}
    \left(s^{-(m+\ell)/2}e^{-\eta/s}\right).
\end{equation*}
The parenthesized factor is uniformly bounded on
the prescribed interval. Hence
\begin{equation*}
    \E\|Y-y_\star\|^\ell
    \le C'_\ell s^{\ell/2}.
\end{equation*}
In particular, $\E\|Y-y_\star\|^2\le Cs$.
Combining this estimate with~\eqref{loc:eq:minimizer-localization} yields
\begin{align*}
    \E\|Y-p\|^2
    &\le
    2\E\|Y-y_\star\|^2
    +2\|y_\star-p\|^2\\
    &\le C\bigl(s+\lambda^2q^2\bigr),
\end{align*}
which proves~\eqref{loc:eq:local-second-moment}.

Finally, applying the energy-gap bound directly
on $\M\setminus V_1$ gives
\begin{equation*}
    \Prob\{Y\notin V_1\}
    =
    \frac{
        \int_{\M\setminus V_1}
        e^{-(\phi(y)-\phi(y_\star))/s}\,\mu(dy)
    }{\mathcal Z}
    \le
    \frac{e^{-\eta/s}}{c_Zs^{m/2}}.
\end{equation*}
This establishes~\eqref{loc:eq:graph-tail} for $V_1$.
Since $V_1\subseteq\psi_p(B_{\R^m}(0,b))$, the
same probability bound holds outside the larger
coordinate neighborhood in the lemma statement.
\end{proof}

Lemma~\ref{loc:lem:localization} bounds both the displacement of the proximal minimizer from $p$ and the sampling
fluctuations around that minimizer. These estimates control the size of an update, but do not yet establish its direction or its effect on the objective.

\noindent\textbf{Mean displacement and expected descent.}
The following lemma connects these localization estimates
to optimization progress. It bounds the mean displacement
and establishes expected descent for current points near
$\M$. When the current point lies on $\M$, it further
identifies the mean displacement as an approximation to
a negative Riemannian-gradient step.

\begin{lemma} \label{loc:lem:ideal-drift}
Under the conditions of Lemma~\ref{loc:lem:localization}, there are uniform
$a_0,C>0$ such that
\begin{align}
 \|\E(Y-p)\|&\le C\lambda,\label{loc:eq:ideal-mean-bound}\\
 \E f(Y)&\le f(p)-a_0\lambda\|g(p)\|^2+C\lambda\tau.
 \label{loc:eq:ideal-descent}
\end{align}
For the feasible anchor $x=p$ one has the sharper mean identity
\begin{equation}\label{loc:eq:ideal-gradient-bias}
 \|\E(Y-p)+\lambda g(p)\|
 \le C\lambda(\lambda+\tau).
\end{equation}
\end{lemma}

\begin{proof}
We first relate the mean displacement to the Riemannian
gradient through an integration-by-parts identity.
We then combine this relation with the moment bounds
from Lemma~\ref{loc:lem:localization} to establish
the mean-displacement and descent estimates.

Retain the notation $p=P(x)$, $n=x-p$, and $s=\lambda\tau$.
Write
\begin{equation*}
    \gamma=E_p^\top g(p),
    \qquad q=\|\gamma\|=\|g(p)\|,
    \qquad A=A_{p,n},
\end{equation*}
where $A_{p,n}$ is defined
in~\eqref{loc:eq:offset-metric}.
Thus, $\gamma$ represents the Riemannian gradient
in the chosen orthonormal tangent coordinates.

Differentiating~\eqref{loc:eq:coordinate-energy} gives
\begin{equation*}
    D\Phi(v)
    =
    v+DN_p(v)^\top N_p(v)-DN_p(v)^\top n
    +\lambda D(f\circ\psi_p)(v).
\end{equation*}
We estimate these terms separately.
First, the coordinate bounds imply
\begin{equation*}
    \|DN_p(v)^\top N_p(v)\|
    \le C\|v\|^3
    \le C\|v\|^2
\end{equation*}
on the fixed coordinate ball.
Next, Taylor's theorem and the uniformly bounded
third derivative of $N_p$ give
\begin{equation*}
    \bigl\|DN_p(v)-D^2N_p(0)[v,\cdot]\bigr\|
    \le C\|v\|^2.
\end{equation*}
Consequently,
\begin{equation*}
    \bigl\|
        DN_p(v)^\top n
        -\langle n,D^2N_p(0)\rangle v
    \bigr\|
    \le C\|n\|\|v\|^2.
\end{equation*}
Finally, $D(f\circ\psi_p)(0)=\gamma$, and the
uniformly bounded Hessian of $f\circ\psi_p$ yields
\begin{equation*}
    \lambda
    \|D(f\circ\psi_p)(v)-\gamma\|
    \le C\lambda\|v\|.
\end{equation*}
Combining these estimates with the definition of $A$
and the bound $\|n\|<r$, we obtain
\begin{equation}\label{loc:eq:coordinate-gradient-expansion}
    D\Phi(v)=Av+\lambda\gamma+R(v),
    \qquad
    \|R(v)\|\le C(\|v\|^2+\lambda\|v\|).
\end{equation}
This expansion uses bounded second derivatives of $f$, while the third-derivative bound is needed only for $N_p$.

We next integrate~\eqref{loc:eq:coordinate-gradient-expansion}
against the local sampling density.
Let $b_1<b_2<b_3$ be the radii used in the preceding proof. Take a smooth function $\chi:\R^m\to[0,1]$ satisfying
\begin{equation*}
    \chi(v)=1 \quad\text{for }\|v\|\le b_1,
    \qquad
    \chi(v)=0 \quad\text{for }\|v\|\ge b_2.
\end{equation*}
The fixed radii allow $\|D\chi\|$ to be bounded uniformly.
Moreover, $\chi$ vanishes in a neighborhood of the
boundary of $B(0,b_3)$.

Write $\Phi_\star=\Phi(v_\star)$.
The function
$\chi a_p e^{-(\Phi-\Phi_\star)/s}$
is $C^1$ and compactly supported inside $B(0,b_3)$.
The integral of each of its partial derivatives is
therefore zero:
\begin{equation*}
    \int_{B(0,b_3)}
    D\!\left[
        \chi(v)a_p(v)
        e^{-(\Phi(v)-\Phi_\star)/s}
    \right]\,dv=0.
\end{equation*}
Expanding the derivative and rearranging gives
\begin{equation*}
    \int_{B(0,b_3)}
    \chi D\Phi\,a_p e^{-(\Phi-\Phi_\star)/s}\,dv =
    s\int_{B(0,b_3)}
    \bigl(D\chi+\chi D\log a_p\bigr)
    a_p e^{-(\Phi-\Phi_\star)/s}\,dv.
\end{equation*}
No boundary term appears because of the support
condition on $\chi$.

Let $\mathcal Z$ be the normalizing integral over
$\M$ defined in the preceding proof.
After division by $\mathcal Z$, the integral of
$a_p e^{-(\Phi-\Phi_\star)/s}$ over $B(0,b_3)$
is the probability of this coordinate neighborhood,
and hence is at most one.
The uniform bounds on $D\chi$ and $D\log a_p$ therefore
imply
\begin{equation*}
    \left\|
        \frac1{\mathcal Z}
        \int_{B(0,b_3)}
        \chi D\Phi\,a_p
        e^{-(\Phi-\Phi_\star)/s}\,dv
    \right\|
    \le Cs.
\end{equation*}

To relate this integral to the full mean displacement,
define
\begin{equation*}
    V=E_p^\top(Y-p),
    \qquad
    M_2=\E\|Y-p\|^2.
\end{equation*}
The random vector $V$ is defined for every $Y\in\M$.
When $Y=\psi_p(v)$, orthogonality gives $V=v$.
Also, $\|V\|\le\|Y-p\|\le D_{\M}$.

The remainder in~\eqref{loc:eq:coordinate-gradient-expansion}
satisfies
\begin{equation*}
    \frac1{\mathcal Z}
    \int_{B(0,b_3)}
    \chi(v)\|R(v)\|a_p(v)
    e^{-(\Phi(v)-\Phi_\star)/s}\,dv \le
    C\bigl(\E\|Y-p\|^2+\lambda\E\|Y-p\|\bigr) \le C(M_2+\lambda\sqrt{M_2})
\end{equation*}
where the last inequality follows from Cauchy--Schwarz.
Substituting the coordinate expansion into the
preceding integration-by-parts estimate thus yields
\begin{equation*}
    \left\|
        \frac1{\mathcal Z}
        \int_{B(0,b_3)}
        \chi(v)(Av+\lambda\gamma)a_p(v)
        e^{-(\Phi(v)-\Phi_\star)/s}\,dv
    \right\|
    \le C(s+M_2+\lambda\sqrt{M_2}).
\end{equation*}

It remains to remove the cutoff from this linear
expression.
The cutoff equals one throughout
$\psi_p(B(0,b_1))$.
The energy gap~\eqref{loc:eq:outside-energy-gap}
and the normalizer bound~\eqref{loc:eq:local-normalizer-lower}
give
\begin{equation*}
    \Prob\{Y\notin\psi_p(B(0,b_1))\}
    \le Cs^{-m/2}e^{-c/s}
    \le Cs.
\end{equation*}
The final inequality holds uniformly on the prescribed
interval for $s$, because
$s^{-m/2-1}e^{-c/s}$ is bounded there.
Since $A$, $V$, and $\lambda\gamma$ are uniformly
bounded, the difference between the cutoff integral
and $A\E V+\lambda\gamma$ is consequently at most
$Cs$.
We conclude that
\begin{equation}\label{loc:eq:mean-coordinate-equation}
    \|A\E V+\lambda\gamma\|
    \le C(s+M_2+\lambda\sqrt{M_2}).
\end{equation}

This estimate controls the tangent component of the
mean displacement.
For the normal component, the global bound
in~\eqref{loc:eq:normal-vector} gives
\begin{equation}\label{loc:eq:mean-normal-bound}
    \|(I_D-\Pi_p)\E(Y-p)\|
    \le
    \E\|(I_D-\Pi_p)(Y-p)\|
    \le CM_2.
\end{equation}
The second-moment estimate~\eqref{loc:eq:local-second-moment}
in Lemma~\ref{loc:lem:localization} gives
\begin{equation}\label{loc:eq:local-moment-substitution}
    M_2=\E\|Y-p\|^2\le C(\lambda^2q^2+s),
    \qquad
    \sqrt{M_2}\le C(\lambda q+\sqrt s).
\end{equation}
The second inequality follows by taking square roots
and using $\sqrt{a+b}\le\sqrt a+\sqrt b$ for $a,b\ge0$.

We can now establish the first mean-displacement bound. Since $q\le G$, $s=\lambda\tau\le\bar\tau\lambda$,
and $0<\lambda\le1$, we have
\begin{equation*}
    M_2
    \le C(\lambda^2q^2+s)
    \le C(G^2\lambda^2+\bar\tau\lambda)
    \le C(G^2+\bar\tau)\lambda.
\end{equation*}
Thus, after enlarging the uniform constant $C$ if
necessary, $M_2\le C\lambda$.
It follows that
\begin{equation*}
    s+M_2+\lambda\sqrt{M_2}
    \le
    \bar\tau\lambda+C\lambda+\sqrt C\,\lambda^{3/2}
    \le
    (\bar\tau+C+\sqrt C)\lambda,
\end{equation*}
where the last inequality uses $\lambda^{3/2}\le\lambda$.
Absorbing the fixed coefficient into $C$ gives
$s+M_2+\lambda\sqrt{M_2}\le C\lambda$.

The uniform bound on $A^{-1}$ and
\eqref{loc:eq:mean-coordinate-equation} therefore give
$\|\E V\|\le C\lambda$.
Using the decomposition
\begin{equation*}
    \E(Y-p)
    =
    E_p\E V+(I_D-\Pi_p)\E(Y-p)
\end{equation*}
together with~\eqref{loc:eq:mean-normal-bound},
we obtain
\begin{equation*}
    \|\E(Y-p)\|\le C\lambda.
\end{equation*}
This proves~\eqref{loc:eq:ideal-mean-bound}.

When $x=p$, we have $n=0$ and $A=I_m$.
Since $g(p)=E_p\gamma$,
\eqref{loc:eq:mean-coordinate-equation} and
\eqref{loc:eq:mean-normal-bound} now imply
\begin{equation*}
    \|\E(Y-p)+\lambda g(p)\|
    \le C(s+M_2+\lambda\sqrt{M_2}).
\end{equation*}
Substituting~\eqref{loc:eq:local-moment-substitution}
and using $q\le G$ yields
\begin{equation*}
    \|\E(Y-p)+\lambda g(p)\|
    \le C(\lambda^2+s+\lambda\sqrt s).
\end{equation*}
The inequality $2\lambda\sqrt s\le\lambda^2+s$
then gives
\begin{equation*}
    \|\E(Y-p)+\lambda g(p)\|
    \le C(\lambda^2+s)
    =C\lambda(\lambda+\tau),
\end{equation*}
which proves~\eqref{loc:eq:ideal-gradient-bias}.

Finally, we establish expected descent.
The objective estimate~\eqref{loc:eq:objective-chord}
implies
\begin{equation*}
    \E f(Y)-f(p)
    \le
    \langle g(p),\E(Y-p)\rangle+CM_2
    =
    \gamma^\top\E V+CM_2.
\end{equation*}
Define the residual vector
\begin{equation*}
    r_s=A\E V+\lambda\gamma.
\end{equation*}
By~\eqref{loc:eq:mean-coordinate-equation},
\begin{equation*}
    \|r_s\|\le C(s+M_2+\lambda\sqrt{M_2}).
\end{equation*}
Solving for $\E V$ gives
\begin{equation*}
    \gamma^\top\E V
    =
    -\lambda\gamma^\top A^{-1}\gamma
    +\gamma^\top A^{-1}r_s.
\end{equation*}
The uniform eigenvalue bounds for $A$ imply that,
for some constant $c_A>0$,
\begin{equation*}
    \gamma^\top A^{-1}\gamma\ge c_Aq^2,
    \qquad
    |\gamma^\top A^{-1}r_s|\le Cq\|r_s\|.
\end{equation*}
Consequently,
\begin{equation*}
    \E f(Y)-f(p)
    \le
    -c_A\lambda q^2
    +Cq(s+M_2+\lambda\sqrt{M_2})
    +CM_2.
\end{equation*}
To simplify the error terms, substitute
\eqref{loc:eq:local-moment-substitution} and use $q\le G$:
\begin{align*}
    qM_2
    &\le C(\lambda^2q^3+qs)
     \le C\lambda^2q^2+Cs,\\
    \lambda q\sqrt{M_2}
    &\le C\lambda^2q^2+C\lambda q\sqrt s,\\
    M_2
    &\le C\lambda^2q^2+Cs.
\end{align*}
Together with $qs\le Gs$, these estimates yield
\begin{equation*}
    \E f(Y)-f(p)
    \le
    -c_A\lambda q^2
    +C\lambda^2q^2
    +C\lambda q\sqrt s
    +Cs.
\end{equation*}
Young's inequality gives
\begin{equation*}
    C\lambda q\sqrt s
    \le
    \frac{c_A}{4}\lambda q^2+C'\lambda s.
\end{equation*}
Choose $\lambda_{\rm loc}$ sufficiently small that
$C\lambda_{\rm loc}\le c_A/4$.
Then
$C\lambda^2q^2\le(c_A/4)\lambda q^2$.
Since $\lambda\le1$, the remaining term
$C'\lambda s$ is bounded by $C's$.
We therefore obtain
\begin{equation*}
    \E f(Y)
    \le
    f(p)-\frac{c_A}{2}\lambda q^2+Cs
    =
    f(p)-\frac{c_A}{2}\lambda\|g(p)\|^2
    +C\lambda\tau.
\end{equation*}
This proves~\eqref{loc:eq:ideal-descent}
with $a_0=c_A/2$.
\end{proof}

From now on, $\lambda_{\rm loc}$ denotes this final choice, for which all the preceding estimates hold.

Lemma~\ref{loc:lem:ideal-drift} establishes a mean-displacement bound and an expected-descent estimate for every current
point $x$ in the specified neighborhood of $\M$. When $x\in\M$, the matrix $A_{p,n}$ becomes the identity,
and~\eqref{loc:eq:ideal-gradient-bias} shows that the mean
displacement approximates $-\lambda g(x)$.

\noindent\textbf{Effect of sampling error.} We next extend these estimates from the ideal law
$Q_{\lambda,\tau}^f(x,\cdot)$ to an approximate law $\widehat Q$.
The Wasserstein error bound allows us to construct
ideal and approximate samples jointly so that their
expected squared distance is controlled.
The estimate~\eqref{loc:eq:shadow-taylor} then relates
this distance to the change in objective value at
the nearest feasible representative.
Because this estimate holds throughout the ambient
space, it also covers approximate samples outside
the neighborhood where $P$ is smooth.
The representative $P(\widehat Y)$ is used only
to analyze the sampled update.

\begin{lemma} \label{loc:lem:transport-descent}
Use the conditions of Lemma~\ref{loc:lem:localization}. Let
$\widehat Y\sim\widehat Q$ satisfy
\begin{equation*}
     W_2(\widehat Q,Q^f_{\lambda,\tau}(x,\cdot))\le \varepsilon.
\end{equation*}
Write $p=P(x)$, $\widehat p=P(\widehat Y)$, and let $Y\sim Q^f_{\lambda,\tau}(x,\cdot)$. Then uniform constants $a_1,C>0$ satisfy
\begin{align}
 \E F(\widehat Y)
 &\le f(p)-a_1\lambda\|g(p)\|^2+C\lambda\tau+C \varepsilon^2/\lambda,
 \label{loc:eq:transport-descent}\\
 \E\|\widehat p-p\|^2
 &\le C\bigl(\lambda^2\|g(p)\|^2+\lambda\tau+\varepsilon^2\bigr),
 \label{loc:eq:shadow-displacement}\\
 \|\E(\widehat p-p)-\E(Y-p)\|&\le2\varepsilon.
 \label{loc:eq:shadow-drift-error}
\end{align}
The last inequality holds at every ambient anchor, without the tube
restriction.
\end{lemma}

\begin{proof}
Use an optimal coupling and abbreviate $q=\|g(p)\|$ and
$M_2=\E\|Y-p\|^2$. The global Taylor bound for $F$ based at $Y\in\M$
gives
\begin{equation*}
     F(\widehat Y)\le f(Y)+\langle g(Y),\widehat Y-Y\rangle
 +C\|\widehat Y-Y\|^2.
\end{equation*}
Split $g(Y)=g(p)+g(Y)-g(p)$. By the Lipschitz bound~\eqref{loc:eq:gradient-chord} and Cauchy--Schwarz,
\begin{equation}\label{loc:eq:transport-taylor}
 \E F(\widehat Y)\le\E f(Y)+q\varepsilon+C\sqrt{M_2}\,\varepsilon+C\varepsilon^2.
\end{equation}
The second-moment bound~\eqref{loc:eq:local-second-moment}
in Lemma~\ref{loc:lem:localization} gives
\begin{equation*}
    \sqrt{M_2}\le C(\lambda q+\sqrt{\lambda\tau}).
\end{equation*}
Substituting this estimate and~\eqref{loc:eq:ideal-descent} into~\eqref{loc:eq:transport-taylor}, we obtain
\begin{equation*}
    \E F(\widehat Y)
    \le
    f(p)-a_0\lambda q^2+C\lambda\tau
    +(1+C\lambda)q\varepsilon
    +C\sqrt{\lambda\tau}\,\varepsilon +C\varepsilon^2.
\end{equation*}
We now bound the two mixed terms.
Since $\lambda\le1$, Young's inequality gives
\begin{equation*}
    (1+C\lambda)q\varepsilon
    \le (1+C)q\varepsilon
    \le
    \frac{a_0}{2}\lambda q^2
    +C'\frac{\varepsilon^2}{\lambda}.
\end{equation*}
Thus, this error term consumes at most half of
the negative term $-a_0\lambda q^2$.
For the other mixed term, the inequality $2uv\le u^2+v^2$ yields
\begin{equation*}
    C\sqrt{\lambda\tau}\,\varepsilon
    \le C'(\lambda\tau+\varepsilon^2).
\end{equation*}
Combining these estimates and using
$\varepsilon^2\le\varepsilon^2/\lambda$, we conclude that
\begin{equation*}
    \E F(\widehat Y)
    \le
    f(p)-\frac{a_0}{2}\lambda q^2
    +C\lambda\tau+C\frac{\varepsilon^2}{\lambda}.
\end{equation*}
This proves~\eqref{loc:eq:transport-descent}
with $a_1=a_0/2$.

Next, let $\widehat p=P(\widehat Y)$.
Since $Y\in\M$, the nearest-point comparison
\eqref{loc:eq:projection-comparison} gives
\begin{equation*}
    \|\widehat p-Y\|
    \le 2\|\widehat Y-Y\|.
\end{equation*}
To bound the displacement of $\widehat p$ from $p$,
write $\widehat p-p=(\widehat p-Y)+(Y-p)$.
Using $\|u+v\|^2\le2\|u\|^2+2\|v\|^2$,
we obtain
\begin{align*}
    \E\|\widehat p-p\|^2
    &\le
    2\E\|\widehat p-Y\|^2
    +2\E\|Y-p\|^2\\
    &\le
    8\E\|\widehat Y-Y\|^2+2M_2\\
    &\le 8\varepsilon^2+2M_2\\
    &\le C(\lambda^2q^2+\lambda\tau+\varepsilon^2),
\end{align*}
where the last step again uses
\eqref{loc:eq:local-second-moment}.
This proves~\eqref{loc:eq:shadow-displacement}.

Finally, the same nearest-point comparison and
Cauchy--Schwarz imply
\begin{align*}
    \|\E(\widehat p-Y)\|
    &\le \E\|\widehat p-Y\|\\
    &\le 2\E\|\widehat Y-Y\|\\
    &\le 2\bigl(\E\|\widehat Y-Y\|^2\bigr)^{1/2}\\
    &\le 2\varepsilon.
\end{align*}
This proves~\eqref{loc:eq:shadow-drift-error}.
The final estimate uses only the feasibility of $Y$
and the sampling-error bound, so it holds for every
current point $x\in\R^D$ satisfying that bound.
\end{proof}

\subsection{Proof of Proposition~\ref{prop:feasibility}} \label{app:proof-feasibility}

Fix $x\in\R^D$ and $\lambda,\tau>0$, and write $Q=Q_{\lambda,\tau}^f(x,\cdot)$, and $\widehat Q= \widehat Q_{\lambda,\tau}^f(x,\cdot)$.  Let $\Pi(\widehat Q,Q)$ denote the set of joint probability
distributions. Thus each $\pi\in\Pi(\widehat Q,Q)$ describes a pair of random variables $\widehat Y\sim\widehat Q$ and $Y\sim Q$. By Assumption~\ref{ass:oracle},
\begin{equation*}
 W_2(\widehat Q,Q)^2
 =
 \inf_{\pi\in\Pi(\widehat Q,Q)}
 \int_{\R^D\times\R^D}
       \|u-v\|^2\,\pi(du,dv) \le \varepsilon^2.
\end{equation*}
Since the distance from $u$ to $\M$ cannot exceed its distance to any $v\in\M$, we have
\begin{equation*}
 \E\dist(\widehat Y,\M)^2
 =
 \int_{\R^D\times\R^D}
       \dist(u,\M)^2\,\pi(du,dv)
 \le
 \int_{\R^D\times\R^D}
       \|u-v\|^2\,\pi(du,dv).
\end{equation*}
Taking the infimum over all $\pi\in\Pi(\widehat Q,Q)$ yields
\begin{equation*}
 \E\dist(\widehat Y,\M)^2
 \le W_2(\widehat Q,Q)^2
 \le \varepsilon^2.
\end{equation*}

\subsection{Proof of Theorem~\ref{thm:mean-step}}
\label{app:proof-mean-step}

The geometric and integration-by-parts estimates above identify the
mean of an ideal local update. Quadratic transport then transfers that
identity to the implemented law without requiring its outputs to be
feasible.

\begin{proof}
Fix a feasible anchor $x\in\M$ and use an optimal coupling of
$\widehat Y\sim\widehat Q$ with
$Y\sim Q_{\lambda,\tau}^f(x,\cdot)$ at cost at most $\varepsilon^2$.
Jensen's inequality gives
\begin{equation*}
     \|\E(\widehat Y-Y)\|\le
 \bigl(\E\|\widehat Y-Y\|^2\bigr)^{1/2}\le \varepsilon.
\end{equation*}
Adding this to the ideal mean estimate~\eqref{loc:eq:ideal-gradient-bias} yields
\begin{equation*}
     \|\E(\widehat Y-x)+\lambda g(x)\|
 \le C\lambda(\lambda+\tau)+\varepsilon.
\end{equation*}
The squared triangle inequality and
Lemma~\ref{loc:lem:localization}, with $p=x$, give
\begin{equation*}
     \E\|\widehat Y-x\|^2
 \le2\E\|Y-x\|^2+2\varepsilon^2
 \le C\bigl(\lambda^2\|g(x)\|^2+\lambda\tau+\varepsilon^2\bigr).
\end{equation*}
Every constant used in the geometric and localization estimates is
uniform over $x\in\M$, proving the asserted uniformity. Division by
$\lambda$ shows that the normalized mean update approaches $-g(x)$
whenever $\lambda,\tau$, and $\varepsilon/\lambda$ all vanish. The second-moment
bound concerns the actual sampled increment, which can fluctuate even
when its mean accurately identifies the gradient direction.
\end{proof}

\subsection{Proof of Theorem~\ref{thm:descent}} \label{app:proof-descent}

An approximate output may leave the manifold, so the objective used in
the proof is the feasible-representative value $F=f\circ P$. The global
Taylor bound for $F$ in Lemma~\ref{loc:lem:geometry} permits such outputs
even when they lie outside the smooth projection tube.

\begin{proof}
Take $r$ from Lemma~\ref{loc:lem:localization}.
For $\dist(x,\M)<r$, write $p=P(x)$.
Applying Lemma~\ref{loc:lem:transport-descent} to
$\widehat Y\sim\widehat Q_{\lambda,\tau}^f(x,\cdot)$ gives
\[
 \E f(P(\widehat Y))
 \le f(p)-a_1\lambda\|g(p)\|^2
      +C\left(\lambda\tau+\frac{\varepsilon^2}{\lambda}\right).
\]
This proves~\eqref{eq:local-descent} with
$C_1=a_1$ and $C_2=C$.
The constants are uniform over the stated range.
\end{proof}

\subsection{Proof of Theorem~\ref{thm:finite-time}} \label{app:proof-finite-time}

The local descent estimate does not apply to a distant ambient anchor.
The first sampling call controls this difficulty, and the same
feasibility bound controls all subsequent excursions. This allows the
iteration to keep its actual sampled anchors throughout.

\begin{proof}
Take $r$ from Theorem~\ref{thm:descent} and write $a=C_1$ and $C=C_2$ for its descent constants.
For $k\ge1$, write
\begin{equation*}
    p_k=P(X_k),
    \qquad F_k=f(p_k),
    \qquad q_k=\|g(p_k)\|,
    \qquad
    B_k=\{\dist(X_k,\M)\ge r\}.
\end{equation*}
These quantities are measurable because $P$ is Borel.
Moreover, compactness gives
\begin{equation*}
    f_\star\le F_k\le f_\star+\Delta_f,
    \qquad q_k\le G.
\end{equation*}

For each refinement step, the next iterate is sampled
from $\widehat Q^f_{\lambda,\tau_k}(X_k,\cdot)$.
On the event $B_k^c$, the current point satisfies
$\dist(X_k,\M)<r$, so
Theorem~\ref{thm:descent} gives
\begin{equation*}
    \E[F_{k+1}\mid X_k]
    \le
    F_k-a\lambda q_k^2
    +C\lambda\tau_k+C\frac{\varepsilon^2}{\lambda}.
\end{equation*}
On $B_k$, we instead use the global bounds on $f$
and $g$.
Set
\begin{equation*}
    B=\Delta_f+a\lambda_{\rm loc}G^2.
\end{equation*}
Since $F_k\ge f_\star$, $\lambda\le\lambda_{\rm loc}$,
and $q_k\le G$, we have
\begin{equation*}
    \E[F_{k+1}\mid X_k]
    \le f_\star+\Delta_f
    \le F_k-a\lambda q_k^2+B.
\end{equation*}
Combining the two cases yields
\begin{equation}\label{loc:eq:excursion-descent}
    \E[F_{k+1}\mid X_k]
    \le
    F_k-a\lambda q_k^2
    +C\lambda\tau_k
    +C\frac{\varepsilon^2}{\lambda}
    +B\mathbf1_{B_k}.
\end{equation}
We next control the probability of $B_k$.
Proposition~\ref{prop:feasibility}, applied to the
call producing $X_k$, gives
\begin{equation*}
    \E\dist(X_k,\M)^2\le\varepsilon^2,
    \qquad k\ge1.
\end{equation*}
For $k=1$, this uses the sampling-error bound for
the initialization call.
Markov's inequality therefore implies
\begin{equation*}
    \Prob(B_k)
    =
    \Prob\{\dist(X_k,\M)^2\ge r^2\}
    \le\frac{\varepsilon^2}{r^2}.
\end{equation*}
Taking expectations
in~\eqref{loc:eq:excursion-descent} and rearranging,
we obtain
\begin{equation*}
    a\lambda\E q_k^2
    \le
    \E F_k-\E F_{k+1}
    +C\lambda\tau_k
    +C\frac{\varepsilon^2}{\lambda}
    +B\frac{\varepsilon^2}{r^2}.
\end{equation*}
Summing from $k=1$ to $N$ makes the objective
differences telescope:
\begin{align*}
    a\lambda\sum_{k=1}^N\E q_k^2
    &\le
    \E F_1-\E F_{N+1}
    +C\lambda\sum_{k=1}^N\tau_k
    +CN\frac{\varepsilon^2}{\lambda}
    +BN\frac{\varepsilon^2}{r^2}\\
    &\le
    \Delta_1
    +C\lambda\sum_{k=1}^N\tau_k
    +CN\frac{\varepsilon^2}{\lambda}
    +BN\frac{\varepsilon^2}{r^2},
\end{align*}
where
$\Delta_1=\E F_1-f_\star$
and the second inequality uses $F_{N+1}\ge f_\star$.

Divide by $aN\lambda$.
The last term contributes
$B\varepsilon^2/(ar^2\lambda)$.
Because $\lambda\le1$, this contribution is bounded
by a constant times $\varepsilon^2/\lambda^2$.
Absorbing the fixed coefficients into $C$ gives
\begin{equation}\label{loc:eq:fixed-accuracy-rate}
    \frac1N\sum_{k=1}^N\E\|g(p_k)\|^2
    \le
    C\left[
        \frac{\Delta_1}{N\lambda}
        +\frac1N\sum_{k=1}^N\tau_k
        +\left(\frac{\varepsilon}{\lambda}\right)^2
    \right].
\end{equation}
All constants depend only on the fixed model
quantities and $\bar\tau$, and are independent of
$N,\lambda,\varepsilon$, and the temperature sequence.

The feasibility conclusion follows directly from
Proposition~\ref{prop:feasibility}:
\begin{equation*}
    \E\dist(X_{k+1},\M)^2\le\varepsilon^2,
    \qquad k\ge0.
\end{equation*}

We now prove the accumulation-point conclusion. Suppose $\tau_k\downarrow0$.
Its averages also converge to zero, so
\eqref{loc:eq:fixed-accuracy-rate} implies
\begin{equation*}
    \limsup_{N\to\infty}
    \frac1N\sum_{k=1}^N\E q_k^2
    \le
    C\left(\frac{\varepsilon}{\lambda}\right)^2.
\end{equation*}
Define the nonnegative random variable
\begin{equation*}
    \ell=\liminf_{k\to\infty}q_k^2.
\end{equation*}
For each trajectory, the lower limit of a sequence
is at most the lower limit of its averages:
\begin{equation*}
    \ell
    \le
    \liminf_{N\to\infty}
    \frac1N\sum_{k=1}^N q_k^2.
\end{equation*}
Indeed, for every $\delta>0$, all sufficiently late
terms satisfy $q_k^2\ge\ell-\delta$.
Taking averages, letting $N\to\infty$, and then
letting $\delta\downarrow0$ proves this inequality.
Fatou's lemma consequently gives
\begin{equation*}
    \E\ell\le
    \E\left[
        \liminf_{N\to\infty}
        \frac1N\sum_{k=1}^N q_k^2
    \right] \le
    \liminf_{N\to\infty}
    \frac1N\sum_{k=1}^N\E q_k^2 \le
    C\left(\frac{\varepsilon}{\lambda}\right)^2.
\end{equation*}
It remains to choose a random accumulation point
at which the squared gradient norm equals $\ell$.
Set $n_0=0$ and, for $j\ge1$, define
\begin{equation*}
    n_j
    =
    \min\left\{
        k>n_{j-1}:
        q_k^2\le\ell+\frac1j
    \right\}.
\end{equation*}
These indices are finite by the definition of
$\ell$, and measurable because they are successive
first indices satisfying a measurable condition.
Moreover,
\begin{equation*}
    \|g(p_{n_j})\|^2\longrightarrow\ell.
\end{equation*}
To choose the limit measurably, we extract subsequences one coordinate at a time.
For a bounded sequence of real random variables $\{u_j\}$, set
$u_*=\liminf_{j\to\infty}u_j$, $m_0=0$, and
\begin{equation*}
    m_j=\min\left\{
        k>m_{j-1}: |u_k-u_*|<\frac1j
    \right\}.
\end{equation*}
There are infinitely many terms within any positive distance of $u_*$,
so these indices are finite. They are measurable by the same first-index
argument as above, and $u_{m_j}\to u_*$.
Apply this construction to the first coordinate of $\{p_{n_j}\}$,
then to the second coordinate of the resulting subsequence, and so on.
Compactness of $\M$ bounds every coordinate. Each selected vector is
measurable, and each extraction preserves the limits of earlier coordinates.
After $D$ extractions, all coordinates converge. Their limit $p_\infty$
is measurable, belongs to $\M$ because $\M$ is closed, and is an
accumulation point of $\{p_k\}$.
Continuity of $g$ gives
\begin{equation*}
    \|g(p_\infty)\|^2=\ell
    \qquad\text{almost surely}.
\end{equation*}
Hence
\begin{equation*}
    \E\|g(p_\infty)\|^2
    =
    \E\ell
    \le
    C\left(\frac{\varepsilon}{\lambda}\right)^2.
\end{equation*}
Finally, Cauchy--Schwarz yields
\begin{equation*}
    \E\|g(p_\infty)\|
    \le
    \bigl(\E\|g(p_\infty)\|^2\bigr)^{1/2}
    \le
    \sqrt C\,\frac{\varepsilon}{\lambda}.
\end{equation*}
For fixed $\lambda$, both upper bounds vanish
as $\varepsilon\downarrow0$.
\end{proof}
\subsection{Proof of Theorem~\ref{thm:boundary}}
\label{app:proof-boundary}
We use the assumptions of Section~\ref{sec:with_boundary_theorems} and first record the properties needed for the proof.
Compactness provides a Borel measurable nearest-point selection $P$ by the selection argument in the proof of Lemma~\ref{loc:lem:geometry}.
Write $D_{\M}=\operatorname{diam}(\M)$ and $\Delta_f=\max_{\M}f-\min_{\M}f$.
Choose $\delta>0$ so that the closed $\delta$-neighborhood of $\M$ lies in the region where $f$ is $C^2$, and let $L_0$ bound $\|\nabla f\|$ there.
For $y,z\in\M$ with $\|y-z\|\le\delta$, the segment between them lies in this neighborhood, so $|f(y)-f(z)|\le L_0\|y-z\|$.
For $\|y-z\|>\delta$, boundedness gives $|f(y)-f(z)|\le\Delta_f\le(\Delta_f/\delta)\|y-z\|$.
Thus we may take $L_f=\max\{1,L_0,\Delta_f/\delta\}$.

The geometric and density assumptions also give
\begin{equation}\label{eq:bd-local-mass}
 \mu(B(y,r)\cap\M)\ge vr^m,
 \qquad y\in\M,\quad 0<r\le r_0,
\end{equation}
for constants $v,r_0>0$. Smooth compact manifolds with smooth boundary
and a density bounded away from zero satisfy this condition: their
interior and boundary charts contain balls and half-balls, respectively,
with uniformly controlled volume distortion. A finite chart cover and
the positive density lower bound give uniform constants. The argument
below uses only~\eqref{eq:bd-local-mass}.

For a fixed $\lambda>0$, define
\begin{equation}\label{eq:bd-thermal-budget}
 K_\lambda=L_f+D_{\M}/\lambda,\quad
 b_{\lambda,\tau}=\min\{r_0,\tau/K_\lambda\}, \quad
 \delta_\partial(\lambda,\tau) =\tau\left[1+\log\frac{1}{v b_{\lambda,\tau}^m}\right].
\end{equation}
The logarithm is nonnegative because
$v b_{\lambda,\tau}^m\le1$ follows from the local-mass bound.
For fixed $\lambda$, this budget is
$O(\tau[1+|\log\tau|])$ as $\tau\downarrow0$.

\begin{proof}[Proof of the one-step bound]
Fix a current point $x\in\R^D$ and positive
parameters $\lambda,\tau$.
Let $p=P(x)$ be its nearest feasible representative,
and write
\begin{equation*}
    d=\|x-p\|=\dist(x,\M).
\end{equation*}
We first compare the ideal response with a proximal
minimizer based at $p$.
For this purpose, define
\begin{equation*}
    h_p(y)
    =
    f(y)+\frac{\|y-p\|^2}{2\lambda},
    \qquad y\in\M.
\end{equation*}
For any $y,z\in\M$, the quadratic term satisfies
\begin{equation*}
        \left|
        \frac{\|y-p\|^2-\|z-p\|^2}{2\lambda}
    \right| \le
    \frac{\|y-p\|+\|z-p\|}{2\lambda}\|y-z\| \le
    \frac{D_{\M}}{\lambda}\|y-z\|.
\end{equation*}
Together with the Lipschitz bound for $f$, this gives
\begin{equation*}
    |h_p(y)-h_p(z)|
    \le K_\lambda\|y-z\|.
\end{equation*}
Because $h_p$ is continuous on compact $\M$, it
attains its minimum at some $z_\star\in\M$.
Consider the neighborhood
\begin{equation*}
    \mathcal B
    =
    B(z_\star,b_{\lambda,\tau})\cap\M.
\end{equation*}
The local-mass bound~\eqref{eq:bd-local-mass} gives
\begin{equation*}
    \mu(\mathcal B)
    \ge v b_{\lambda,\tau}^m>0.
\end{equation*}
We can therefore define a probability measure
supported in this neighborhood by
\begin{equation*}
    \nu(dy)
    =
    \frac{\mathbf1_{\mathcal B}(y)}
         {\mu(\mathcal B)}\,\mu(dy).
\end{equation*}
Every point in $\mathcal B$ has proximal energy
close to the minimum. Indeed,
$K_\lambda b_{\lambda,\tau}\le\tau$, so
\begin{equation*}
    \int_{\M}h_p(y)\,\nu(dy)
    \le
    h_p(z_\star)+K_\lambda b_{\lambda,\tau}
    \le
    \min_{\M}h_p+\tau.
\end{equation*}
Moreover, the density of $\nu$ with respect to $\mu$
is constant on $\mathcal B$, which gives
\begin{equation*}
    \operatorname{KL}(\nu\|\mu)
    =
    \log\frac1{\mu(\mathcal B)}
    \le
    \log\frac1{v b_{\lambda,\tau}^m}.
\end{equation*}
The ideal response uses the current point $x$,
whereas $h_p$ uses its representative $p$.
To compare these two energies, expand the quadratic
penalty:
\begin{equation}\label{eq:bd-anchor-decomposition}
\begin{aligned}
    E_{\lambda,x}^f(y)
    &=
    h_p(y)+\frac{d^2}{2\lambda}+t_x(y),\\
    t_x(y)
    &=
    -\frac{\langle x-p,y-p\rangle}{\lambda}.
\end{aligned}
\end{equation}
For any $y,z\in\M$, Cauchy--Schwarz gives
\begin{equation*}
    |t_x(y)-t_x(z)|
    =
    \frac{|\langle x-p,y-z\rangle|}{\lambda}
    \le
    \frac{dD_{\M}}{\lambda}.
\end{equation*}
Consequently,
\begin{equation*}
    \sup_{\M}t_x-\inf_{\M}t_x
    \le\frac{dD_{\M}}{\lambda}.
\end{equation*}

Write $Q=Q_{\lambda,\tau}^f(x,\cdot)$.
The variational characterization in
Proposition~\ref{prop:landing} states that
\begin{equation*}
    \int_{\M}E_{\lambda,x}^f\,dQ
    +\tau\operatorname{KL}(Q\|\mu)
    \le
    \int_{\M}E_{\lambda,x}^f\,d\nu
    +\tau\operatorname{KL}(\nu\|\mu).
\end{equation*}
Substituting~\eqref{eq:bd-anchor-decomposition}
cancels the constant $d^2/(2\lambda)$ on both sides.
Since $\operatorname{KL}(Q\|\mu)\ge0$, we obtain
\begin{align*}
    \int_{\M}h_p\,dQ
    &\le
    \int_{\M}h_p\,d\nu
    +\tau\operatorname{KL}(\nu\|\mu)
    +\int_{\M}t_x\,d\nu-\int_{\M}t_x\,dQ\\
    &\le
    \min_{\M}h_p
    +\tau
    +\tau\log\frac1{v b_{\lambda,\tau}^m}
    +\frac{dD_{\M}}{\lambda}.
\end{align*}
Thus, for $Y\sim Q$,
\begin{equation}\label{eq:bd-ideal-comparison}
    \E h_p(Y)
    \le
    \min_{\M}h_p
    +\delta_\partial(\lambda,\tau)
    +\frac{D_{\M}}{\lambda}\dist(x,\M).
\end{equation}

We now account for the approximation error.
Let $\widehat Y\sim\widehat Q$ satisfy Assumption~\ref{ass:oracle}. Choose a joint distribution of $\widehat Y$ and $Y$
with these marginals that attains the quadratic
Wasserstein distance. Then
\begin{equation*}
    \E\|\widehat Y-Y\|^2
    =
    W_2(\widehat Q,Q)^2
    \le\varepsilon^2.
\end{equation*}
Since $Y\in\M$ and $P(\widehat Y)$ is a nearest
point on $\M$,
\begin{equation*}
    \|P(\widehat Y)-Y\| \le
    \|P(\widehat Y)-\widehat Y\|
    +\|\widehat Y-Y\| \le 2\|\widehat Y-Y\|.
\end{equation*}
Both $P(\widehat Y)$ and $Y$ lie on $\M$.
The Lipschitz bound for $h_p$ and
Cauchy--Schwarz therefore imply
\begin{align*}
    \E h_p(P(\widehat Y))
    &\le
    \E h_p(Y)
    +K_\lambda\E\|P(\widehat Y)-Y\|\\
    &\le
    \E h_p(Y)
    +2K_\lambda
      \bigl(\E\|\widehat Y-Y\|^2\bigr)^{1/2}\\
    &\le
    \E h_p(Y)+2K_\lambda\varepsilon.
\end{align*}
Finally, the definition of the proximal gap gives
\begin{equation*}
    \min_{\M}h_p=f(p)-r_\lambda(p).
\end{equation*}
Combining this identity with
\eqref{eq:bd-ideal-comparison} and expanding
$h_p(P(\widehat Y))$ yields
\begin{equation}\label{eq:bd-one-step}
    \E f(P(\widehat Y))
    +\frac{\E\|P(\widehat Y)-p\|^2}{2\lambda}\le
    f(p)-r_\lambda(p)
    +\delta_\partial(\lambda,\tau)
    +\frac{D_{\M}}{\lambda}\dist(x,\M)
    +2K_\lambda\varepsilon.
\end{equation}
This establishes the required comparison for a single approximate update.
\end{proof}

\begin{proof}[Proof of the finite-time and exact-sampling conclusions]
Write $p_k=P(X_k)$.
Proposition~\ref{prop:feasibility}, applied to every
call including initialization, gives
\begin{equation}\label{eq:bd-feasibility-transfer}
    \E\dist(X_{k+1},\M)^2\le\varepsilon^2,
    \qquad k\ge0.
\end{equation}
Hence, by Cauchy--Schwarz,
\begin{equation*}
    \E\dist(X_k,\M)
    \le
    \bigl(\E\dist(X_k,\M)^2\bigr)^{1/2}
    \le\varepsilon,
    \qquad k\ge1.
\end{equation*}

Apply~\eqref{eq:bd-one-step} with $x=X_k$,
$p=p_k$, and $\tau=\tau_k$.
The approximate output is $X_{k+1}$, whose nearest
feasible representative is $p_{k+1}$.
Averaging over the distribution of $X_k$ gives
\begin{align*}
    &\E f(p_{k+1})
    +\E r_\lambda(p_k)
    +\frac{\E\|p_{k+1}-p_k\|^2}{2\lambda}\\
    &\qquad\le
    \E f(p_k)
    +\delta_\partial(\lambda,\tau_k)
    +\frac{D_{\M}}{\lambda}\E\dist(X_k,\M)
    +2K_\lambda\varepsilon\\
    &\qquad\le
    \E f(p_k)
    +\delta_\partial(\lambda,\tau_k)
    +\left(\frac{D_{\M}}{\lambda}+2K_\lambda\right)
        \varepsilon.
\end{align*}
Rearranging and summing from $k=1$ to $N$,
we obtain
\begin{align*}
    &\sum_{k=1}^N
    \E\left[
        r_\lambda(p_k)
        +\frac{\|p_{k+1}-p_k\|^2}{2\lambda}
    \right]\\
    &\qquad\le
    \E f(p_1)-\E f(p_{N+1})
    +\sum_{k=1}^N\delta_\partial(\lambda,\tau_k)
    +N\left(\frac{D_{\M}}{\lambda}+2K_\lambda\right)
        \varepsilon.
\end{align*}
Let
\begin{equation*}
    f_\star=\min_{\M}f,
    \qquad
    \Delta_1=\E f(p_1)-f_\star.
\end{equation*}
Since $\E f(p_{N+1})\ge f_\star$, division by $N$
yields
\begin{equation*}
\begin{aligned}
    &\frac1N\sum_{k=1}^N
    \E\left[
        r_\lambda(p_k)
        +\frac{\|p_{k+1}-p_k\|^2}{2\lambda}
    \right]\\
    &\qquad\le
    \frac{\Delta_1}{N}
    +\frac1N\sum_{k=1}^N
        \delta_\partial(\lambda,\tau_k)
    +\left(\frac{D_{\M}}{\lambda}+2K_\lambda\right)
        \varepsilon.
\end{aligned}
\end{equation*}

To express this bound in the form stated in the
theorem, observe that
\begin{equation*}
    \log\frac1{b_{\lambda,\tau}}
    =
    \max\left\{
        \log\frac1{r_0},
        \log K_\lambda-\log\tau
    \right\}.
\end{equation*}
For fixed $\lambda$, the definition
in~\eqref{eq:bd-thermal-budget} therefore implies
\begin{equation*}
    \delta_\partial(\lambda,\tau)
    \le C_\lambda\tau(1+|\log\tau|),
    \qquad \tau>0,
\end{equation*}
where $C_\lambda$ is independent of $\tau$.
Substituting this estimate into the one-step and
averaged inequalities proves the finite-time
conclusions.

We next consider exact sampling, with
$\varepsilon=0$.
Equation~\eqref{eq:bd-feasibility-transfer} implies
that $X_k=p_k\in\M$ almost surely for all $k\ge1$.
Suppose
\begin{equation*}
    \sum_{k=1}^\infty
    \delta_\partial(\lambda,\tau_k)<\infty.
\end{equation*}
Define the excess proximal energy of the sampled
point by
\begin{equation*}
    \eta_k
    =
    h_{p_k}(p_{k+1})
    -\min_{y\in\M}h_{p_k}(y).
\end{equation*}
This quantity is nonnegative.
Because the current point is feasible and the
sampling law is exact, applying
\eqref{eq:bd-ideal-comparison} at $x=p_k$ and
averaging over $p_k$ gives
\begin{equation*}
    \E\eta_k
    \le\delta_\partial(\lambda,\tau_k).
\end{equation*}
Tonelli's theorem consequently yields
\begin{equation*}
    \E\left[\sum_{k=1}^\infty\eta_k\right]
    =
    \sum_{k=1}^\infty\E\eta_k
    \le
    \sum_{k=1}^\infty
        \delta_\partial(\lambda,\tau_k)
    <\infty.
\end{equation*}
It follows that
\begin{equation*}
    \sum_{k=1}^\infty\eta_k<\infty
    \qquad\text{almost surely}.
\end{equation*}

The definition of $\eta_k$ also gives a direct
relation between successive objective values.
Indeed,
\begin{equation*}
    h_{p_k}(p_{k+1})
    =
    f(p_{k+1})
    +\frac{\|p_{k+1}-p_k\|^2}{2\lambda},
    \qquad
    \min_{\M}h_{p_k}
    =
    f(p_k)-r_\lambda(p_k).
\end{equation*}
Subtracting these two expressions and rearranging
gives
\begin{equation*}
    f(p_{k+1})-f(p_k)
    =
    \eta_k-r_\lambda(p_k)
    -\frac{\|p_{k+1}-p_k\|^2}{2\lambda}.
\end{equation*}
Summing along the trajectory,
\begin{equation*}
    f(p_{N+1})
    +\sum_{k=1}^N
    \left[
        r_\lambda(p_k)
        +\frac{\|p_{k+1}-p_k\|^2}{2\lambda}
    \right]
    =
    f(p_1)+\sum_{k=1}^N\eta_k.
\end{equation*}
Since $f(p_{N+1})\ge f_\star$, the nonnegative
partial sums on the left satisfy
\begin{equation*}
    \sum_{k=1}^N
    \left[
        r_\lambda(p_k)
        +\frac{\|p_{k+1}-p_k\|^2}{2\lambda}
    \right]
    \le
    f(p_1)-f_\star+\sum_{k=1}^\infty\eta_k.
\end{equation*}
The right-hand side is finite almost surely and
does not depend on $N$.
Therefore
\begin{equation*}
    \sum_{k=1}^\infty r_\lambda(p_k)<\infty,
    \qquad
    \sum_{k=1}^\infty\|p_{k+1}-p_k\|^2<\infty
    \qquad\text{almost surely}.
\end{equation*}
In particular, $r_\lambda(p_k)\to0$.
Furthermore,
\begin{equation*}
    f(p_{N+1})
    =
    f(p_1)+\sum_{k=1}^N\eta_k
    -\sum_{k=1}^N
    \left[
        r_\lambda(p_k)
        +\frac{\|p_{k+1}-p_k\|^2}{2\lambda}
    \right].
\end{equation*}
Both partial sums on the right converge almost
surely, so $f(p_k)$ converges almost surely.

Finally, we characterize the accumulation points.
The function $(p,y)\mapsto h_p(y)$ is continuous
on the compact set $\M\times\M$.
Its minimum over $y$ therefore depends continuously
on $p$, and hence
\begin{equation*}
    r_\lambda(p)=f(p)-\min_{y\in\M}h_p(y)
\end{equation*}
is continuous.
Compactness of $\M$ guarantees the existence
of accumulation points of $\{p_k\}$.
On every trajectory for which
$r_\lambda(p_k)\to0$, any accumulation point
$p_\infty$ satisfies $r_\lambda(p_\infty)=0$.
Equivalently,
\begin{equation}\label{eq:bd-limit-proximal}
    f(p_\infty)
    \le
    f(y)+\frac{\|y-p_\infty\|^2}{2\lambda},
    \qquad y\in\M.
\end{equation}

Let $w\in\mathcal T_{\M}(p_\infty)$.
By the definition of the tangent cone, there exist
$t_j\downarrow0$ and $y_j\in\M$ such that
\begin{equation*}
    \frac{y_j-p_\infty}{t_j}\longrightarrow w.
\end{equation*}
Applying~\eqref{eq:bd-limit-proximal} at $y=y_j$
and dividing by $t_j$ gives
\begin{equation*}
    0
    \le
    \frac{f(y_j)-f(p_\infty)}{t_j}
    +\frac{\|y_j-p_\infty\|^2}{2\lambda t_j}.
\end{equation*}
Differentiability of $f$ implies
\begin{equation*}
    \frac{f(y_j)-f(p_\infty)}{t_j}
    \longrightarrow
    \langle\nabla f(p_\infty),w\rangle.
\end{equation*}
The quadratic term vanishes because
\begin{equation*}
    \frac{\|y_j-p_\infty\|^2}{2\lambda t_j}
    =
    \frac{t_j}{2\lambda}
    \left\|\frac{y_j-p_\infty}{t_j}\right\|^2
    \longrightarrow0.
\end{equation*}
Passing to the limit proves
\begin{equation*}
    \langle\nabla f(p_\infty),w\rangle\ge0,
    \qquad w\in\mathcal T_{\M}(p_\infty).
\end{equation*}
Thus every accumulation point satisfies the
constrained first-order stationarity condition.
\end{proof}
\section{Constructing Accurate Conditional Samplers}\label{app:statistical}
The error bound requirement in Assumption~\ref{ass:oracle} separates optimization analysis from the choice of sampler. This appendix gives a concrete finite-data construction and proves a sufficient condition to satisfy the error bound in~\eqref{eq:oracle}. The construction serves as an explicit example showing that the accuracy requirement can be met. We construct $\widehat Q$ using a uniform-ball kernel density estimator
and rejection sampling, and bound its error relative to
$Q_{\lambda,\tau}^{f}$. We also analyze direct sampling from the
observations.

\subsection{From Feasible Observations to a Conditional Sampler}\label{app:conditional-construction}
Write the empirical reference distribution constructed directly from the dataset $\mathcal{D}_n$ as
\begin{equation*}
 \mu_n=\frac1n\sum_{i=1}^n\delta_{Y^{(i)}},
\end{equation*}
where $\delta$ denotes Dirac measure. Replacing $\mu$ by $\mu_n$ in~\eqref{eq:kernel} gives an implementable categorical distribution on observation indices:
\begin{equation}\label{eq:stat-categorical}
 \pi_i(x;\lambda,\tau)
 \ \propto\ 
 \exp\!\left[-\frac{f(Y^{(i)})+
            \|Y^{(i)}-x\|^2/(2\lambda)}{\tau}\right].
\end{equation}
The returned point has law
\begin{equation}\label{eq:stat-indexed-sampler}
 \widehat Q_{\lambda,\tau}^{\,n}(x,\cdot)
 =\sum_{i=1}^n\pi_i(x;\lambda,\tau)\delta_{Y^{(i)}}
 =Q_{\lambda,\tau}^{f,\mu_n}(x,\cdot).
\end{equation}
Here the weights are assigned to observation indices, with weights of repeated observations combined. The outputs of $\pi_i(x; \lambda, \tau)$ are feasible because they are observations, but it cannot produce a point with an objective below the best observed value.

To allow new outputs, one can instead smooth the empirical reference. Let $U$ be uniform on the ambient unit ball and, independently, choose an observation index $I$ uniformly. For bandwidth $0<h\le1$, set
\begin{equation}\label{eq:stat-smoothed-reference}
 V=Y^{(I)}+hU,
 \qquad
 \nu_{n,h}(B)=\frac1n\sum_{i=1}^n
                  \Prob\{Y^{(i)}+hU\in B\},
\end{equation}
for Borel sets $B\subseteq\R^D$. This model generates continuous ambient outputs without using any explicit information of $\M$. Though it may generate infeasible points, the conditional accuracy bound in~\eqref{eq:stat-uniform-bound}, together with Proposition~\ref{prop:feasibility}, controls their distance to $\M$. The model-dependent kernel is written as:
\begin{equation}\label{eq:stat-learned-target}
 Q_{\lambda,\tau}^{f,\nu_{n,h}}(x,dy)
 =\frac{\exp[-E_{\lambda,x}^{f}(y)/\tau]\,\nu_{n,h}(dy)}
        {\int\exp[-E_{\lambda,x}^{f}(z)/\tau]\,\nu_{n,h}(dz)}.
\end{equation}
The superscript following the comma identifies the reference distribution, the original ideal kernel remains $Q_{\lambda,\tau}^{f}=Q_{\lambda,\tau}^{f,\mu}$. 

We provide an exact sampler for~\eqref{eq:stat-learned-target} in Algorithm~\ref{alg:stat-exact-rejection}. Distribution $\nu_{n, h}$ has the explicitly known support
\begin{equation*}
 S_{n,h}
 \defeq \supp(\nu_{n,h})
 =\bigcup_{i=1}^{n}\overline B(Y^{(i)},h).
\end{equation*}
To construct a valid acceptance probability, we need to find a lower bound $b\le\inf_{S_{n,h}}f$. Such a bound need not require minimizing $f$ over $S_{n, h}$. For example, when a Lipschitz constant $L_f$ for $f$ on this region is available,
the choice $b=\min_{1\le i\le n}f(Y^{(i)})-L_f h$ is valid because every proposal lies within distance $h$ of a
training observation. If $f$ is known to be nonnegative over $S_{n, h}$, one may simply take $b=0$. For $\lambda,\tau>0$, define the acceptance probability
\begin{equation}\label{eq:stat-acceptance}
 a_{\lambda,\tau,x}(v) \defeq \exp \left[-\frac{E_{\lambda,x}^{f}(v)-b}{\tau}\right].
\end{equation}
Recall the construction of $E_{\lambda,x}$~\eqref{eq:prox-energy}, the nonnegative quadratic penalty ensures that $E_{\lambda,x}^{f}(v)\ge b$, so $a_{\lambda,\tau,x}(v)\in(0,1]$.

\begin{algorithm}[t]
\caption{Exact rejection sampler for the learned conditional law}
\label{alg:stat-exact-rejection}
\begin{algorithmic}[1]
\Require Anchor $x$, locality $\lambda>0$, temperature $\tau>0$,
         proposal law $\nu_{n,h}$, and valid lower bound $b$
\Repeat
    \State Draw $V\sim\nu_{n,h}$
    \State Draw $U\sim\operatorname{Unif}(0,1)$ independently
\Until{$U\le a_{\lambda,\tau,x}(V)$}
\State \Return $\widehat Y=V$
\end{algorithmic}
\end{algorithm}

To characterize the sampling cost, fix the query $(x,\lambda,\tau)$. The expectation of the number $T$ of needed proposals over the proposal and acceptance draws satisfies
\begin{equation*}
 \E T
 =\frac{1}{\alpha_{\lambda,\tau}(x)}
 =\frac{e^{-b/\tau}}
        {\int\exp[-E_{\lambda,x}^{f}(v)/\tau]\,\nu_{n,h}(dv)}.
\end{equation*}
Consequently, the algorithm terminates almost surely with finite
expected proposal count. Theorem~\ref{thm:finite-data} provides
the uniform upper bound~\eqref{eq:stat-rejection-cost} over its
prescribed query and parameter ranges.


\subsection{Uniform Accuracy Guarantees}
\label{app:conditional-statistics}
\label{app:stat-empirical-certificate}
\label{app:stat-scope}

The following theorem bounds the error from finite data and smoothing,
assuming $W_2(\mu_n,\mu)\le d$. Lemma~\ref{lem:stat-empirical} gives a
sufficient sample size for this event.

\begin{theorem}[Uniform accuracy of the smoothed sampler]\label{thm:finite-data}
Suppose $\M\subseteq K_0=[-R,R]^D$, where $R>0$. Fix observations
$Y^{(1)},\ldots,Y^{(n)}\in\M$ satisfying
\begin{equation}\label{eq:stat-deterministic-premise}
 W_2(\mu_n,\mu)\le d.
\end{equation}
For the smoothed distribution~\eqref{eq:stat-smoothed-reference}, assume
$f$ is $L_f$-Lipschitz on $K=[-R-1,R+1]^D$. Define the energy on
$K$ by the same formula. Choose a compact set $A$ containing $K$ and
a prescribed initial point $X_0\in\R^D$. Fix
\begin{equation}\label{eq:stat-parameter-rectangle}
 \Theta=[\lambda_{\min},\lambda_{\max}]
        \times[\tau_{\min},\tau_{\max}],
 \qquad \lambda_{\min},\tau_{\min}>0.
\end{equation}
Define
\begin{equation*}
 \begin{gathered}
 R_A=\sup_{x\in A,\,y\in K}\|x-y\|,
 \qquad D_K=\operatorname{diam}K,\\
 L_*=L_f+R_A/\lambda_{\min},
 \qquad B_*=\osc_K f+R_A^2/(2\lambda_{\min}).
 \end{gathered}
\end{equation*}
Then, for every $0<h\le1$,
\begin{equation}\label{eq:stat-uniform-bound}
 \sup_{\substack{x\in A\\(\lambda,\tau)\in\Theta}}
 W_2\!\left(Q_{\lambda,\tau}^{f,\nu_{n,h}}(x,\cdot),
            Q_{\lambda,\tau}^{f}(x,\cdot)\right)
 \le\left[
 e^{B_*/\tau_{\min}}
 \left((d+h)^2+\frac{D_K^2L_*}{\tau_{\min}}(d+h)\right)
 \right]^{1/2}.
\end{equation}
Given a finite lower bound $b\le\inf_{S_{n,h}}f$, rejection sampling
using~\eqref{eq:stat-acceptance} returns the law
in~\eqref{eq:stat-learned-target} exactly. This is a Borel probability
kernel with finite second moments and outputs in $K$. The expected
number of proposals is at most
\begin{equation}\label{eq:stat-rejection-cost}
 \exp(H_*/\tau_{\min}),
 \qquad
 H_*=\sup_K f-b+R_A^2/(2\lambda_{\min}).
\end{equation}
If an approximate solver has uniform $W_2$ error at most $s$ relative to
the target~\eqref{eq:stat-learned-target}, its total error is bounded by
the right-hand side of~\eqref{eq:stat-uniform-bound} plus $s$. To use
this bound at every iteration, keep the parameters in $\Theta$ and
the solver's outputs in $A$.
\end{theorem}

For any $\varepsilon>0$, sufficiently small $d$ and $h$ make the bound
at most $\varepsilon$. The proof gives explicit choices and a sufficient
sample size using Lemma~\ref{lem:stat-empirical}.

\begin{corollary}[Uniform accuracy of empirical conditional sampling]\label{cor:stat-empirical-conditional}
Under all regularity assumptions in Section~\ref{sec:problem}, fix observations
$Y^{(1)},\ldots,Y^{(n)}\in\M$ satisfying $W_2(\mu_n,\mu)\le d$.
Choose a compact set $A$ containing $\M$ and the prescribed initial
point $X_0\in\R^D$, and use the parameter set $\Theta$
in~\eqref{eq:stat-parameter-rectangle}. Let $L_f^{\M}<\infty$ satisfy
\[
 |f(y)-f(z)|\le L_f^{\M}\|y-z\|,
 \qquad y,z\in\M.
\]
Such a constant exists under assumptions. Define
\begin{equation*}
 \begin{gathered}
 D_{\M}=\operatorname{diam}\M,\qquad
 R_{A,\M}=\sup_{x\in A,\,y\in\M}\|x-y\|,\\
 L_A=L_f^{\M}+R_{A,\M}/\lambda_{\min},
 \qquad
 B_A=\osc_{\M} f+R_{A,\M}^2/(2\lambda_{\min}).
 \end{gathered}
\end{equation*}
The empirical sampler~\eqref{eq:stat-categorical} is a Borel kernel
with outputs in $\M$ and satisfies
\begin{equation}\label{eq:stat-empirical-conditional-bound}
 \sup_{\substack{x\in A\\(\lambda,\tau)\in\Theta}}
 W_2\!\left(\widehat Q_{\lambda,\tau}^{\,n}(x,\cdot),
            Q_{\lambda,\tau}^{f}(x,\cdot)\right)
 \le\left[
 e^{B_A/\tau_{\min}}
 \left(d^2+\frac{D_{\M}^2L_A}{\tau_{\min}}d\right)
 \right]^{1/2}.
\end{equation}
\end{corollary}

With the sample size in Lemma~\ref{lem:stat-empirical}, the required
data event has probability at least $1-\delta$. Conditional on that
event, a bound of at most $\varepsilon$ gives
Assumption~\ref{ass:oracle} for the fixed dataset, simultaneously for
all $x\in A$ and $(\lambda,\tau)\in\Theta$. Both samplers keep their
outputs in $A$, so this also covers adaptively chosen inputs.

The constants depend on the positive lower bounds for $\lambda$ and
$\tau$. Schedules in which either parameter approaches zero require
separate error bounds. Appendix~\ref{app:hatQ_implementations}
describes other implementations.

\subsection{Auxiliary Lemmas and Proofs}
\label{app:stat-proofs}
\label{app:stat-proof-finite-data}

The first lemma bounds $W_2(\mu_n,\mu)$. The second bounds the error
after exponential reweighting.

\begin{lemma}[A sufficient sample count for reference accuracy]\label{lem:stat-empirical}
Suppose $\mu$ is supported in a known cube $K_0=[-R,R]^D$, where
$R>0$, and let $Y^{(1)},\ldots,Y^{(n)}$ be independent observations
with law $\mu$. Put $D_0=2R\sqrt D$. For $d>0$ and
$\delta\in(0,1)$, define
\begin{equation*}
 \eta=\frac{d}{\sqrt{2D}},
 \qquad
 J=\left\lceil\frac{2R}{\eta}\right\rceil^D.
\end{equation*}
If the sample size satisfies
\begin{equation}\label{eq:stat-sample-count}
 n\ge\frac{2D_0^4}{d^4}
         \left[J\log 2+\log(1/\delta)\right],
\end{equation}
then
\begin{equation}\label{eq:stat-data-event}
 \Prob\{W_2(\mu_n,\mu)\le d\}\ge1-\delta.
\end{equation}
\end{lemma}

\begin{proof}
Partition $K_0$ into at most $J$ Borel cells of side length at most
$\eta$, assigning each boundary point to one cell. Let $p$ and
$\widehat p$ be their population and empirical mass vectors. For each
union $B$ of cells, Hoeffding's inequality gives
\[
 \Prob\{\mu_n(B)-\mu(B)>t\}\le e^{-2nt^2}.
\]
Their total variation distance is
\[
 \operatorname{TV}(\widehat p,p)
 =\max_{B\text{ a union of cells}}\{\mu_n(B)-\mu(B)\}.
\]
There are at most $2^J$ unions. A union bound gives, with probability
at least $1-\delta$,
\begin{equation}\label{eq:stat-cell-mass-error}
 \operatorname{TV}(\widehat p,p)
 \le\sqrt{\frac{J\log2+\log(1/\delta)}{2n}}.
\end{equation}
In each cell $j$, couple mass $\min\{p_j,\widehat p_j\}$ using the
normalized restrictions of the two measures whenever this mass is
positive. Its squared transport cost is at most $D\eta^2$ per unit
mass. Since $\sum_j\min\{p_j,\widehat p_j\}
=1-\operatorname{TV}(\widehat p,p)$, the two remaining marginals
have mass $\operatorname{TV}(\widehat p,p)$. Coupling them across
$K_0$ costs at most $D_0^2$ per unit mass. Hence
\[
 W_2(\mu_n,\mu)^2
 \le D\eta^2+D_0^2\operatorname{TV}(\widehat p,p).
\]
The first term is $d^2/2$. By~\eqref{eq:stat-sample-count}
and~\eqref{eq:stat-cell-mass-error}, the second is at most $d^2/2$
on the stated event, proving~\eqref{eq:stat-data-event}.
\end{proof}

\begin{lemma}[Stability of exponential reweighting]\label{lem:stat-tilt}
Let $K\subseteq\R^D$ be compact with diameter $D_K$. Let $H:K\to\R$
be $L_H$-Lipschitz with $\osc_K H\le B$. For a probability measure
$\xi$ on $K$, define
\[
 T_{H,\tau}\xi(dy)
 =\frac{e^{-H(y)/\tau}}{\int e^{-H(z)/\tau}\,\xi(dz)}\,\xi(dy).
\]
If probability measures $\nu$ and $\mu$ supported in $K$ satisfy
$W_2(\nu,\mu)\le d$, then, for every $\tau>0$,
\begin{equation}\label{eq:stat-tilt-bound}
 W_2(T_{H,\tau}\nu,T_{H,\tau}\mu)^2
 \le e^{B/\tau}
       \left[d^2+\frac{D_K^2L_H}{\tau}d\right].
\end{equation}
\end{lemma}
The lemma also applies to discrete measures and measures supported on
a manifold.

\begin{proof}
Subtracting $\min_K H$ does not change the reweighted laws. After
this shift, $w=e^{-H/\tau}$ satisfies
\[
 a:=e^{-B/\tau}\le w\le1,
 \qquad \operatorname{Lip}(w)\le L_H/\tau.
\]
Take an optimal coupling $\gamma$ of $U\sim\nu$ and $V\sim\mu$ with
$\E\|U-V\|^2\le d^2$. Set $Z_\nu=\E w(U)$ and $Z_\mu=\E w(V)$;
both normalizing constants are at least $a$. Define
\[
 \gamma_0(dU,dV)
 =\min\!\left\{\frac{w(U)}{Z_\nu},\frac{w(V)}{Z_\mu}\right\}
 \gamma(dU,dV).
\]
Its marginals are bounded by the two reweighted laws. Since the
multiplier is at most $1/a$, its transport cost is at most $d^2/a$.
Both normalized weights have expectation one, so the missing mass is
\[
 t=1-\gamma_0(K\times K)
  =\frac12\E\left|
             \frac{w(U)}{Z_\nu}-\frac{w(V)}{Z_\mu}
             \right|.
\]
Since $|Z_\mu-Z_\nu|\le\E|w(U)-w(V)|$,
\begin{align*}
 2t
 &\le\frac{\E|w(U)-w(V)|+|Z_\mu-Z_\nu|}{Z_\nu}\\
 &\le\frac{2L_H}{a\tau}\E\|U-V\|
 \le\frac{2L_Hd}{a\tau}.
\end{align*}
Let $r_\nu$ and $r_\mu$ be the remaining marginals, each of mass
$t$. If $t>0$, adding $(r_\nu\otimes r_\mu)/t$ to $\gamma_0$
completes the coupling at additional cost at most $D_K^2t$.
If $t=0$, $\gamma_0$ is already a coupling. The total cost proves
\eqref{eq:stat-tilt-bound}.
\end{proof}

\begin{proof}[Proof of Theorem~\ref{thm:finite-data}]
Couple each smoothed observation with the original observation to get
\[
 W_2(\nu_{n,h},\mu_n)\le h.
\]
Using~\eqref{eq:stat-deterministic-premise} and the triangle inequality gives
\begin{equation}\label{eq:stat-reference-triangle}
 W_2(\nu_{n,h},\mu)\le d+h.
\end{equation}
Both reference measures are supported in $K$. For $y,z\in K$, $x\in A$,
and $(\lambda,\tau)\in\Theta$,
\[
 \left|\frac{\|y-x\|^2-\|z-x\|^2}{2\lambda}\right|
 \le\frac{\|y-x\|+\|z-x\|}{2\lambda}\|y-z\|
 \le\frac{R_A}{\lambda_{\min}}\|y-z\|.
\]
Thus $E_{\lambda,x}^{f}$ is $L_*$-Lipschitz on $K$. The quadratic
term lies between $0$ and $R_A^2/(2\lambda_{\min})$, so
$\osc_K E_{\lambda,x}^{f}\le\osc_K f+R_A^2/(2\lambda_{\min})=B_*$.
Apply Lemma~\ref{lem:stat-tilt}
to~\eqref{eq:stat-reference-triangle} and use $\tau\ge\tau_{\min}$.
The constants are uniform over $A\times\Theta$,
which proves~\eqref{eq:stat-uniform-bound}. The triangle inequality
adds any solver error $s$.

Every proposal lies in $S_{n,h}\subseteq K$, where $b\le f$, so
$0\le E_{\lambda,x}^{f}(V)-b\le H_*$. Each proposal is therefore
accepted with probability at least $e^{-H_*/\tau_{\min}}>0$.
Independent trials terminate almost surely, and their geometric
count gives~\eqref{eq:stat-rejection-cost}. The accepted law
is~\eqref{eq:stat-learned-target}, as shown in
Section~\ref{app:conditional-construction}. Since $X_0\in A$ and
every output lies in $K\subseteq A$, the bound applies along any
parameter schedule in $\Theta$.

For any Borel output set, the numerator
in~\eqref{eq:stat-learned-target} is a Borel function of
$(x,\lambda,\tau)$ because the integrand is jointly measurable.
The denominator is positive and finite. Thus the target is a Borel
probability kernel, with finite second moments because its support
is compact.

\paragraph{Choosing the sample accuracy and bandwidth.}
For a requested error $\varepsilon>0$, define
\[
 A_* = e^{B_*/\tau_{\min}},
 \qquad C_*=\frac{D_K^2L_*}{\tau_{\min}},
 \qquad
 t_\varepsilon
 =\frac{\sqrt{C_*^2+\varepsilon^2/A_*}-C_*}{2}.
\]
Choose
\begin{equation*}
 d=h=\min\{1/2,t_\varepsilon/2\},
 \qquad s\le\varepsilon/2.
\end{equation*}
Then $d+h\le t_\varepsilon$ and
$t_\varepsilon^2+C_*t_\varepsilon=\varepsilon^2/(4A_*)$.
The right-hand side of~\eqref{eq:stat-uniform-bound} is at most
$\varepsilon/2$. Adding a solver error $s\le\varepsilon/2$ gives a
total error at most $\varepsilon$; exact rejection sampling has $s=0$.
Using this $d$ in~\eqref{eq:stat-sample-count} ensures the required
reference accuracy with probability at least $1-\delta$.

\end{proof}

\begin{proof}[Proof of Corollary~\ref{cor:stat-empirical-conditional}]
First establish the Lipschitz bound on $\M$. By compactness, choose
$r>0$ so that the closed $r$-neighborhood of $\M$ lies in the
open set where $f$ is $C^2$. This neighborhood is compact, so the
gradient has a finite bound $G_r$ there. If $y,z\in\M$ satisfy
$\|y-z\|\le r$, every point on their line segment is within distance
$r$ of $y$. The segment therefore stays in this neighborhood, giving
$|f(y)-f(z)|\le G_r\|y-z\|$. If
$\|y-z\|>r$, then $|f(y)-f(z)|\le(\osc_{\M} f/r)\|y-z\|$.
Thus $L_f^{\M}=\max\{G_r,\osc_{\M} f/r\}$ is finite.

The quadratic difference estimate in the proof of
Theorem~\ref{thm:finite-data} shows that the energy on $\M$ is
$L_A$-Lipschitz with oscillation at most $B_A$.
The law in~\eqref{eq:stat-indexed-sampler} is the exponential
reweighting of $\mu_n$. Apply Lemma~\ref{lem:stat-tilt} with $K=\M$
and use $\tau\ge\tau_{\min}$ to
get~\eqref{eq:stat-empirical-conditional-bound}.
The normalized weights are positive and continuous in the input and
parameters whenever the parameters are positive, so they define a
Borel probability kernel.
Its outputs lie in $\M\subseteq A$, which keeps every subsequent
input in the region covered by the bound.

To choose the required reference accuracy, let
$a_A=e^{B_A/\tau_{\min}}$ and $c_A=D_{\M}^2L_A/\tau_{\min}$. Given
$\varepsilon>0$, it suffices to take
\begin{equation*}
 d\le d_\varepsilon
 :=\frac{\sqrt{c_A^2+4\varepsilon^2/a_A}-c_A}{2}.
\end{equation*}
Indeed, $a_A(d_\varepsilon^2+c_Ad_\varepsilon)=\varepsilon^2$.
Lemma~\ref{lem:stat-empirical} with $d=d_\varepsilon$ then gives a
sufficient sample size.

\end{proof}
\section{More Implementations of \texorpdfstring{$\widehat Q$}{Q-hat}}
\label{app:hatQ_implementations}

Other than the approach provided in Appendix~\ref{app:statistical},
several standard methods can construct $\widehat Q$ through reference
estimation and conditional sampling. Density estimators and generators
use $\mathcal D_n$ to approximate $\mu$ by a fitted distribution $\nu$.
Score-based methods instead estimate the scores of Gaussian-smoothed
$\mu$ and use them in reverse diffusion. We describe these estimation
methods first, followed by methods for sampling the next iterate.

For a reference probability measure $\nu$ on $\R^D$, a current point
$x$, and $\lambda,\tau>0$, the corresponding conditional target is
\begin{equation*}
 \begin{aligned}
Q_{\lambda,\tau}^{f,\nu}(x,dy)
 &=\frac{w_{x,\lambda,\tau}(y)\nu(dy)}{Z_{x,\lambda,\tau}^{\nu}},
 &w_{x,\lambda,\tau}(y)
 &=\exp\!\left[-\frac{E_{\lambda,x}^{f}(y)}{\tau}\right],\\
E_{\lambda,x}^{f}(y)&=f(y)+\frac{\|y-x\|^2}{2\lambda},
 &Z_{x,\lambda,\tau}^{\nu}&=\int w_{x,\lambda,\tau}(y)\nu(dy).
 \end{aligned}
\end{equation*}
We assume $f$ is measurable and can be evaluated on the reference
support, and $0<Z_{x,\lambda,\tau}^{\nu}<\infty$.
The weight favors smaller objective values and shorter distances to
$x$, while preserving the support of $\nu$. Adding Gaussian noise to
construct a smoothed reference is a separate step in some methods below.

Taking $\nu=\mu$ recovers the ideal kernel $Q_{\lambda,\tau}^{f}$;
a fitted $\nu$ gives a target based on the estimated reference.
With the data and fitted model fixed, $\widehat Q_{\lambda,\tau}^{f}$
is the law of the point returned by the sampling procedure.
It equals $Q_{\lambda,\tau}^{f,\nu}$ when that target is sampled exactly.
Its accuracy relative to the ideal kernel therefore depends on both
reference estimation and conditional sampling.

\subsection{Estimating the Reference Distribution}
\label{app:impl-reference-estimation}

A density estimate or generator supplies samples from a reference
distribution $\nu$. A score model supplies the gradients used in
reverse diffusion. These models are fitted once and reused as
$x$, $\lambda$, and $\tau$ change.

\subsubsection{Gaussian Kernel Density Estimation}
\label{app:impl-kde}

Gaussian kernel density estimation places a Gaussian distribution at
each observation~\cite{rosenblatt1956remarks,parzen1962estimation}.
Choose a bandwidth $h>0$ and give the $n$ components equal weights:
\begin{equation}\label{eq:impl-kde}
p_{n,h}^{\mathrm G}(y)=\frac{1}{n}\sum_{i=1}^{n}
 \phi_{h^2I_D}(y-Y^{(i)}),
 \qquad \nu_{n,h}^{\mathrm G}(dy)=p_{n,h}^{\mathrm G}(y)\,dy,
\end{equation}
where $\phi_{\Sigma}$ is the centered Gaussian density with covariance
$\Sigma$. To generate a candidate, choose an observation uniformly,
draw $\xi\sim\mathcal N(0,I_D)$, and return $Y^{(i)}+h\xi$.
Thus $h$ controls how far candidates spread around the data.
The finite sum gives the density, and differentiating its logarithm
gives the score needed by the Langevin methods below. For conditional
sampling, these candidates can be weighted by
$w_{x,\lambda,\tau}$, or drawn closer to $x$ using
Section~\ref{app:impl-gaussian-proposals}.

\subsubsection{Generative Models and Normalizing Flows}
\label{app:impl-generators}

A trained generator $G_\theta:\R^d\to\R^D$ maps a simple random
input into a candidate point. To sample, draw $Z\sim\gamma$, such as
a standard Gaussian, and compute $Y=G_\theta(Z)$. The resulting
reference distribution is $\nu_\theta=(G_\theta)_\#\gamma$.
Independent inputs give independent candidates, which can then be
weighted by $w_{x,\lambda,\tau}(Y)$.

A normalizing flow gives a concrete way to fit such a generator and
evaluate its density~\cite{rezende2015flows}. Choose $d=D$ and an
invertible differentiable $G_\theta$ with nonsingular derivative.
If $\gamma$ has density $p_0$, compute
\[
p_{\nu_\theta}(y)
 =p_0(G_\theta^{-1}(y))
   \left|\det DG_\theta^{-1}(y)\right|.
\]
Fit $\theta$ by maximizing
$\sum_{i=1}^n\log p_{\nu_\theta}(Y^{(i)})$, then keep it fixed during
optimization. The fitted density can be used by Metropolis or
gradient-based samplers. For a generator without an explicit density,
Section~\ref{app:impl-latent-sampling} describes sampling through $Z$.

\subsubsection{Score Estimation}
\label{app:impl-scores}

Reverse diffusion needs the score at several noise levels. For noise
variance $t>0$, the density after adding Gaussian noise to $Y\sim\nu$
and its score are
\begin{equation*}
p_t^\nu(z)=\int\phi_{tI_D}(z-y)\nu(dy),
 \qquad s_t^\nu(z)=\nabla_z\log p_t^\nu(z).
\end{equation*}
The Gaussian smoothing gives a density even when $\nu$ is supported
on a manifold.

To estimate $s_t^\mu$ from the observations, train a network by
denoising score matching~\cite{vincent2011connection,song2021sde}.
At each training step, choose an observation uniformly, draw a noise
variance $t$ and independent $\xi\sim\mathcal N(0,I_D)$, and form
$z=Y^{(i)}+\sqrt t\,\xi$. Give the network $(z,\sqrt t)$ and use
$-\xi/\sqrt t$ as the regression target. The resulting loss is
\begin{equation*}
 \frac{1}{n}\sum_{i=1}^{n}\E_{t,\xi}\!\left[
a(t)\left\|
s_\theta(Y^{(i)}+\sqrt t\,\xi,\sqrt t)
 +\frac{\xi}{\sqrt t}\right\|^2\right].
\end{equation*}
Here $a(t)>0$ weights the noise levels; choose it and the distribution
of $t$ so that $\E_t[a(t)/t]<\infty$.
The fitted network estimates the score at each noise level in its
training range. Section~\ref{app:impl-diffusion} uses it to generate
candidates from the current point $x$.

\subsection{Sampling the Conditional Distribution}
\label{app:impl-conditional-sampling}

Each method below returns one point for a fixed call $(x,\lambda,\tau)$.
All its internal sampling steps use the same $x$, $\lambda$, and $\tau$.
The returned point becomes the next iterate of PRISM.

Candidates drawn from $\nu$ use the full weight
$w_{x,\lambda,\tau}$. Alternatively, we can account for the distance
to $x$ when drawing candidates. Put $v=\lambda\tau$ and define
\begin{equation*}
K_v^{\nu}(x,dy)
 =\frac{\exp[-\|y-x\|^2/(2v)]\nu(dy)}
 {\int\exp[-\|z-x\|^2/(2v)]\nu(dz)}.
\end{equation*}
This distribution favors reference points near $x$. It leaves only
the objective to be included in the weights:
\begin{equation}\label{eq:impl-factorization}
Q_{\lambda,\tau}^{f,\nu}(x,dy)
 =\frac{e^{-f(y)/\tau}K_{\lambda\tau}^{\nu}(x,dy)}
 {\int e^{-f(z)/\tau}K_{\lambda\tau}^{\nu}(x,dz)}.
\end{equation}
Gaussian mixtures and reverse diffusion provide two ways to draw
from $K_v^\nu(x,\cdot)$.

\subsubsection{Gaussian Mixture Proposals}
\label{app:impl-gaussian-proposals}

For the Gaussian KDE in~\eqref{eq:impl-kde}, we can draw candidates
near $x$ by changing the component weights, means, and variance.
Completing the square gives
\begin{equation*}
 \begin{aligned}
K_v^{\nu_{n,h}^{\mathrm G}}(x,\cdot)
 &=\sum_{i=1}^{n}\pi_i(x)\,\mathcal N(m_i(x),c^2I_D),\\
 \pi_i(x)&=\frac{\exp[-\|Y^{(i)}-x\|^2/(2(h^2+v))]}
 {\sum_{j=1}^{n}\exp[-\|Y^{(j)}-x\|^2/(2(h^2+v))]},\\
m_i(x)&=\frac{vY^{(i)}+h^2x}{h^2+v},
 \qquad c^2=\frac{h^2v}{h^2+v}.
 \end{aligned}
\end{equation*}
Compute these quantities once for the call. Choose an index $i$
with probability $\pi_i(x)$, draw $\xi\sim\mathcal N(0,I_D)$, and
set $V=m_i(x)+c\xi$. Repeat to obtain $B$ independent candidates,
evaluate $f$ at each, and return one with probability proportional
to $e^{-f(V)/\tau}$. This gives an approximate conditional sample.
If a lower bound $L\le f$ is known on the proposal support, an exact
alternative is to draw candidates one at a time, accept each with
probability $\exp[-(f(V)-L)/\tau]$, and return the first accepted point.

\subsubsection{Importance Resampling and Metropolis Sampling}
\label{app:impl-resampling}

\emph{Importance resampling.}
This method draws a batch of candidates and selects one using their
weights~\cite{smith1992sampling}. Draw $V_1,\ldots,V_B$ independently
from $\nu$, compute $w_{x,\lambda,\tau}(V_b)$ for each candidate,
and divide by the sum to obtain selection probabilities:
\begin{equation*}
 \Pr(J=b\mid V_1,\ldots,V_B)
 =\frac{w_{x,\lambda,\tau}(V_b)}
 {\sum_{a=1}^{B}w_{x,\lambda,\tau}(V_a)}.
\end{equation*}
Draw $J$ with these probabilities and return $V_J$.
Only reference samples and objective values are needed.
For candidates from $K_{\lambda\tau}^\nu(x,\cdot)$, use
$e^{-f(V_b)/\tau}$ instead. A finite batch gives an approximation;
if most weight falls on a few candidates, a larger batch may be needed.

If candidates instead come from a density $r_x$, assume that $r_x$
and the reference density $p_\nu$ are defined with respect to the same
measure, with $r_x>0$ wherever $p_\nu>0$. Replace each weight by
$w_{x,\lambda,\tau}(V_b)p_\nu(V_b)/r_x(V_b)$.
Draw a new batch if the total weight is zero. This ratio corrects for
the different candidate distribution.

\emph{Independence Metropolis.}
This method updates one candidate at a time~\cite{hastings1970montecarlo}.
Draw $V_0\sim\nu$ and choose a number of steps $N$.
At each step, draw $V'\sim\nu$ independently and compare its energy
with that of the current point $V$. Accept $V'$ with probability
\begin{equation}\label{eq:impl-independence-mh}
 \alpha(V,V')=\min\left\{1,
 \exp\!\left[-\frac{E_{\lambda,x}^{f}(V')-E_{\lambda,x}^{f}(V)}{\tau}\right]
 \right\}.
\end{equation}
Lower-energy proposals are always accepted. On rejection, keep $V$.
After $N$ steps, return the current point.
The reference cancels from the acceptance ratio, so its density is
not needed. The chain has $Q_{\lambda,\tau}^{f,\nu}(x,\cdot)$ as
an invariant distribution.

\subsubsection{Langevin and Hamiltonian Monte Carlo}
\label{app:impl-langevin}

These methods use gradients to move candidates. Suppose $\nu$ has
a positive differentiable density $p_\nu$ on $\R^D$ and $f$ is
differentiable there. Compute the target log density, up to a constant,
and its gradient:
\[
 \ell_x(y)=\log p_\nu(y)-\frac{f(y)}{\tau}
              -\frac{\|y-x\|^2}{2\lambda\tau},
 \qquad
 \nabla\ell_x(y)=\nabla\log p_\nu(y)
              -\frac{\nabla f(y)}{\tau}-\frac{y-x}{\lambda\tau}.
\]
The three gradient terms favor regions of high reference density,
smaller objective values, and proximity to $x$, respectively.

\emph{Unadjusted Langevin algorithm (ULA).}
Start at $V_0=x$, choose a step size $\eta>0$ and a number of steps
$N$, and repeat
\begin{equation}\label{eq:impl-ula}
V_{\ell+1}=V_\ell+\eta\nabla\ell_x(V_\ell)
                      +\sqrt{2\eta}\,\xi_\ell,
 \qquad \xi_\ell\sim\mathcal N(0,I_D)
\end{equation}
with independent noise at each step. Each update takes a step along
the target score and adds Gaussian noise. Return $V_N$.
The step size and number of steps determine the discretization and
finite-run errors~\cite{roberts1996exponential}.

\emph{Metropolis-adjusted Langevin algorithm (MALA).}
Use the right-hand side of~\eqref{eq:impl-ula} to propose $V'$ from
the current $V$, then accept or reject it. The proposal density is
$g_\eta(v'\mid v)=\phi_{2\eta I_D}(v'-v-\eta\nabla\ell_x(v))$;
accept with probability
\[
 \min\left\{1,
 \frac{\exp[\ell_x(V')]g_\eta(V\mid V')}
 {\exp[\ell_x(V)]g_\eta(V'\mid V)}\right\}.
\]
Keep $V$ on rejection and return the state after $N$ steps.
This correction requires density values and makes the conditional
target invariant. Gaussian KDE supplies both the density and score.

\emph{Hamiltonian Monte Carlo.}
Starting from $y=x$, use an auxiliary momentum to make longer
proposals~\cite{duane1987hybrid,neal2011mcmc}.
At each iteration, draw $p\sim\mathcal N(0,I_D)$ and set
$H_x(y,p)=-\ell_x(y)+\|p\|^2/2$.
Record $H_{\rm old}=H_x(y,p)$, then take a chosen number of leapfrog
steps of size $\eta$. Each step consists of
\[
p\leftarrow p+\tfrac{\eta}{2}\nabla\ell_x(y),\qquad
y\leftarrow y+\eta p,\qquad
p\leftarrow p+\tfrac{\eta}{2}\nabla\ell_x(y).
\]
Reverse the final momentum to obtain $(y',p')$ and accept the new
position with probability $\min\{1,\exp[H_{\rm old}-H_x(y',p')]\}$.
Otherwise keep the old position. Repeat for $N$ iterations and return the position.
The conditional target is invariant under these updates.

\subsubsection{Sampling in Latent Variables}
\label{app:impl-latent-sampling}

For a generator $Y=G_\theta(Z)$, define the target distribution of
its input by
\begin{equation*}
 \pi_{\theta,x}(dz)
 =\frac{\exp[-E_{\lambda,x}^f(G_\theta(z))/\tau]\gamma(dz)}
 {\int\exp[-E_{\lambda,x}^f(G_\theta(u))/\tau]\gamma(du)}.
\end{equation*}
If $Z\sim\pi_{\theta,x}$, then
$G_\theta(Z)\sim Q_{\lambda,\tau}^{f,\nu_\theta}(x,\cdot)$,
even when $G_\theta$ is not invertible. Thus we can approximate the
conditional target by sampling in the input space and mapping the
result through $G_\theta$.

For weighted resampling, draw $Z_1,\ldots,Z_B$ independently from
$\gamma$ and compute $Y_b=G_\theta(Z_b)$. Select one $Y_b$ with
probability proportional to $e^{-E_{\lambda,x}^f(Y_b)/\tau}$.

Alternatively, start at $Z_0\sim\gamma$ and run independence Metropolis
with proposals from $\gamma$. Apply~\eqref{eq:impl-independence-mh}
to their images under $G_\theta$ and return $G_\theta(Z_N)$ after
$N$ steps. Both procedures need only samples from $\gamma$ and
evaluations of $G_\theta$ and $f$.

If $\gamma$ has a positive differentiable density $p_0$ on $\R^d$
and $G_\theta$ and $f$ are differentiable, Langevin or Hamiltonian
sampling can instead use the log density
$\log p_0(z)-E_{\lambda,x}^f(G_\theta(z))/\tau$.

\subsubsection{Reverse Diffusion}
\label{app:impl-diffusion}

Reverse diffusion uses estimates of the scores $s_t^\nu$ to generate
approximate samples from $K_v^\nu(x,\cdot)$. The network in
Section~\ref{app:impl-scores} estimates the scores of Gaussian-smoothed
$\mu$ from $\mathcal D_n$, so its intended reference is $\nu=\mu$.
Set $v=\lambda\tau$ and choose variances
$v=t_0>\cdots>t_L>0$ within the training range. Start at $Z_0=x$
and, for $j=0,\ldots,L-1$, use
\begin{equation}\label{eq:impl-reverse-euler}
 \begin{aligned}
Z_{j+1}&=Z_j+(t_j-t_{j+1})s_\theta(Z_j,\sqrt{t_j})
                  +\sqrt{t_j-t_{j+1}}\,\xi_j,\\
Z_0&=x,\qquad \xi_j\sim\mathcal N(0,I_D).
 \end{aligned}
\end{equation}
At each step, evaluate the network at the current point and noise
level, move in the score direction, and add independent Gaussian
noise with the displayed variance. Run $B$ independent trajectories
from the same $x$ to obtain endpoints $V_1,\ldots,V_B$.
Stopping at $t_L>0$ leaves residual noise in these approximate proposals.
Evaluate $f(V_b)$, select one endpoint with probability proportional
to $e^{-f(V_b)/\tau}$, and return it. By the factorization
in~\eqref{eq:impl-factorization}, this objective resampling gives an
approximate draw from $Q_{\lambda,\tau}^{f,\nu}(x,\cdot)$.

To see why the proposals have this form, let $Y_0\sim\nu$ and
$Y_t=Y_0+B_t$, with independent standard Brownian motion $B$.
Bayes' formula gives
$\operatorname{Law}(Y_0\mid Y_v=x)=K_v^\nu(x,\cdot)$.
The corresponding reverse diffusion~\cite{song2021sde} is
\begin{equation*}
dZ_u=s_{v-u}^\nu(Z_u)\,du+d\overline B_u,
 \qquad Z_0=x,\quad 0\le u<v,
\end{equation*}
where $\overline B$ is standard Brownian motion.
For $0<u<v$, its law is
$p_{v-u}^\nu(z)\phi_{uI_D}(x-z)\,dz/p_v^\nu(x)$.
As $u\uparrow v$, the Gaussian smoothing vanishes and this law
converges weakly to $K_v^\nu(x,\cdot)$.
Equation~\eqref{eq:impl-reverse-euler} discretizes this process
and replaces its score by the fitted network.

The TFBind8 and Superconductor implementations further denoise the endpoints,
then decode or clip them before objective resampling;
Appendix~\ref{app:experiment_appendix} gives the task-specific details.
Related uses of diffusion models for objective-dependent sampling
appear in \citet{kong2025diffusion}.

\subsubsection{Sequential Monte Carlo}
\label{app:impl-smc}

Sequential Monte Carlo introduces the energy in stages~\cite{delmoral2006smc}.
Choose $0=\beta_0<\beta_1<\cdots<\beta_L=1$, with stage targets
\begin{equation*}
 \pi_\ell(dy)\propto
 \exp[-\beta_\ell E_{\lambda,x}^f(y)/\tau]\nu(dy).
\end{equation*}
These run from the reference $\nu$ to the conditional target.
Draw $B$ independent points from $\nu$. At each stage
$\ell=1,\ldots,L$, first compute weights
$\exp[-(\beta_\ell-\beta_{\ell-1})E_{\lambda,x}^f(V_b)/\tau]$
and normalize them to sum to one. Next, draw $B$ points independently
with replacement using these probabilities and give them equal weights.
Finally, move each selected point using one or more independence
Metropolis steps: propose $V'\sim\nu$ and use
\eqref{eq:impl-independence-mh} with energy $\beta_\ell E_{\lambda,x}^f$.
These moves preserve $\pi_\ell$ and allow repeated copies to separate.
After stage $L$, return a uniformly selected point as the approximate
sample.

\section{Supplemental Experiments and Implementation Details}\label{app:experiment_appendix}

\subsection{Detailed Experimental Setup for Section~\ref{sec:experiment1}}\label{app:experiment1_appendix}

\paragraph{Construction of the PCA task.}
We consider a Gaussian population $Z\sim\mathcal N(0,C)$ with
\begin{equation*}
 C=R\diag(10,8,6,5,4,3,2,1,0.5,0.2)R^\top,
\end{equation*}
where $R=[r_1,\ldots,r_{10}]$ is obtained by QR factorization of a standard Gaussian matrix with seed $42$. A two-dimensional representation has reconstruction $PZ$, where $P=UU^\top$ and $U^\top U=I_2$. Each such projector represents a point on $\mathrm{Gr}(2,10)$. Its reconstruction error is the following objective:
\begin{equation*}
 f(X)=\E\|Z-XZ\|_2^2
     =\tr\!\big((I-X)C(I-X)^\top\big).
\end{equation*}
We use this population objective directly, including its full expression for ambient candidates.

\paragraph{Reference dataset.}
We generate $n=5{,}000$ feasible projector observations using seed $42$.
With probability $0.003$, an observation is uniform on $\M$, otherwise, it follows
\begin{equation*}
 \nu_c(dP)\propto
 \exp\{2000\,\tr(P_cP)\}\,\sigma(dP),
\end{equation*}
where $\sigma$ is uniform probability measure on $\M$. The observations therefore concentrate around a suboptimal subspace, with sparse coverage elsewhere. These observations form the reference dataset $\mathcal D_n$. The learner receives the observations without their source mixture labels or parameters.

\paragraph{Estimating the refinement kernel.}
Let $y_i=\operatorname{vec}(P_i)\in\R^{100}$.
We center the observations and use singular value decomposition to estimate an affine basis $B$, giving coordinates $z_i=B^\top(y_i-\bar y)$.
We then fit quadratic relations from the numerical null space:
\begin{equation*}
 h_\ell(z)=a_\ell+b_\ell^\top z+z^\top H_\ell z=0,
 \qquad
 \widehat\M=\{\bar y+Bz:h(z)=0\}.
\end{equation*}
Both fits use all $5000$ observations and relative singular-value threshold $10^{-10}$, while polynomial features are scaled by their root mean square. On this fitted $\widehat \M$, we fit a uniform--radial mixture with density
\begin{equation*}
 \widehat\rho(z)=\widehat\pi+
 \frac{1-\widehat\pi}{Z_{\widehat\kappa}}
 \exp\!\left\{-\frac{\widehat\kappa}{2}
                    \|z-\widehat m\|_2^2\right\}
\end{equation*}
relative to normalized induced volume, where $Z_{\widehat\kappa}$ normalizes the radial component. Five trimming rounds retain observations within the median radius plus ten interquartile ranges of the current center. The retained-point mean, projected onto the learned support, estimates $\widehat m$, while the excluded fraction estimates $\widehat\pi$.
We estimate $\widehat\kappa$ by dividing the inferred intrinsic dimension by the mean squared retained-point radius.
The normalizer is approximated by thermodynamic integration using Metropolis-adjusted Langevin chains on the estimated support. The fitted density, mapped back by $z\mapsto\bar y+Bz$, defines $\widehat\nu$. Using what we have constructed, the fitted target is
\begin{equation*}
 Q_{\lambda,\tau}^{f,\widehat\nu}(x,dy)
 \propto
 \exp\!\left\{-\frac{f(y)}{\tau}
              -\frac{\|y-x\|_2^2}{2\lambda\tau}\right\}
 \widehat\nu(dy).
\end{equation*}

\begin{figure}[t]
 \centering
 \includegraphics[width=\linewidth]{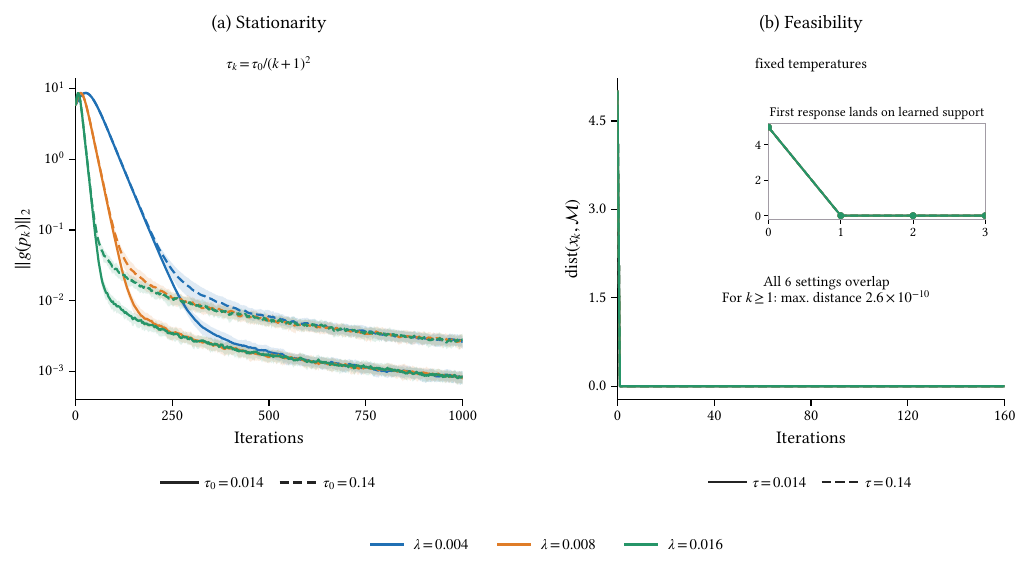}
 \caption{Stationarity and feasibility results.
 (a) Current gradient norm over $1{,}000$ updates with $\tau_k=\tau_0/(k+1)^2$.
 (b) Raw distance to the true manifold over $160$ updates at fixed temperature, the zoomed-in figure shows the first three updates. Colors distinguish $\lambda$, while solid and dashed lines denote temperature settings $0.014$ and $0.14$, respectively. Lines and shading show means and $\pm$ one sample SD across $50$ runs.}
 \label{fig:exp-stationarity-landing}
\end{figure}

\paragraph{Additional stationarity and feasibility results.}
Figure~\ref{fig:exp-stationarity-landing} uses the same fitted refinement kernel in Section~\ref{sec:experiment1} and $50$ trajectories for each $(\lambda, \tau)$ pair.

Figure~\ref{fig:exp-stationarity-landing}(a) shows the current Riemannian gradient norm $\|g(p_k)\|_2$. Here, we use $1000$ updates with $\tau_k=\tau_0/(k+1)^2$ as suggested in~\eqref{eq:temperature-choice}. This behavior qualitatively illustrates the effect of decreasing temperature in the stationarity results presented in Theorem~\ref{thm:finite-time}.

Figure~\ref{fig:exp-stationarity-landing}(b) shows $\dist(X_k,\M)$ over $160$ updates with fixed $\tau=\tau_0$.
The maximum distance after the first response is $2.59\times10^{-10}$ across all settings and runs, further illustrating the results presented in Proposition~\ref{prop:feasibility}.

\subsection{Detailed Experimental Setup for Section~\ref{sec:experiment_building_control}}
\label{app:exp_building}

\paragraph{Physical simulator and operating conditions.}
We use the public BOPTEST \texttt{bestest\_\allowbreak hydronic\_\allowbreak heat\_\allowbreak pump}
test case~\cite{blum2021building}, comprising a single-zone residential
building, hydronic floor heating, and an air-to-water heat pump.
The Modelica model is executed as a Functional Mock-up Unit (FMU).
We retain its thermal dynamics, embedded Brussels weather,
occupancy inputs, and equipment interlocks, and control the
compressor through its modulation override.

Data collection and deployment begin at zero-based weather day 23,
following seven days of default-controller warmup.
All complete collected records start from the same warmed simulator
state. Windows extracted later within these records have different
initial thermal states.
The controllers observe 119 internal simulator variables, the current
temperature and power measurements, and exact weather and occupancy
forecasts.

Deployment lasts 48 hours, with 96 half-hour control commands.
Two quarter-hour observations are recorded during each control
interval, following the grouped notation
$y_t=(y_{t,1},y_{t,2})$ in the main text.
Electrical power is averaged over each observation interval
using integrated simulator power.

\paragraph{Reference temperature, electricity price, and constraints.}
Writing $h$ for elapsed time in hours, the prescribed temperature is
\[
 r(h)=21.3316628756+1.2\sin^2(\pi h/24),
 \qquad 0\le h\le48.
\]
Thus, the discrete reference satisfies
$r_{t,q}=r(t/2+q/4)$ for $q\in\{1,2\}$.
The electricity price is $0.2535$ currency units/kWh, increasing
to $0.4535$ during hours $[17,20)$ and $[41,44)$.

All controllers use the objective and MPC formulation in
\eqref{eq:exp-building-objective} and
\eqref{eq:exp-building-mpc}.
The prediction horizon contains four commands, or two hours,
and shrinks to three, two, and one commands at the final decisions.
Room temperature is constrained to $[18,26]\,{}^\circ\mathrm C$.
Each planning problem includes its endpoint tracking penalty,
whereas the reported deployment objective includes this penalty
only at hour 48.
There is no recovery period or terminal water-temperature
constraint; hidden wall and floor thermal energy is not equalized.

\paragraph{Collection of operating records.}
We use the first 16 trajectories, in generation order, from an
existing collection of 64 trajectories.
Each trajectory spans eight simulated days and contains 384
half-hour transitions.
For trajectory $i\in\{0,\ldots,15\}$, the sampling seed is $3101+i$.
We draw a background operating level
$b\sim\operatorname{Unif}(0.34,0.46)$ and fluctuation scale
$\sigma\sim\operatorname{Unif}(0.15,0.35)$, then generate
\[
 \begin{aligned}
 \eta_0&\sim\mathcal N(0,1),\\
 \eta_t&=0.95\eta_{t-1}
          +\sqrt{1-0.95^2}\,\epsilon_t,
 \qquad \epsilon_t\overset{\mathrm{iid}}\sim\mathcal N(0,1),\\
 u_t&=\operatorname{sigmoid}
       \bigl(\operatorname{logit}(b)+\sigma\eta_t\bigr).
 \end{aligned}
\]

Each day contains a randomly positioned excitation window whose
length is uniformly selected from 12--20 control intervals
(6--10 hours).
Within this window, commands are constant over blocks of
1--4 intervals, truncated at the window boundary.
With probability $0.6$, a block uses a uniformly selected level
from $\{0.08,0.2,0.4,0.6,0.85,1\}$.
Otherwise, its level is zero with probability $0.25$ and
log-uniform on $[10^{-4},0.03]$ with probability $0.75$.
We also insert the six-command sequence
$(0,0.003,0.006,0.009,0.012,0.025)$ at a random position each day,
reversing its order with probability $0.5$.
These small commands expose compressor switching behavior.

Every simulated trajectory is retained.
The sampling procedure receives neither the target temperature
nor electricity prices and performs no objective-based rejection.

\paragraph{Window extraction and fitting split.}
We separate complete trajectories into 13 fitting records and
three validation records before extracting windows.
The validation record indices are $\{2,7,14\}$; the remaining
indices among $\{0,\ldots,15\}$ are used for fitting.
Each two-hour window contains four commands, their eight
quarter-hour observations, its initial simulator state and
measurements, and the corresponding external inputs.
Windows are extracted with a stride of one control interval.

The fitting and validation subsets contain 4,992 and 1,152
original transitions, yielding 4,953 and 1,143 overlapping windows,
respectively.
Validation selects model settings and is not subsequently
included in fitting.
All normalization statistics are estimated from fitting records.
Overlapping windows require no additional simulator calls and
are not treated as independent physical samples.
The target temperature, electricity prices, and control-objective
values are not used to fit the reference model.

\paragraph{Conditional response model.}
Let $c_t$ contain the current full state $z_t$, current measured
outputs, time, and known external inputs over the prediction horizon.
The response predictor $\widehat R(c_t,u)$ maps a candidate command
sequence directly to its measured responses.
It predicts the temperature and power outputs rather than
recursively predicting the complete internal simulator state.

For each of the eight quarter-hour observation positions, we fit
a causal ridge-regression head for the three temperature changes
relative to the current measurements.
A separate ridge head predicts a power amplitude, which is
transformed by softplus and multiplied by a learned compressor
switching gate.
Features include initial state and measurements, external inputs,
time-of-day features, current and preceding commands within the
window, squared and logarithmic commands, switching indicators,
and command--context interactions.
Features for an observation do not include commands applied
after that observation.

Ridge penalties are selected from
$\{10^{-6},10^{-4},10^{-2}\}$ using validation prediction errors.
The switching thresholds are estimated from observed command
transitions and compressor states.
The fitted turn-on and turn-off thresholds are approximately
$0.0099361$ and $0.0051124$.
Validation selects sigmoid width $10^{-5}$ from
$\{10^{-5},5\cdot10^{-5},2\cdot10^{-4}\}$.
The gate is initialized using the observed compressor state
at the current decision time.

The response heads contain 6,192 fitted coefficients and
intercepts. No neural network or physical differential equations
are supplied to this model.
Validation RMSEs are $0.01649$, $0.36324$, and
$0.46188\,{}^\circ\mathrm C$ for room, supply-water, and
return-water temperatures, respectively, and $0.09689$ kW
for electrical power.

\paragraph{Continuous reference law.}
For a decision with $H=H_t$ planned commands, define
\begin{equation}
 \begin{gathered}
 w\sim\gamma_H=\mathcal N(0,I_H),\qquad
 u(w)=\operatorname{sigmoid}(m_H+L_Hw),\\
 G_{c_t}(w)=\bigl(u(w),\widehat R(c_t,u(w))\bigr),\qquad
 \widehat\nu_{c_t}=(G_{c_t})_\#\gamma_H.
 \end{gathered}
 \label{eq:exp-building-generator}
\end{equation}
The sigmoid acts componentwise.
For the four-command model, we clip fitting commands to
$[10^{-6},1-10^{-6}]$, transform them to logits, and estimate
their mean $m$ and covariance $\Sigma$.
We set
\[
 LL^\top=0.95\Sigma+0.05D+10^{-5}I,
 \qquad
 D_{ii}=\max\{\Sigma_{ii},0.03^2\}.
\]
For shorter final horizons, $m_H$ and $L_H$ are the leading
$H$ entries and leading $H\times H$ Cholesky block.
The predictor retains the corresponding first $2H$ observation
heads. No additional fitting is required.

The control prior is shared across contexts; dependence on the
current state and forecasts enters through $\widehat R$.
Given the context and commands, predicted responses are deterministic.
The continuous reference can generate command sequences absent
from the collected records.

\paragraph{Finite implementation of the refinement kernel.}
A window contains $H$ commands and $2H$ four-channel observations,
giving $9H$ joint coordinates.
Let $S_H(u,y)$ standardize these coordinates using fitting
statistics and divide the resulting vector by $\sqrt{9H}$.
Command means and scales are position-specific, with scale floor
$0.05$.
Response means are observation-position-specific, while response
scales are pooled by channel, with floors
$(0.1,0.1,0.1,0.05)$ in the corresponding physical units.

For compactness, define $J_t(u,y)$ to be the complete planning
objective displayed in~\eqref{eq:exp-building-mpc}, evaluated
on the candidate commands and responses at decision time $t$.
Prices, targets, the preceding command, and the prediction
horizon are fixed by that decision.
For standardized anchor $a_\ell$, the latent target of outer
update $\ell$ is
\begin{equation}
 \pi_\ell(dw\mid c_t)\propto
 \mathbf 1_{\widehat{\mathcal C}_{c_t}}(w)
 \exp\!\left[
 -\frac{
 J_t(G_{c_t}(w))
 +\|S_H(G_{c_t}(w))-a_\ell\|^2/(2\lambda)}
 {\tau_\ell}
 \right]\gamma_H(dw).
 \label{eq:exp-building-latent-target}
\end{equation}
Here $\widehat{\mathcal C}_{c_t}$ imposes predicted room-temperature
constraints and excludes two uncertain switching intervals from
every planned command.
These intervals are inferred from fitting records and padded
by eight fitted sigmoid widths, giving approximately
\[
 [0.00489884,0.00532598],
 \qquad
 [0.00939341,0.01047873].
\]
The learned-feasibility tolerance is $10^{-7}$.
These restrictions concern the learned model and do not certify
physical feasibility.

At each decision, learned-model SLSQP~(Sequential Least Squares Programming) is initialized from six
ranked candidate sequences and all $2^H$ binary OFF/ON patterns,
with at most 180 iterations per start. The OFF level is $10^{-6}$ and the ON level is $0.3984719283$,
the median observed active fitting command above the excluded
switching intervals. The best learned-feasible solution initializes the sampler;
up to four distinct solutions supply local Gaussian proposal centers.

We use five outer updates with $\lambda=10$ and geometrically
decreasing temperatures $\tau$ from $0.02$ to $10^{-5}$.
Each update performs 256 Metropolis transitions.
With probability $0.25$, the proposal is a symmetric latent
random walk with standard deviation
$0.15\sqrt{\tau_\ell/0.02}+0.005$.
Otherwise, it is drawn from a Gaussian mixture whose covariance
matrices are constructed from local target curvature.
Acceptance probabilities include the appropriate proposal-density
correction.
The final chain state determines the next anchor.
After the fifth update, the controller executes the first command
of the final sampled plan.

The implemented refinement kernel $\widehat Q$ is obtained by
mapping the finite-chain output through $G_{c_t}$; it is distinct
from the response predictor $\widehat R$ and the reference law
$\widehat\nu_{c_t}$.
No physical queries are used to select or repair candidate plans.
Local optimization and finite sampling both contribute to this
implementation; exact sampling or global mixing is not assumed.

\paragraph{Physical-query budget and comparators.}
One oracle call advances the physical simulator by one half-hour
control interval.
The 16 selected records contain
$16\times8\times48=6{,}144$ original transitions, including
both fitting and validation records.
This is the acquisition allowance charged to \ouralg-MPC.
The records are reused from the historical 64-record collection,
which required $24{,}576$ transitions in total.
Thus, the matched allowance measures the transitions attributable
to the selected data, not the entire experimental-development cost.

Budgeted CEM-MPC receives $6{,}144$ online planning calls.
At each ordinary four-command decision, 64 calls support
16 candidate rollouts, evaluated in four generations of four.
The final three decisions have horizons $3$, $2$, and $1$ and
use $63$, $64$, and $65$ calls, respectively.
The allowance is therefore distributed throughout deployment.

CEM uses a temporally correlated Gaussian proposal with initial
command standard deviation $0.5$ and correlation
$0.7^{|i-j|}$ between command positions.
Its elite fraction is $0.2$, with at least two elite candidates.
The previous plan is shifted to initialize the next decision,
and evaluated incumbents are retained across generations.
All candidate responses are obtained from the physical FMU.

Reference MPC uses the same physical FMU and prediction horizon
without the matched query cap.
Its numerical search combines CEM with population 64 for eight
generations, binary switching patterns at several amplitudes,
and finite-difference SLSQP on four promising patterns.
SLSQP uses at most 40 iterations and difference step $0.001$.
The reported deployment requires 263,843 planning calls.
This comparator uses no learned response model and represents
a best-found numerical solution, rather than a certified optimum.

The collected-trajectory comparator is the feasible first-48-hour
trajectory with the lowest common objective among the same
16 records.
Record 006 is selected.
Its initial state and external inputs match deployment.
The comparator is one complete recorded trajectory, not a
combination of windows or a pointwise selection across records.

Initialization, execution, evaluation replays, and development
queries are recorded separately from the matched acquisition
and planning allowances.
Controller settings were selected using physical pilots and
then fixed before the reported optimizer seeds were evaluated.

\paragraph{Physical evaluation and reported statistics.}
We evaluate five runs each of \ouralg-MPC and Budgeted CEM-MPC
using seeds 8501--8505, with one fitted reference model,
initial state, and external-input realization held fixed.
Reference MPC is evaluated once using seed 8401.
All runs are retained.

Controller commands are executed using 60-second FMU communication
steps and independently replayed using 15-second steps.
All evaluated controller runs pass room-temperature safety checks
at the recorded integration reads, with tolerance
$10^{-4}\,{}^\circ\mathrm C$.
The collected comparator passes its existing 60-second checks;
it does not receive a separate 15-second replay.
These checks do not constitute a continuous-time safety certificate.

Reported costs use the 60-second execution, quarter-hour tracking
quadrature, and integrated electrical power.
Tracking RMSE is
\[
 \operatorname{RMSE}
 =
 \left[
 \frac{1}{2K}
 \sum_{t=0}^{K-1}\sum_{q=1}^{2}
 (T_{t,q}-r_{t,q})^2
 \right]^{1/2}.
\]
Means and sample standard deviations are computed across the
five optimizer runs, with denominator four for the sample variance.
Figure shading denotes pointwise standard deviation of the
individually executed trajectories, rather than the response
to an averaged command sequence.

\begin{table}[h]
 \centering
 \caption{Executed 48-hour performance.
 Stochastic controllers report mean $\pm$ sample standard deviation
 across five runs; the other comparators each represent one trajectory.}
 \label{tab:exp-building-results}
 \small
 \setlength{\tabcolsep}{4pt}
 \begin{tabular}{@{}lcc@{}}
  \toprule
  Method & Total objective & Tracking RMSE (${}^\circ\mathrm C$) \\
  \midrule
  \ouralg-MPC
   & $26.969\pm0.054$
   & $0.07245\pm0.00043$ \\
  Budgeted CEM-MPC
   & $32.128\pm1.102$
   & $0.10922\pm0.00725$ \\
  Reference MPC
   & $26.672$
   & $0.06744$ \\
  Lowest-cost collected trajectory
   & $409.903$
   & $0.89068$ \\
  \bottomrule
 \end{tabular}
\end{table}

The collected trajectory's cumulative cost exceeds the displayed
vertical range in Figure~\ref{fig:exp-building-control};
the underlying values are retained without modification.
The comparison concerns the weighted control objective,
not an equivalent percentage reduction in electricity consumption.
The repeated runs measure optimizer variation in one operating
scenario; they do not establish generalization to different
buildings, weather conditions, or fitted datasets.

\subsection{Detailed Experimental Setup for Section~\ref{sec:experiment2}} \label{app:experiments2}

This study extends the Grassmann experiment in Section~\ref{sec:experiment1} to heterogeneous offline design tasks.
It retains the proximal-energy construction~\eqref{eq:prox-energy} and recursive anchoring of Algorithm~\ref{alg:prism}, while estimating objective guidance from recorded labels.

\paragraph{Data and evaluation.}
The fixed exports contain 32,898 TFBind8 sequences, 10,000 TFBind10 sequences, 17,014 Superconductor descriptor vectors, 10,004 Ant designs, 10,004 DKitty designs, and 441 ChEMBL token sequences.
Their raw dimensions are respectively 8, 10, 86, 60, 56, and 31. Both TFBind tasks use four symbols per position.
Model and sampling choices use seeds 19--28, and reported scores use seeds 29--33 on the same datasets.
For each method and task, we select the eligible configuration with the highest mean batch maximum across seeds 19 to 28 and fix it before obtaining scores on the next five seeds.
Oracle values are used for this selection and final evaluation, but not for model fitting or checkpoint selection.
Each method submits 256 designs per task and confirmation seed, including repeated designs.
We retain the official Ant oracle's coordinate clipping before simulation.

\paragraph{Offline learning.}
\ouralg uses grouped 80/10/10 training, checkpoint-selection, and diagnostic folds, keeping identical raw designs together.
Continuous coordinates are standardized, and categorical sequences use standardized log-odds coordinates for learned models and raw symbols for categorical transitions.
Transform and label-normalization statistics are fitted on the training fold.
The reward surrogate $\widehat r$ is the mean of three multilayer perceptrons, with rewards expressed in standardized training-label units.
The score model is parameterized through a noise-conditioned denoiser $D_\theta$, with
\begin{equation*}
  s_\theta(z,\sigma)=\frac{D_\theta(z,\sigma)-z}{\sigma^2},
\end{equation*}
and trained by Gaussian denoising regression~\cite{vincent2011connection}.
All networks have two hidden layers of width 128 and use AdamW with learning rate $10^{-3}$, zero weight decay, and batch size 1,024.
Checkpoints are selected using offline validation only.
Rewards use squared error except on TFBind10, where pairwise ranking plus $0.1$ times squared error is fitted uniformly over distinct designs with duplicate-mean labels. Checkpoint selection minimizes top-decile regret.

\begin{table}[t]
  \centering
  \caption{Allocated optimizer updates per seed. Reward updates apply to each of the three ensemble members. Broad score noise is log-uniform on $[0.005,4]$, mixture noise combines that law equally with log-uniform noise on $[0.01,0.5]$. The selected TFBind10, Ant, and ChEMBL samplers do not use the trained neural score model.}
  \label{tab:matched-training}
  \small
  \begin{tabular}{@{}lrrl@{}}
    \toprule
    Task & Score updates & Reward updates/member & Score noise \\
    \midrule
    TFBind8 & 9,227 & 2,632 & Broad \\
    TFBind10 & 8,847 & 2,561 & Mixture \\
    Superconductor & 6,728 & 2,089 & Broad \\
    Ant & 7,482 & 2,273 & Mixture \\
    DKitty & 7,626 & 2,306 & Broad \\
    ChEMBL & 4,386 & 1,355 & Broad \\
    \bottomrule
  \end{tabular}
\end{table}

\paragraph{Kernel approximation.}
Let $r$ denote reward in standardized training-label units, $f=-r$, and $\widehat f=-\widehat r$.
For recipes using Euclidean proximity, the intended model-based target has the form
\begin{equation*}
  Q_{\lambda,\tau}^{\widehat f,\nu}(x,dz)
  \propto \exp\!\left\{\frac{\widehat r(z)-\|z-x\|^2/(2\lambda)}{\tau}\right\}
  \nu(dz),
\end{equation*}
where $\nu$ is the reference law, using the notation of~\eqref{eq:stat-learned-target}.
For reverse proposals, the learned scores approximate Gaussian-noised scores of the empirical training law. For Langevin refinement, the reference is Gaussian-smoothed.
The implemented law $\widehat Q$ additionally reflects finite sampling and numerical approximations.
Initial anchors follow a random permutation with the highest-label fifth of distinct training designs placed first, using duplicate-mean labels.
We take the required batch from this ordering and repeat it only after exhausting the distinct designs.
The anchor-only control submits the initial batch without updates.
Each selected recipe performs four outer anchor updates, with $\tau=0.1$ throughout.
References are distinct training designs weighted by their duplicate-row frequencies.

\emph{TFBind8 and Superconductor: reverse Euler.}
The kernel factorization
\begin{equation*}
\begin{aligned}
  Q_{\lambda,\tau}^{\widehat f,\nu}(x,dz)
  &\propto e^{\widehat r(z)/\tau}K_{\lambda\tau}^{\nu}(x,dz),\\
  K_v^{\nu}(x,dz)&\propto e^{-\|z-x\|^2/(2v)}\nu(dz),
\end{aligned}
\end{equation*}
places the quadratic anchoring in a conditional-reference proposal and the objective tilt in endpoint resampling.
Let $v=\lambda\tau$ and $t_0=v>\cdots>t_{64}=10^{-4}$ be a geometric variance grid.
For each anchor, 32 independent particles start at $z_0=x$ and follow
\begin{equation*}
  z_{j+1}=z_j+(t_j-t_{j+1})s_\theta(z_j,\sqrt{t_j})
       +\sqrt{t_j-t_{j+1}}\,\xi_j,
  \qquad \xi_j\sim N(0,I).
\end{equation*}
In continuous time down to zero variance, reversing Brownian noising with the exact reference score samples $K_v^{\nu}(x,\cdot)$, the reference law conditioned on a noisy observation equal to the anchor~\cite{song2021sde}.
Learned scores, finite integration, and the positive terminal variance approximate this construction.
A terminal denoising step uses $D_\theta(z_{64},0.01)$, motivated by the Gaussian posterior-mean identity~\cite{efron2011tweedie}.
One endpoint is resampled with weights proportional to $\exp(\widehat r/\tau)$.
TFBind8 endpoints are decoded before weighting and forming the next anchor.
TFBind8 uses $\lambda=40$ ($\sigma_{\rm start}=2$), and Superconductor uses $\lambda=10$ ($\sigma_{\rm start}=1$) and clips denoiser predictions to the training coordinate box.
The box constraint alone does not ensure physical feasibility.

\emph{TFBind10: categorical MH and portfolio selection.}
For references $r_i$ with masses $w_i$, the mutation-mixture law is
\begin{equation*}
  p_\epsilon(z)=\sum_i w_i\prod_{\ell=1}^{10}
       \left[(1-\epsilon)\mathbf 1\{z_\ell=r_{i\ell}\}+\epsilon/4\right].
\end{equation*}
We use $\epsilon=0.05$, $\lambda=2.5$, and the target
$p_\epsilon(z)\exp(\widehat r(z)/\tau-2d_H(z,x))$.
Eight chains per anchor make 64 symmetric single-site substitutions per outer round, with the Metropolis acceptance rule~\cite{hastings1970montecarlo}.
States are archived every four steps, retaining at most 4,096 distinct designs by surrogate rank.
The output reserves 128 distinct training designs with the highest duplicate-mean labels and fills 128 slots from the archive by surrogate rank.
Thus the final portfolio is not an unmodified sample from the MH target.

\emph{Ant and DKitty: Langevin refinement.}
For Ant, $\widehat p_h(z)=\sum_i w_i\phi_{h^2I}(z-r_i)$ is a Gaussian KDE over 4,096 randomly selected distinct training references~\cite{rosenblatt1956remarks}.
At each bandwidth, the anchored log-target gradient is
\begin{equation}
  b_{h,x}(z)=\nabla\log\widehat p_h(z)
       +\frac{\nabla\widehat r(z)}{\tau}
       -\frac{z-x}{\lambda\tau}.
  \label{eq:matched-langevin}
\end{equation}
With $\lambda=60$, eight geometric bandwidths decrease from $0.5$ to $0.05$, using 16 MALA steps per bandwidth and one chain per anchor.
The proposal is $z'=z+\eta b_{h,x}(z)+\sqrt{2\eta}\,\xi$, where $\eta=0.1\min(h^2,\lambda\tau)$, followed by the usual proposal-density acceptance correction~\cite{roberts1996exponential}.
For DKitty, the KDE score in~\eqref{eq:matched-langevin} is replaced by $s_\theta(z,0.01)$, with $\lambda=2.5$.
Each of 32 chains per anchor makes 64 ULA steps of size $10^{-5}$ per round, without acceptance correction.
Both implementations select terminal chains uniformly, since objective guidance is already included in their transitions.

\emph{ChEMBL: exact empirical reference.}
This recipe has the form of the index sampler~\eqref{eq:stat-indexed-sampler}, aggregating duplicate observations, and substituting the learned objective for recorded reward.
For $\widehat\mu=\sum_i w_i\delta_{r_i}$, the component probabilities are
\begin{equation*}
  \pi_i(x)\propto w_i\exp\!\left\{
      \frac{\widehat r(r_i)-\|r_i-x\|^2/(2\lambda)}{\tau}\right\}.
\end{equation*}
The implementation uses the full training reference, $\lambda=2.5$, and 32 independent component draws per anchor with uniform terminal selection. Every generated molecule belongs to the observed training support.

\paragraph{Connection to the analytical guarantees.}
For a Euclidean recipe with a reference law $\nu$, at any fixed anchor and parameter pair, the triangle inequality separates the discrepancies as
\begin{equation*}
\begin{aligned}
 W_2(\widehat Q,Q^{f,\mu})
 &\le \underbrace{W_2(\widehat Q,Q^{\widehat f,\nu})}_{\text{sampling and numerical error}}\\
 &\quad+\underbrace{W_2(Q^{\widehat f,\nu},Q^{\widehat f,\mu})}_{\text{reference estimation error}}\\
 &\quad+\underbrace{W_2(Q^{\widehat f,\mu},Q^{f,\mu})}_{\text{objective prediction error}},
\end{aligned}
\end{equation*}
with the shared $(x,\lambda,\tau)$ suppressed.
Theorem~\ref{thm:finite-data} and Corollary~\ref{cor:stat-empirical-conditional} give sufficient reference-to-kernel accuracy conditions for their specified constructions.
The Gaussian KDE is different from the compactly supported smoothing used in that theorem.
Theorem~\ref{thm:descent} explains why temperature and conditional error limit descent resolution, while Proposition~\ref{prop:feasibility} links that error to manifold adherence.
We do not estimate a uniform $W_2$ error or establish the required smooth-manifold assumptions on these benchmarks.
TFBind10 also changes proximity to Hamming distance and adds portfolio selection.
Our fixed positive temperature and four-round runs therefore measure finite-budget design quality, rather than the cooling and stationarity regime of Theorem~\ref{thm:finite-time}.

\paragraph{Baselines and scope.}
We rerun gradient ascent, CMA-ES~\cite{hansen1996covariance}, GP-qEI~\cite{ginsbourger2010kriging,balandat2020botorch}, REINFORCE~\cite{williams1992reinforce}, CbAS~\cite{brookes2019conditioning}, MINs~\cite{kumar2020modelinversion}, COMs~\cite{trabucco2021conservative}, DDOM~\cite{krishnamoorthy2023diffusion}, and DiffOPT~\cite{kong2024diffusion}.
These are pinned reruns with runtime compatibility adaptations.
Author-default models for classical methods and DDOM use seeded 90/10 row splits. DiffOPT retains its release's ordered 90/10 split and full-export normalization. Shared models use \ouralg's folds and transforms.
DDOM uses 1,000 epochs and batch size 128 with baseline optimization and generation hyperparameters retaining their documented native configurations.
COMs uses the author README's concrete configuration for all tasks.
DDOM follows the author-code weighting and sampling formulas with the stated training budget. DiffOPT follows the released guided reverse sampler.
Offline exports, oracle assets, confirmation seed IDs, final query counts, and compatible model options are shared to ensure fairness.
The runs use eight RTX 5090 GPUs, \ouralg training uses PyTorch 2.11 with CUDA 12.8, while classical baseline ports use TensorFlow.

\paragraph{Shared approximation models.}
For each task, the baseline menu includes author-default models and all compatible \ouralg reward and reference-model options.
Gradient ascent, CMA-ES, REINFORCE, CbAS, and MINs can substitute the frozen reward ensemble. GP-qEI uses it for surrogate evaluations within its Bayesian-optimization loop.
For methods requiring predictive probabilities, a Gaussian scale combines ensemble variance and offline validation error.
COMs can initialize its predictor from the shared ensemble before conservative training.
DDOM can initialize its conditional denoiser from the shared unconditional score before conditional training, or use a label-conditioned empirical Gaussian-mixture score, retaining classifier-free guidance.
DiffOPT can replace its reward, reference score, or both, using learned or exact empirical Gaussian-mixture scores converted to its variance-preserving noise convention, and reward guidance on clean or denoised inputs.
Shared neural models are fitted independently for each confirmation seed.

\end{document}